# Neural Methods for Effective, Efficient, and Exposure-Aware Information Retrieval

*Bhaskar Mitra*

A dissertation submitted in partial fulfillment

of the requirements for the degree of

**Doctor of Philosophy**

of

**University College London**.

Department of Computer Science

University College London

March 19, 2021

I, Bhaskar Mitra, confirm that the work presented in this thesis is my own. Where information has been derived from other sources, I confirm that this has been indicated in the work.

*Dedicated to Ma and Bapu*

# Abstract

Neural networks with deep architectures have demonstrated significant performance improvements in computer vision, speech recognition, and natural language processing. The challenges in information retrieval (IR), however, are different from these other application areas. A common form of IR involves ranking of documents—or short passages—in response to keyword-based queries. *Effective* IR systems must deal with query-document vocabulary mismatch problem, by modeling relationships between different query and document terms and how they indicate relevance. Models should also consider lexical matches when the query contains rare terms—such as a person's name or a product model number—not seen during training, and to avoid retrieving semantically related but irrelevant results. In many real-life IR tasks, the retrieval involves extremely large collections—such as the document index of a commercial Web search engine—containing billions of documents. *Efficient* IR methods should take advantage of specialized IR data structures, such as inverted index, to efficiently retrieve from large collections. Given an information need, the IR system also mediates how much exposure an information artifact receives by deciding whether it should be displayed, and where it should be positioned, among other results. *Exposure-aware* IR systems may optimize for additional objectives, besides relevance, such as parity of exposure for retrieved items and content publishers.

    In this thesis, we present novel neural architectures and methods motivated by the specific needs and challenges of IR tasks. We ground our contributions with a detailed survey of the growing body of neural IR literature. Our key contribution towards improving the *effectiveness* of deep ranking models is developing the

Duet principle which emphasizes the importance of incorporating evidence based on both patterns of exact term matches and similarities between learned latent representations of query and document. To *efficiently* retrieve from large collections, we develop a framework to incorporate query term independence into any arbitrary deep model that enables large-scale precomputation and the use of inverted index for fast retrieval. In the context of stochastic ranking, we further develop optimization strategies for exposure-based objectives. Finally, this dissertation also summarizes our contributions towards benchmarking neural IR models in the presence of large training datasets and explores the application of neural methods to other IR tasks, such as query auto-completion.

# Impact Statement

The research presented in this thesis was conducted while the author was employed at Bing, a commercial web search engine. Many of the research questions investigated here have consequently been motivated by real world challenges in building large scale retrieval systems. The insights gained from these studies have informed, and continues to influence, the design of retrieval models in both industry and academia.

Duet was the first to demonstrate the usefulness of deep representation learning models for document ranking. Since then, the academic community has continued to build on those early results, culminating in large improvements in retrieval quality from neural methods over traditional IR approaches at the TREC 2019 Deep Learning track. Many of these deep models have subsequently been deployed at the likes of Bing and Google. Similarly, our foray into efficient neural architectures is both of academic interest and critical to increasing the scope of impact of these computation-heavy models.

Search systems do not just exist in laboratory environments but are inherently sociotechnical instruments that mediate what information is accessible and consumed. This presents enormous responsibility on these systems to ensure that retrieved results are representative of the collections being searched, and that the retrieval is performed in a manner fair to both content producers and consumers. Lack of attention to these facets may lead to serious negative consequences—*e.g.*, the formation of information filter bubbles or visible demographic bias in search results. The capability to directly optimize for expected exposure may be key to addressing some of these concerns.

Finally, to accurately qualify the impact of our research, we must look beyond published empirical and theoretical results for contributions. Science is not an individual endeavour, and therefore any evaluation of research must also encompass the impact of the artifacts produced by said research on its immediate field of study and the academic community around it. A more meaningful evaluation of our contributions, therefore, requires juxtaposition of how the field of neural information retrieval has evolved during the course of this research and have been supported directly by our work. In 2016, when neural IR was still an emerging area, we organized the first workshop focused on this topic at the ACM SIGIR conference. That year, approximately 8% of the published papers at SIGIR were related to this topic. In contrast, this year at the same conference about 79% of the publications employed neural methods. Our research has directly contributed to this momentum in several ways—including, building standard task definition and benchmarks in the form of MS MARCO and the TREC Deep Learning track, which has been widely adopted as the primary benchmark by the community working on deep learning for search. More recently we have also released a large user behavior dataset, called ORCAS, that may enable training even larger and more sophisticated neural retrieval models. We have organized several workshops and tutorials to bring together researchers whose work cuts across information retrieval, natural language processing, and machine learning domains to build a community around this topic. Our early survey of neural IR methods also resulted in an instructive manuscript that both summarized important progress and charted out key directions for the field.

# Acknowledgements

It takes a proverbial village to teach the skills necessary to do good science and become an independent researcher. So, when I pause and look back at my academic journey thus far, I am filled with gratitude towards every person who professionally and personally supported, mentored, and inspired me on this journey. This thesis is as much a product of the last four years of my own research as it is the fruit of the time and labor that others invested in me on the way.

I am grateful to my supervisor Emine Yilmaz not just for her mentorship and guidance during my PhD but also for her initial encouragement to pursue a doctorate. I remember my apprehension about pursuing a part-time degree while employed full-time. Emine did not just believe that I could succeed in this endeavour, but also helped me find the confidence to pursue it. My PhD journey would not even have started without her support and guidance. I am grateful to her for placing that enormous trust in me and then guiding me with patience and kindness. I thank her for the countless technical discussions and insights, and for all the close collaboration including organizing the TREC Deep Learning track.

A few years before the start of my PhD journey, while I was still a ranking engineer at Bing, I walked into Nick Craswell's office one day and expressed my interest to join his applied science team in Cambridge (UK) to "learn the ropes" for how to do research. The trust Nick placed in me that day when he encouraged me to pursue that dream, altered the course of my professional career. Seven years later, I am still proud to be part of Nick's team and grateful for having grown under his mentorship. Nick has constantly encouraged and supported, and often gone out of his way to do so, my dreams of pursuing a career in research. I co-authored my first book with Nick. We co-organized the first deep learning workshop at SIGIR together. We have collaborated on numerous projects and papers over the years. His insightful comments and thoughtful feedback have always served as important lessons for me on how to do good science.

I moved to Cambridge in the summer of 2013 with limited understanding of what it takes to be a researcher or to publish. I am grateful that in those early formative years of my research career, Milad Shokouhi, Filip Radlinkski, and Katja Hofmann took me under their wings. I co-authored my first two peer reviewed publications with the three of them. All the advice and lessons I received during that time stuck with me and continues to influence and guide my research to this day. I am grateful to all three of them for their patient and thoughtful mentorship.

I am indebted to Fernando Diaz for his invaluable mentorship over the years and for the many critical lessons on doing good science and good scholarship. All the projects or publications we have collaborated on have been incredibly instructive. Fernando continues to shape my research agenda and my personal vision of what kind of a researcher I want to be. I want to thank Fernando for the enormous trust that he placed in me by taking me under his wings. I have cherished him both as a mentor and a collaborator, and hope for continued collaborations in the future.

David Hawking deserves a special mention in this thesis. I was fortunate to receive David's mentorship around the time when I was still juggling personal doubts on whether (and how) to do a PhD. His recommendations and guidance had a strong influence on my eventual plans for the doctorate and even how I structure this thesis. David has made no secret of his eagerness for the day when he can finally congratulate me on a successful completion of my PhD. In fact, he has expressed that excitement since the day University College London (UCL) accepted my PhD application. So, not only am I filled with immense gratitude for David's guidance in helping me realize this important professional milestone, but I am also excited about finally celebrating this day with him that has been four years in the making.

Over the years, I have been privileged to collaborate with and receive coaching from many other incredible members of the research community. Susan Dumais is thanks goes out to Sarah Turnbull for her vital support in enabling me to pursue this doctorate as a part-time student, and as a remote student for the final year.

I am thankful to Claudia Hauff and Mounia Lalmas, who served as examiners for my final viva examination, for their invaluable feedback and advice that helped me to significantly improve this manuscript.

I am grateful to Mona Soliman Habib and Kathleen McGrow for their vital encouragement to pursue a doctorate long before I started on this journey.

Finally, my parents have sacrificed more towards my education and academic training than I personally can in this lifetime. I am grateful for the gifts of patience, kindness, and love that I received from them while growing up. My parents are without doubt the two happiest people I have ever met and I am grateful that they infected me (a little) with their undiluted enthusiasm for life. I thank them for their love and support, but most importantly for teaching me what is important in life.

I want to conclude by acknowledging the unprecedented circumstances facing the world at the time of writing this thesis. The world is still gripped with the coronavirus disease 2019 (COVID-19) pandemic. I am grateful to all the medical and essential workers who are risking their lives daily to keep everyone safe and alive, and I grieve with everyone who have lost loved ones during this pandemic. At the same time, millions of voices are protesting on the streets calling for gender equality, racial justice, immigration justice, and climate justice in the face of unprecedented existential threats to our planet. Many of us who work in science and technology, are reevaluating the impact and influence of our own work on society and the world. These are uncertain times but also a historic moment when people around the world are coming together in unprecedented numbers fighting to build a kinder and more just world for everyone, and for that I am most grateful.

# Contents

# List of Figures

# List of Tables

# Chapter 1

# Introduction

Over the last decade, there have been dramatic improvements in performance in computer vision, speech recognition, and machine translation tasks, witnessed in research and in real-world applications [20–24]. These breakthroughs were largely fuelled by recent advances in neural network models, usually with multiple hidden layers, known as deep architectures combined with the availability of large datasets [25] and cheap compute power for model training. Exciting novel applications, such as conversational agents [26, 27], have also emerged, as well as game-playing agents with human-level performance [28, 29]. Work has now begun in the information retrieval (IR) community to apply these neural methods, leading to the possibility of advancing the state of the art or even achieving breakthrough performance as in these other fields.

Retrieval of information can take many forms [30]. Users can express their information need in the form of a text query—by typing on a keyboard, by selecting a query suggestion, or by voice recognition—or the query can be in the form of an image, or in some cases the need can be implicit. Retrieval can involve ranking existing pieces of content, such as documents or short-text answers, or composing new responses incorporating retrieved information. Both the information need and the retrieved results may use the same modality (*e.g.*, retrieving text documents in response to keyword queries), or be different (*e.g.*, image search using text queries). The information within the document text may be semi-structured, and the organization scheme may be shared between groups of documents in the collection—*e.g.*, web pages from the same domain [31]. If the query is ambiguous, retrieval system may consider user history, physical location, temporal changes in information, or other context when ranking results. IR systems may also help users formulate their intent (*e.g.*, via query auto-completion or query suggestion) and can extract succinct summaries of results that take the user's query into account. *Neural IR* refers to the application of shallow or deep neural networks to these retrieval tasks.

We note that many natural language processing (NLP) tasks exist that are not IR. Machine translation of text from one human language to another is not an IR task. However, translation could be used in an IR system, to enable cross-language retrieval on a multilingual corpus [32]. Inferring attributes of a named entity [33], from text or graph-structured data, is not an IR task in itself. However, an IR system could use inferred entity attributes to enhance its performance on IR tasks. In general, many NLP tasks do not involve information access and retrieval, so are not IR tasks, but some can still be useful as part of a larger IR system.

In this thesis, we focus on neural methods that employ deep architectures to retrieve and rank documents in response to a query, an important IR task. A search query may typically contain a few terms, while the document length, depending on the scenario, may range from a few terms to hundreds of sentences or more. Neural models for IR use learned latent representations of text, and usually contain a large number of parameters that need to be tuned. ML models with large set of parameters typically benefit from large quantity of training data [34–38]. Unlike traditional *learning to rank* (LTR) approaches [39] that train ML models over a set of hand-crafted features, recent neural models for IR typically accept the raw text of a query and document as input. Learning suitable representations of text also demands large-scale datasets for training [7]. Therefore, unlike classical IR models, these neural approaches tend to be data hungry, with performance that improves with more training data.

In other fields, the design of neural network models has been informed by characteristics of the application and data. For example, the datasets and successful architectures are quite different in visual object recognition, speech recognition, and game playing agents. While IR shares some common attributes with the field of NLP, it also comes with its own set of unique challenges. *Effective* IR systems must deal with query-document vocabulary mismatch problem, by modeling relationships between different query and document terms and how they indicate relevance. Models should also consider lexical matches when the query contains rare terms—such as a person's name or a product model number—not seen during training, and to avoid retrieving semantically related but irrelevant results. In many real-life IR tasks, the retrieval involves extremely large collections—such as the document index of a commercial Web search engine—containing billions of documents. *Efficient* IR methods should take advantage of specialized IR data structures, such as inverted index, to efficiently retrieve from large collections. Given an information need, the IR system also mediates how much exposure an information artifact receives by deciding whether it should be displayed, and where it should be positioned, among other results. *Exposure-aware* IR systems may optimize for additional objectives, besides relevance, such as parity of exposure for retrieved items and content publishers.

In our work, we focus on methods using deep neural networks for document ranking, and to a lesser extent other retrieval tasks. We identify key challenges and principles which motivate our design of novel neural approaches to ranking. We study these proposed methods with respect to retrieval quality, query response time, and exposure disparity. In Section 1.1, we summarize our key contributions. The remainder of this chapter is dedicated to describing the problem formulation. In Section 1.2, we provide an overview of the IR tasks that we use for evaluation. In Section 1.3, we describe a set of common notations that we use in the remainder of this thesis. Finally, we describe relevant IR metrics in Section 1.4.

## 1.1 Contributions

In this thesis, we explore neural network based approaches to IR. This section summarizes the key research contributions of this thesis by chapter. Where appropriate, we also cite the publications that forms the basis of that chapter.

- **Chapter 1-3** are based on the book by Mitra and Craswell [1]—and corresponding tutorials [2–6]. The current chapter introduces key IR tasks, evaluation metrics, and mathematical notations that are referenced throughout in this thesis. Chapter 2 presents our motivation for exploring neural IR methods. Chapter 3 provides a survey of existing literature on neural and traditional non-neural approaches to IR. Key concepts related to IR models and neural representation learning are explained. These three chapters have no novel theoretical or algorithmic contributions, but provides a detailed overview of the field that also serves as the background for the remaining sections.

- **Chapter 4** is based on [7–10] and emphasizes the importance of incorporating evidence based on both patterns of exact query term matches in the document as well as the similarity between query and document text based on learned latent representations for retrieval. We operationalize this principle by proposing a deep neural network architecture, called Duet, that jointly learns two deep neural networks focused on matching using lexical and latent representations of text, respectively. We benchmark the proposed model on: (i) Bing document ranking task, (ii) TREC Complex Answer Retrieval task, (iii) MS MARCO passage ranking task, and (iv) TREC 2019 Deep Learning track document and passage ranking tasks and demonstrate that estimating relevance by inspecting both lexical and latent matches performs better than considering only one of those aspects for retrieval.

- **Chapter 5** is based on [13, 40] and studies neural methods in the context of retrieval from the full collection, instead of just reranking. In particular, we study the impact of incorporating the query term independence (QTI) assumption in neural architectures. We find that incorporating QTI assumption in several deep neural ranking models results in minimal (or no) degradation in ranking effectiveness. However, under the QTI assumption, the learned ranking functions can be combined with specialised IR data structures, *e.g.*, inverted index, for fast and scalable candidate generation in the first stage of retrieval. We benchmark on: (i) MS MARCO passage ranking task and

(ii) TREC 2019 Deep Learning track to demonstrate that neural methods can be employed for more effective but also efficient candidate generation.

- **Chapter 6** is based on [14] and studies learning to rank with neural networks in the context of stochastic ranking. Due to presentation bias, a static ranked list of results may cause large difference in exposure of items with similar relevance. We present a stochastic ranking framework that can optimize towards exposure targets under different constraints—*e.g.*, individual and group exposure parity. While the original study [14] is a collaborative project focusing on the expected exposure metric, this chapter summarizes our key contributions related to the framework of model optimization for individual and groupwise parity of expected exposure.

- **Chapter 7**, based on [18, 19], looks at the application of neural IR beyond ad hoc retrieval—to the query auto-completion (QAC) task. The ranking task, in case of QAC, poses challenges that are different from those in query-document or query-passage matching. In this chapter, we study two applications of deep models for QAC: (i) Recommending completions for rare query prefixes and (ii) modeling query reformulations for session context-aware QAC.

- **Chapter 8** summarizes findings from our recent efforts on large-scale benchmarking of deep neural IR methods at TREC [15]. The TREC Deep Learning track [15, 16] provides a strict blind evaluation for IR methods that take advantage of large supervised training datasets, and have been instrumental in demonstrating the superior retrieval quality for many recent neural methods proposed by the research community.

- Finally, in **Chapter 9**, we conclude with a discussion on the future of neural IR research. In this chapter, we reflect on the progress we have already made as a field and provide some personal perspectives on the road ahead.

## 1.2 Evaluation tasks

We focus on text retrieval in IR, where the user enters a text query and the system returns a ranked list of search results. Search results may be passages of text or full text documents. The system's goal is to rank the user's preferred search results at the top. This problem is a central one in the IR literature, with well-understood challenges and solutions.

Text retrieval methods for full text documents and for short text passages have application in ad hoc retrieval systems and question answering systems, respectively. We describe these two tasks in this section.

### 1.2.1 Ad hoc retrieval

Ranked document retrieval is a classic problem in information retrieval, as in the main task of the Text Retrieval Conference [41], and performed by commercial search engines such as Google, Bing, Baidu, and Yandex. TREC tasks may offer a choice of query length, ranging from a few terms to a few sentences, whereas search engine queries tend to be at the shorter end of the range. In an operational search engine, the retrieval system uses specialized index structures to search potentially billions of documents. The results ranking is presented in a search engine results page (SERP), with each result appearing as a summary and a hyperlink. The engine can instrument the SERP, gathering implicit feedback on the quality of search results such as click decisions and dwell times.

A ranking model can take a variety of input features. Some ranking features may depend on the document alone, such as how popular the document is with users, how many incoming links it has, or to what extent document seems problematic according to a Web spam classifier. Other features depend on how the query matches the text content of the document. Still more features match the query against document metadata, such as referred text of incoming hyperlink anchors, or the text of queries from previous users that led to clicks on this document. Because anchors and click queries are a succinct description of the document, they can be a useful source of ranking evidence, but they are not always available. A newly created document would not have much link or click text. Also, not every document is popular enough to have past links and clicks, but it still may be the best search result for a user's rare or tail query. In such cases, when text metadata is unavailable, it is crucial to estimate the document's relevance primarily based on its text content.

In the text retrieval community, retrieving documents for short-text queries by considering the long body text of the document is an important challenge. These *ad hoc* retrieval tasks have been an important part of the Text REtrieval Conference (TREC) [42], starting with the original tasks searching newswire and government documents, and later with the Web track[1] among others. The TREC participants are provided a set of, say fifty, search queries and a document collection containing 500-700K newswire and other documents. Top ranked documents retrieved for each query from the collection by different competing retrieval systems are assessed by human annotators based on their relevance to the query. Given a query, the goal of the IR model is to rank documents with better assessor ratings higher than the rest of the documents in the collection. In Section 1.4, we describe standard IR metrics for quantifying model performance given the ranked documents retrieved by the model and the corresponding assessor judgments for a given query.

### 1.2.2 Question-answering

Question-answering tasks may range from choosing between multiple choices (typically entities or binary true-or-false decisions) [43–46] to ranking spans of text or passages [47–51], and may even include synthesizing textual responses by gathering evidence from one or more sources [52, 53]. TREC question-answering experiments [47] has participating IR systems retrieve spans of text, rather than documents, in response to questions. IBM's DeepQA [51] system—behind the Watson project that famously demonstrated human-level performance on the American TV quiz show, "Jeopardy!"—also has a primary search phase, whose goal is to find as many potentially answer-bearing passages of text as possible. With respect to the question-answering task, the scope of this thesis is limited to ranking answer containing passages in response to natural language questions or short query texts.

Retrieving short spans of text pose different challenges than ranking docu-ments. Unlike the long body text of documents, single sentences or short passages tend to be on point with respect to a single topic. However, answers often tend to use different vocabulary than the one used to frame the question. For example, the span of text that contains the answer to the question "what year was Martin Luther King Jr. born?" may not contain the term "year". However, the phrase "what year" implies that the correct answer text should contain a year—such as '1929' in this case. Therefore, IR systems that focus on the question-answering task need to model the patterns expected in the answer passage based on the intent of the question.

---

[1] `http://www10.wwwconference.org/cdrom/papers/317/node2.html`

The focus of this thesis is on ad hoc retrieval, and to a lesser extent on question-answering. However, neural approaches have shown interesting applications to other existing retrieval scenarios, including query recommendation [54], modelling diversity [55], modelling user click behaviours [56], entity ranking [57, 58], knowledge-based IR [59], and even optimizing for multiple IR tasks [60]. In addition, recent trends suggest that advancements in deep neural networks methods are also fuelling emerging IR scenarios such as proactive recommendations [61–63], conversational IR [64, 65], and multi-modal retrieval [66]. Neural methods may have an even bigger impact on some of these other IR tasks. To demonstrate that neural methods are useful in IR—beyond the document and passage ranking tasks—we also present, in this thesis, a brief study on employing deep models for the QAC task in Chapter 7.

## 1.3  Notation

We adopt some common notation for this thesis shown in Table 1.1. We use lower-case to denote vectors (*e.g.*, $\vec{x}$) and upper-case for tensors of higher dimensions (*e.g.*, $X$). The ground truth $rel_q(d)$ in Table 1.1 may be based on either manual relevance annotations or be implicitly derived from user behaviour on SERP (*e.g.*, from clicks).

**Table 1.1:** Notation used in this thesis.

| Meaning | Notation |
| --- | --- |
| Single query | $q$ |
| Single document | $d$ |
| Set of queries | $Q$ |
| Collection of documents | $D$ |
| Term in query $q$ | $t_q$ |
| Term in document $d$ | $t_d$ |
| Full vocabulary of all terms | $T$ |
| Set of ranked results retrieved for query $q$ | $R_q$ |
| Result tuple (document $d$ at rank $i$) | $\langle i,d \rangle$, where $\langle i,d \rangle \in R_q$ |
| Relevance label of document $d$ for query $q$ | $rel_q(d)$ |
| $d_i$ is more relevant than $d_j$ for query $q$ | $rel_q(d_i) > rel_q(d_j)$, or $d_i \succ_q d_j$ |
| Frequency of term $t$ in document $d$ | $tf(t,d)$ |
| Number of documents that contain term $t$ | $df(t)$ |
| Vector representation of text $z$ | $\vec{v}_z$ |
| Probability function for an event $\mathcal{E}$ | $p(\mathcal{E})$ |

## 1.4 Metrics

A large number of IR studies [67–74] have demonstrated that users of retrieval systems tend to pay attention mostly to top-ranked results. IR metrics, therefore, focus on rank-based comparisons of the retrieved result set $R$ to an ideal ranking of documents, as determined by manual judgments or implicit feedback from user behaviour data. These metrics are typically computed at a rank position, say $k$, and then averaged over all queries in the test set. Unless otherwise specified, $R$ refers to the top-$k$ results retrieved by the model. Next, we describe a few standard metrics used in IR evaluations.

**Precision and recall** Precision and recall both compute the fraction of relevant documents retrieved for a query $q$, but with respect to the total number of documents in the retrieved set $R_q$ and the total number of relevant documents in the collection $D$, respectively. Both metrics assume that the relevance labels are binary.

$$Precision_q = \frac{\sum_{\langle i,d \rangle \in R_q} rel_q(d)}{|R_q|} \quad (1.1)$$

$$Recall_q = \frac{\sum_{\langle i,d \rangle \in R_q} rel_q(d)}{\sum_{d \in D} rel_q(d)} \quad (1.2)$$

**Mean reciprocal rank (MRR)** Mean reciprocal rank [75] is also computed over binary relevance judgments. It is given as the reciprocal rank of the first relevant document averaged over all queries.

$$RR_q = \max_{\langle i,d \rangle \in R_q} \frac{rel_q(d)}{i} \quad (1.3)$$

**Mean average precision (MAP)** The average precision [76] for a ranked list of documents $R$ is given by,

$$AveP_q = \frac{\sum_{\langle i,d \rangle \in R_q} Precision_{q,i} \times rel_q(d)}{\sum_{d \in D} rel_q(d)} \quad (1.4)$$

Where, $Precision_{q,i}$ is the precision computed at rank $i$ for the query $q$. The average precision metric is generally used when relevance judgments are binary, although variants using graded judgments have also been proposed [77]. The mean of the average precision over all queries gives the MAP score for the whole set.

**Normalized discounted cumulative gain (NDCG)** There are a few different variants of the discounted cumulative gain ($DCG_q$) metric [78] which can be used when graded relevance judgments are available for a query $q$—say, on a five-point scale between zero to four. One incarnation of this metric is as follows.

$$DCG_q = \sum_{\langle i,d \rangle \in R_q} \frac{gain_q(d)}{\log_2(i+1)} \quad (1.5)$$

The ideal DCG (*IDCG$_q$*) is computed the same way but by assuming an ideal rank order for the documents up to rank *k*. The normalized DCG (*NDCG$_q$*) is then given by,

$$\text{N}DCG_q = \frac{DCG_q}{IDCG_q} \tag{1.6}$$

**Normalized cumulative gain (NCG)** A metric related to NDCG but suitable for evaluating the quality of retrieval for first stage candidate generation methods is NCG—*i.e.*, NDCG without the position discounting.

$$\text{C}G_q = \sum_{\langle i,d \rangle \in R_q} gain_q(d) \tag{1.7}$$

$$\text{N}CG_q = \frac{CG_q}{ICG_q} \tag{1.8}$$

NCG has been employed in the literature [15, 16, 79] to measure how much relevant items are recalled as part of candidate generation without paying attention to the exact order in which the candidates appear in the retrieved set.

# Chapter 2

# Motivation

We expect a good retrieval system to exhibit certain general attributes. We highlight some of them in this chapter. The design of any neural methods for IR should be informed by these desired properties. We operationalize these intuitions later in Chapters 4-7. In this chapter, we also introduce a general taxonomy of neural approaches for document ranking by categorizing them based on the step of the retrieval process they influence. This discussion on a general taxonomy should serve as a common lens through which we can inspect both existing neural IR approaches as well as the new deep neural ranking models described in the rest of this thesis.

## 2.1 Desiderata of IR models

For any IR system, the relevance of the retrieved items to the input query is of foremost importance. But to evaluate the effectiveness of an IR system in isolation without considering critical dimensions, such as the efficiency of the system or its robustness to collections with different properties, can be tantamount to a theoretical exercise without practical usefulness. An IR system mediates what information its users are exposed to and consume. It is, therefore, also important to quantify and limit any systematic disparity that the retrieval system may inadvertently cause with respect to exposure of information artifacts of similar relevance, or their publishers. These concerns not only serve as yard sticks for comparing the different neural and non-neural approaches but also guide our model designs. Where appropriate, we connect these motivations to our contributions in this area, some of which form the basis for subsequent chapters in this thesis.

### 2.1.1 Semantic matching

Most traditional approaches to ad hoc retrieval count repetitions of the query terms in the document text. *Exact term matching* between query and document text, while simple, serves as a foundation for many IR systems. Different weighting and normalization schemes over these counts leads to a variety of TF-IDF models, such as BM25 [80]. However, by only inspecting the query terms the IR model ignores all the evidence of *aboutness* from the rest of the document. So, when ranking for the query "Australia" only the occurrences of "Australia" in the document are considered—although the frequency of other terms like "Sydney" or "kangaroo" may be highly informative [81, 82]. In the case of the query "what channel are the seahawks on today", the query term "channel" provides hints to the IR model to pay attention to the occurrences of "ESPN" or "Sky Sports" in the document text—none of which appears in the query itself.

For IR tasks, such as QAC, the lexical similarity between the input (*e.g.*, the query prefix) and candidate items (*e.g.*, the possible completions) is minimal. In such scenarios, understanding the relationship between the query prefix and suffix requires going beyond inspecting lexical overlap.

Semantic understanding, however, goes further than mapping query terms to document terms [83]. A good IR model may consider the terms "hot" and "warm" related, as well as the terms "dog" and "puppy"—but must also distinguish that a user who submits the query "hot dog" is not looking for a "warm puppy" [84]. At the more ambitious end of the spectrum, semantic understanding would involve logical reasoning by the IR system—so for the query "concerts during SIGIR" it associates a specific edition of the conference (the upcoming one) and considers both its location and dates when recommending concerts nearby during the correct week. These examples motivate that IR models should have some latent representations of intent as expressed by the query and of the different topics in the document text—so that *inexact matching* can be performed that goes beyond lexical term counting.

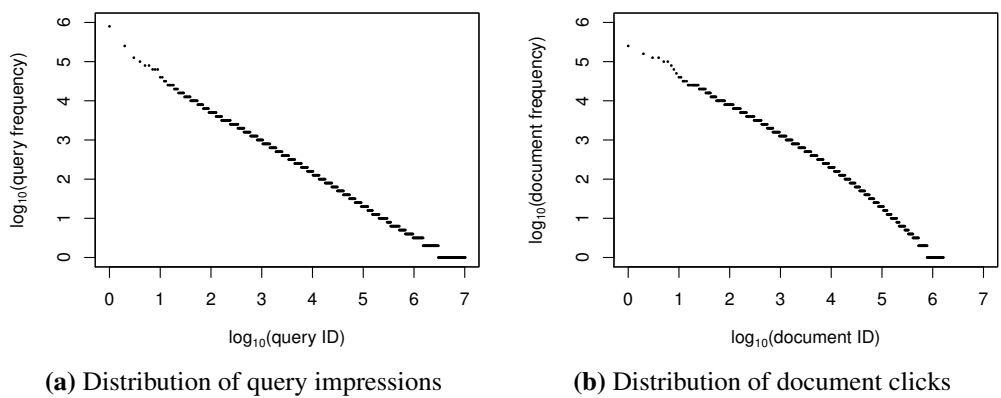

(a) Distribution of query impressions  (b) Distribution of document clicks

**Figure 2.1:** A Log-Log plot of frequency versus rank for query impressions and document clicks in the AOL query logs [85]. The plots highlight that these quantities follow a Zipfian distribution.

## 2.1.2 Robustness to rare inputs

Query frequencies in most IR tasks follow a Zipfian distribution [86] (see Figure 2.1). In the publicly available AOL query logs [85], for example, more than 70% of the distinct queries are seen only once in the period of three months from which the queries are sampled. In the same dataset, more than 50% of the distinct documents are clicked only once. Downey et al. [87] demonstrate that typical web search engines may struggle to retrieve these infrequently searched-for documents and perform poorer on queries containing terms that appear extremely rarely in its historical search logs. Improving the robustness of retrieval systems in the context of rare queries and documents is an important challenge in IR.

Many IR models that learn latent representations of text from data often naively assume a fixed size vocabulary. These models perform poorly when the query consists of terms rarely (or never) seen during training. Even if the model does not assume a fixed vocabulary, the quality of the latent representations may depend heavily on how often the terms under consideration appear in the training dataset. In contrast, *exact matching* models, like BM25 [80], can take advantage of the specificity [88–93] of rare terms to retrieve the few documents from the index that contains these terms. Kangassalo et al. [94] demonstrated that users of search systems also tend to consider term specificity in their query formulation process and may add rare terms to help discriminate between relevant and nonrelevant documents.

Semantic understanding in an IR model cannot come at the cost of poor retrieval performance on queries containing rare terms. When dealing with a query such as "pekarovic land company" the IR model will benefit from considering exact matches of the rare term "pekarovic". In practice an IR model may need to effectively trade-off exact and inexact matching for a query term. However, the decision of when to perform exact matching can itself be informed by semantic understanding of the context in which the terms appear in addition to the terms themselves.

### 2.1.3 Robustness to variable length text

Depending on the task, the IR system may be expected to retrieve documents, passages, or even short sequences consisting of only a few terms. The design of a retrieval model for long documents is likely to share some similarities to a passage or short text retrieval system, but also be different to accommodate distinct challenges associated with retrieving long text. For example, the challenge of vocabulary mismatch, and hence the importance of semantic matching, may be amplified when retrieving shorter text [95–97]. Similarly, when matching the query against longer text, it is informative to consider the positions of the matches [98–100], but may be less so in the case of short text matching. When specifically dealing with long text, the compute and memory requirements may be significantly higher for machine learned systems (*e.g.*, [101]) and require careful design choices for mitigation.

Typical text collections contain documents of varied lengths (see Figure 2.2). Even when constrained to document retrieval, a good IR system must be able to deal with documents of different lengths without over-retrieving either long or short documents [102, 103]. Relevant documents may also contain irrelevant sections, and the relevant content may either be localized, or spread over multiple sections in the document [104]. Document length normalization is well-studied in the context of IR models (*e.g.*, [92, 93, 102, 105, 106]), and this existing research should inform the design of any new IR models.

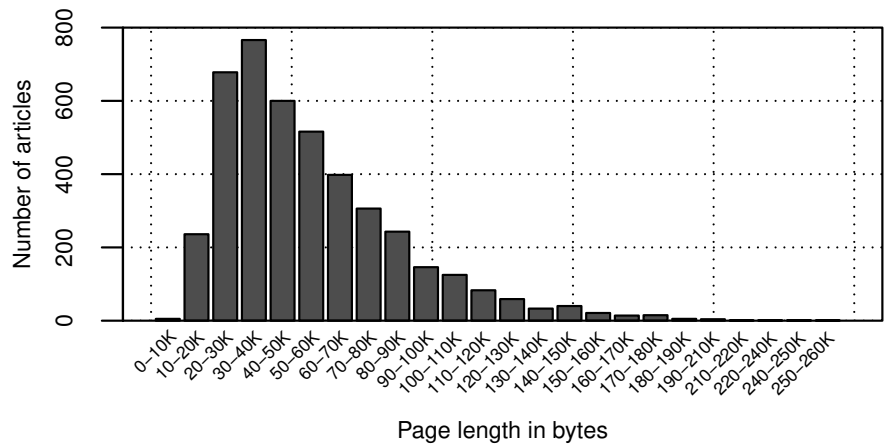

**Figure 2.2:** Distribution of document length (in bytes) of Wikipedia featured articles as of June 30, 2014. Source: `https://en.wikipedia.org/wiki/Wikipedia:Featured_articles/By_length`.

### 2.1.4 Efficiency

Efficiency is one of the salient points of any retrieval system [107–109]. A typical commercial Web search engine may deal with tens of thousands of queries per second[1]—retrieving results for each query from an index containing billions of documents. Search engines typically involve specialised data structures, such as inverted index, and large multi-tier architectures—and the retrieval process generally consists of multiple stages of pruning the candidate set of documents [110–112]. The IR model at the bottom of this *telescoping* setup may need to sift through billions of documents—while the model at the top may only need to re-rank between tens of promising documents. The retrieval approaches that are suitable at one level of the stack may be highly impractical at a different step—models at the bottom need to be *fast* but mostly focus on eliminating irrelevant or junk results, while models at the top tend to develop more sophisticated notions of *relevance*, and focus on distinguishing between documents that are much closer on the relevance scale. So far, much of the focus on neural IR approaches have been limited to re-ranking top-*n* documents which considerably constrains the impact of these methods.

---

[1] `http://www.internetlivestats.com/one-second/#google-band`

### 2.1.5 Parity of exposure

IR systems mediate what information its users are exposed to. Under presentation bias [67, 68, 72, 73, 113, 114], a static ranking may disproportionately distribute exposure between items of similar relevance raising concerns about producer-side fairness [115–118]. Exposure optimization has been proposed as a means of achieving fairness in ranking for individuals [116] or groups defined by sensitive attributes such as gender or race [117, 118]. Stochastic ranking policies that optimize for individual or group parity of exposure in expectation may be more appropriate under these settings.

### 2.1.6 Sensitivity to context

Retrieval in the wild can leverage many implicit and explicit context information [119–132]. The query "weather" may refer to the weather in Seattle or in London depending on where the user is located. An IR model may retrieve different results for the query "decorations" depending on the current season. The query "giants match highlights" may be better disambiguated if the system knows whether the user is a fan of baseball or American football, whether she is located on the East or the West coast of USA, or if the model has knowledge of recent sport fixtures. In conversational IR systems [133], the correct response to the question "When did she become the prime minister?" would depend on disambiguating the correct entity based on the context of references made in the previous turns of the conversation. In *proactive retrieval* scenarios [134–137], the retrieval can even be triggered based solely on implicit context without any explicit query submission from the user. Relevance in many applications is, therefore, situated in the user and task context, and is an important consideration in the design of IR systems.

### 2.1.7 Robustness to corpus variance

An interesting consideration for IR models is how well they perform on corpora whose distributions are different from the data that the model was trained on. Models like BM25 [80] have very few parameters and often demonstrate reasonable performance "out of the box" on new corpora with little or no additional tuning of parameters. Supervised deep learning models containing millions (or even billions) of parameters, on the other hand, are known to be more sensitive to distributional differences between training and evaluation data, and have been shown to be especially vulnerable to adversarial inputs [138]. The application of unsupervised term embeddings on collections and tasks that are different from the original data the representations were trained on is common in the literature. While these can be seen as examples of successful transfer learning, we also find evidence [139] that term embeddings trained on collections distributionally closer to the test samples perform significantly better.

Some of the variances in performance of deep models on new corpora is offset by better retrieval on the test corpus that is distributionally closer to the training data, where the model may have picked up crucial corpus specific patterns. For example, it may be understandable if a model that learns term representations based on the text of Shakespeare's Hamlet is effective at retrieving passages relevant to a search query from The Bard's other works, but performs poorly when the retrieval task involves a corpus of song lyrics by Jay-Z. However, the poor performances on new corpus can also be indicative that the model is overfitting, or suffering from the Clever Hans[2] effect [140]. For example, an IR model trained on recent news corpus may learn to associate "Theresa May" with the query "uk prime minister" and as a consequence may perform poorly on older TREC datasets where the connection to "John Major" may be more appropriate.

ML models that are hyper-sensitive to corpus distributions may be vulnerable when faced with unexpected changes in distributions in the test data. This can be particularly problematic when the test distributions naturally evolve over time due to underlying changes in the user population or behaviour. The models may need to be re-trained periodically or designed to be invariant to such changes (*e.g.*, [141]).

While this list of desired attributes of an IR model is in no way complete, it serves as a reference for comparing many of the neural and non-neural approaches described in the rest of this thesis.

---

[2]https://en.wikipedia.org/wiki/Clever_Hans

## 2.2 Designing neural models for IR

In the previous section, we discuss several important desiderata of IR models. These expectations inform the design of neural architectures described in this thesis.

Machine learning models—including neural networks—are employed for learning to rank [39] in IR, as we discuss in Section 3.4. However, unlike traditional LTR methods that depend on manually crafted features, the focus of our work is on neural ranking models that accept raw text as input—and focus on learning latent representations of text appropriate for the ranking task. Learning good representations of text is key to effective semantic matching in IR and is a key ingredient for all methods proposed in Chapters 4-7.

Section 2.1.2 highlights the importance of exact matching when dealing with rare terms. In Chapter 4 and Section 7.1, we operationalize this intuition and demonstrate that neural methods that combine lexical and semantic matching achive more robustness to rare inputs for different retrieval tasks.

An important IR task, in the context of this thesis, is ranking documents that may be hundreds of sentences long. As far as we are aware, the Duet model—described in Chapter 4—is the first to consider deep neural network based representation learning to rank documents. The different shift-invariant architectures discussed in Section 3.5.2 may also be appropriate for dealing with documents of different lengths. In more recent work [103], we have specifically emphasized on the challenges of dealing with long document text and demonstrated that without careful design neural models can under-retrieve longer documents.

In most real IR tasks—such as Web search—retrieval involves collections with billions of documents. In traditional IR, efficient data structures such as inverted index [142] or prefix-trees [143] are commonly employed. When designing neural ranking models, it is important to consider how they may interact with traditional IR data structures, such as inverted index. In Chapter 5, We propose a strategy that allows using deep networks—in combination with standard inverted index—to retrieve from the full collection using predominantly offline precomputation without sacrificing fast query response time. We show that this strategy generalizes effec-tively to several recent state-of-the-art deep architectures for IR.

The learning to rank literature has traditionally focused on generating static rankings of items given user intent. In Chapter 6, we argue that stochastic ranking policies are crucial when optimizing for fair distribution of exposure over items (or groups of items) of similar relevance. We demonstrate that learning to rank models can be trained towards exposure parity objectives.

Neural models can incorporate side-information on the task or user context. In Section 7.2, we explore neural representation learning in the context of session modeling for more effective QAC.

Generalizing neural models across different corpora continues to be an important open problem. Neural models with large number of learnable parameters risk overfitting to the distributions observed in the training data. These models may underperform when the properties of the test dataset is significantly different from the training corpus. While we do not discuss this particular topic in this thesis, we refer the interested reader to our recent work [141, 144] related to regularization of neural ranking models.

# Chapter 3

# Background

In this chapter, we introduce the fundamentals of neural IR, in context of traditional retrieval research, with visual examples to illustrate key concepts and a consistent mathematical notation for describing key models. Section 3.1 presents a survey of IR models. Section 3.2 introduces neural and non-neural methods for learning term embeddings, without the use of supervision from IR labels, and with a focus on the notion of similarity. Section 3.3 surveys some specific approaches for incorporating such embeddings in IR. Section 3.4 introduces supervised learning to rank models. Section 3.5 introduces the fundamentals of deep models—including standard architectures and toolkits—before Section 3.6 surveys some specific approaches for incorporating deep neural networks (DNNs) in IR.

## 3.1 IR Models

### 3.1.1 Traditional IR models

In this section, we introduce a few of the traditional IR approaches. The decades of insights from these IR models not only inform the design of our new neural based approaches, but these models also serve as important baselines for comparison. They also highlight the various desiderata that we expect the neural IR models to incorporate.

**BM25** There is a broad family of statistical functions in IR that consider the number of occurrences of each query term in the document—*i.e.*, term-frequency (TF)—and the corresponding inverse document frequency (IDF) of the same terms in the full collection (as an indicator of the informativeness of the term). One theoretical basis for such formulations is the probabilistic model of IR that yielded the BM25 [80] ranking function.

$$BM25(q,d) = \sum_{t_q \in q} idf(t_q) \cdot \frac{tf(t_q,d) \cdot (k_1 + 1)}{tf(t_q,d) + k_1 \cdot \left(1 - b + b \cdot \frac{|d|}{avgdl}\right)} \tag{3.1}$$

Where, *avgdl* is the average length of documents in the collection $D$, and $k_1$ and $b$ are parameters that are usually tuned on a validation dataset. In practice, $k_1$ is sometimes set to some default value in the range $[1.2, 2.0]$ and $b$ as $0.75$. The $idf(t)$ is computed as,

$$idf(t) = \log \frac{|D| - df(t) + 0.5}{df(t) + 0.5} \tag{3.2}$$

BM25 aggregates the contributions from individual terms but ignores any phrasal or proximity signals between the occurrences of the different query terms in the document. A variant of BM25 [145, 146] also considers documents as composed of several fields (such as, title, body, and anchor texts).

**Language modelling (LM)** In the language modelling based approach [90, 147, 148], documents are ranked by the posterior probability $p(d|q)$.

$$p(d|q) = \frac{p(q|d).p(d)}{\sum_{\bar{d} \in D} p(q|\bar{d}).p(\bar{d})} \tag{3.3}$$

$$\propto p(q|d).p(d) \tag{3.4}$$

$$= p(q|d) \quad \text{, assuming p(d) is uniform} \tag{3.5}$$

$$= \prod_{t_q \in q} p(t_q|d) \tag{3.6}$$

$\hat{p}(\mathscr{E})$ is the maximum likelihood estimate (MLE) of the probability of event $\mathscr{E}$, and $p(q|d)$ indicates the probability of generating query $q$ by randomly sampling terms from document $d$. In its simplest form, we can estimate $p(t_q|d)$ by,

$$p(t_q|d) = \frac{tf(t_q,d)}{|d|} \tag{3.7}$$

However, most formulations of language modelling based retrieval typically employ some form of smoothing [148] by sampling terms from both the document $d$ and the full collection $D$. The two common smoothing methods are:

1. Jelinek-Mercer smoothing [149]

$$p(t_q|d) = \left(\lambda \frac{tf(t_q,d)}{|d|} + (1-\lambda)\frac{\sum_{\bar{d}\in D} tf(t_q,\bar{d})}{\sum_{\bar{d}\in D} |\bar{d}|}\right) \tag{3.8}$$

2. Dirichlet Prior Smoothing [150]

$$p(t_q|d) = \left(tf(t_q,d) + \mu \frac{\sum_{\bar{d}\in D} tf(t_q,\bar{d})}{\sum_{\bar{d}\in D} |\bar{d}|}\right) / \left(|d| + \mu\right) \tag{3.9}$$

Both TF-IDF and language modelling based approaches estimate document relevance based on the count of only the query terms in the document. The position of these occurrences and the relationship with other terms in the document are ignored.

**Translation models** Berger and Lafferty [151] proposed an alternative method to estimate $p(t_q|d)$ in the language modelling based IR approach (Equation 3.6), by assuming that the query $q$ is being generated via a "translation" process from the document $d$.

$$p(t_q|d) = \sum_{t_d \in d} p(t_q|t_d) \cdot p(t_d|d) \tag{3.10}$$

The $p(t_q|t_d)$ component allows the model to garner evidence of relevance from non-query terms in the document. Berger and Lafferty [151] propose to estimate $p(t_q|t_d)$ from query-document paired data similar to techniques in statistical machine translation [152, 153]—but other approaches for estimation have also been explored [154].

**Dependence model** None of the three IR models described so far consider proximity between query terms. To address this, Metzler and Croft [155] proposed a linear model over proximity-based features.

$$
\begin{aligned}
DM(q,d) = (1 - \lambda_{ow} - \lambda_{uw}) \sum_{t_q \in q} log\left((1-\alpha_d)\frac{tf(t_q,d)}{|d|} + \alpha_d \frac{\sum_{\bar{d} \in D} tf(t_q,\bar{d})}{\sum_{\bar{d} \in D} |\bar{d}|}\right) \\
+ \lambda_{ow} \sum_{c_q \in ow(q)} log\left((1-\alpha_d)\frac{tf_{\#1}(c_q,d)}{|d|} + \alpha_d \frac{\sum_{\bar{d} \in D} tf_{\#1}(c_q,\bar{d})}{\sum_{\bar{d} \in D} |\bar{d}|}\right) \\
+ \lambda_{uw} \sum_{c_q \in uw(q)} log\left((1-\alpha_d)\frac{tf_{\#uwN}(c_q,d)}{|d|} + \alpha_d \frac{\sum_{\bar{d} \in D} tf_{\#uwN}(c_q,\bar{d})}{\sum_{\bar{d} \in D} |\bar{d}|}\right)
\end{aligned}
\quad (3.11)
$$

Where, $ow(q)$ and $uw(q)$ are the set of all contiguous *n*-grams (or phrases) and the set of all bags of terms that can be generated from query $q$. $tf_{\#1}$ and $tf_{\#uwN}$ are the ordered-window and unordered-window operators from Indri [156]. Finally, $\lambda_{ow}$ and $\lambda_{uw}$ are the tuneable parameters of the model.

**Pseudo relevance feedback (PRF)** PRF-based methods—*e.g.*, Relevance Models (RM) [157, 158]—typically demonstrate strong performance at the cost of executing an additional round of retrieval. The set of ranked documents $R_1$ from the first round of retrieval is used to select expansion terms to augment the query which is used to retrieve a new ranked set of documents $R_2$ that is presented to the user.

The underlying approach to scoring a document in RM is by computing the KL divergence [159] between the query language model $\theta_q$ and the document language model $\theta_d$.

$$
score(q,d) = -\sum_{t \in T} p(t|\theta_q) log \frac{p(t|\theta_q)}{p(t|\theta_d)} \quad (3.12)
$$

Without PRF,

$$
p(t|\theta_q) = \frac{tf(t,q)}{|q|} \quad (3.13)
$$

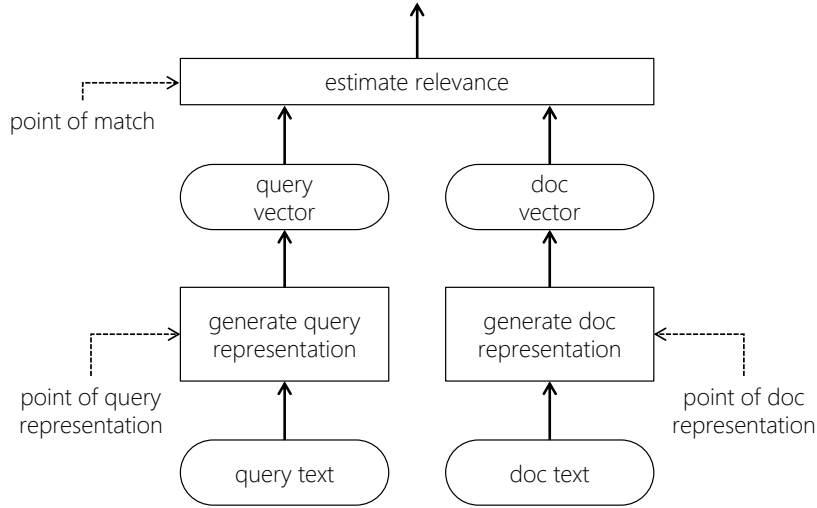

**Figure 3.1:** Document ranking typically involves a query and a document representation steps, followed by a matching stage. Neural models can be useful either for generating good representations or in estimating relevance, or both.

But under the RM3 [160] formulation the new query language model $\bar{\theta}_q$ is estimated by,

$$p(t|\bar{\theta}_q) = \alpha \frac{tf(t,q)}{|q|} + (1-\alpha) \sum_{d \in R_1} p(t|\theta_d) p(d) \prod_{\bar{t} \in q} p(\bar{t}|\theta_d) \quad (3.14)$$

Besides language models, PRF based query expansion has also been explored in the context of other retrieval approaches (*e.g.*, [161, 162]). By expanding the query using the results from the first round of retrieval PRF based approaches tend to be more robust to the vocabulary mismatch problem plaguing many other traditional IR methods.

### 3.1.2 Anatomy of neural IR models

Document ranking comprises of performing three primary steps—generate a representation of the query that specifies the information need, generate a representation of the document that captures the distribution over the information contained, and match the query and the document representations to estimate their mutual rele-

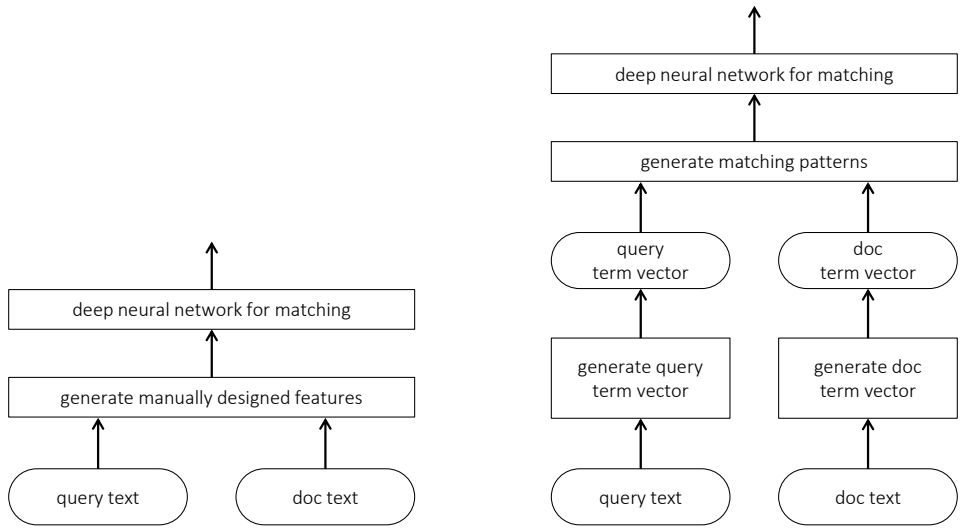

**(a)** Learning to rank using manually designed features (e.g., Liu [39])

**(b)** Estimating relevance from patterns of exact matches (e.g., [163])

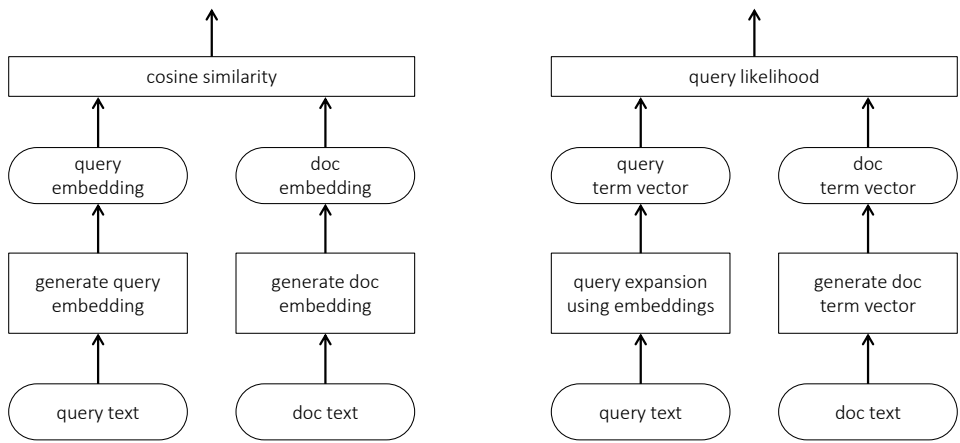

**(c)** Learning query and document representations for matching (e.g., [164, 165])

**(d)** Query expansion using neural embeddings (e.g., [139, 166])

**Figure 3.2:** Examples of different neural approaches to IR. In (a) and (b) the neural network is only used at the point of matching, whereas in (c) the focus is on learning effective representations of text using neural methods. Neural models can also be used to expand or augment the query before applying traditional IR techniques, as shown in (d).

vance. All existing neural approaches to IR can be broadly categorized based on whether they influence the query representation, the document representation, or in estimating relevance. A neural approach may impact one or more of these stages shown in Figure 3.1.

Neural networks are useful as learning to rank models as we will discuss in Section 3.4. In these models, a joint representation of query and document is generated using manually designed features and the neural network is used only at the *point of match* to estimate relevance, as shown in Figure 3.2a. In Section 3.6.4, we will discuss DNN models, such as [7, 163], that estimate relevance based on patterns of exact query term matches in the document. Unlike traditional learning to rank models, however, these architectures (shown in Figure 3.2b) depend less on manual feature engineering and more on automatically detecting regularities in good matching patterns. More recent deep learning methods, such as [167], consume query and document as single concatenated sequence of terms, instead of representing them as separate term vectors.

In contrast, many (shallow and deep) neural IR models depend on learning useful low-dimensional vector representations—or *embeddings*—of query and document text, and using them within traditional IR models or in conjunction with simple similarity metrics (*e.g.*, cosine similarity). These models shown in Figure 3.2c may learn the embeddings by optimizing directly for the IR task (*e.g.*, [164]), or in an unsupervised setting (*e.g.*, [165]). Finally, Figure 3.2d shows IR approaches where the neural models are used for query expansion [139, 166].

While the taxonomy of neural approaches described in this section is rather simple, it does provide an intuitive framework for comparing the different neural approaches in IR and highlights the similarities and distinctions between these different techniques.

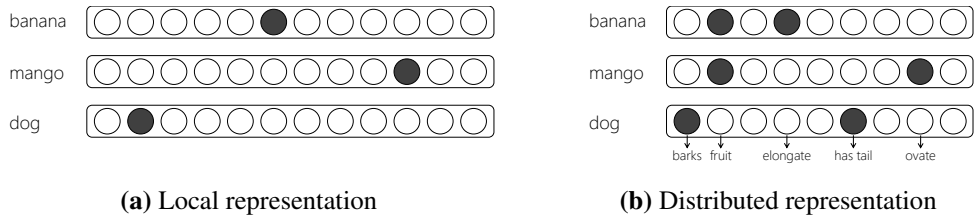

(a) Local representation　　　　　　(b) Distributed representation

**Figure 3.3:** Under local representations the terms "banana", "mango", and "dog" are distinct items. But distributed vector representations may recognize that "banana" and "mango" are both fruits, but "dog" is different.

## 3.2 Unsupervised learning of term representations

### 3.2.1 A tale of two representations

Vector representations are fundamental to both information retrieval and machine learning. In IR, terms are typically the smallest unit of representation for indexing and retrieval. Therefore, many IR models—both non-neural and neural—focus on learning good vector representations of terms. Different vector representations exhibit different levels of generalization—some consider every term as a distinct entity while others learn to identify common attributes. Different representation schemes derive different notions of similarity between terms from the definition of the corresponding vector spaces. Some representations operate over fixed-size vocabularies, while the design of others obviate such constraints. They also differ on the properties of compositionality that defines how representations for larger units of information, such as passages and documents, can be derived from individual term vectors. These are some of the important considerations for choosing a term representation suitable for a specific task.

**Local representations** Under local (or *one-hot*) representations, every term in a fixed size vocabulary $T$ is represented by a binary vector $\vec{v} \in \{0, 1\}^{|T|}$, where only one of the values in the vector is one and all the others are set to zero. Each position in the vector $\vec{v}$ corresponds to a term. The term "banana", under this representation, is given by a vector that has the value one in the position corresponding to "banana" and zero everywhere else. Similarly, the terms "mango" and "dog" are represented by setting different positions in the vector to one. Figure 3.3a highlights that under this scheme each term is a unique entity, and "banana" is as distinct from "dog" as it is from "mango". Terms outside of the vocabulary either have no representation or are denoted by a special "UNK" symbol under this scheme.

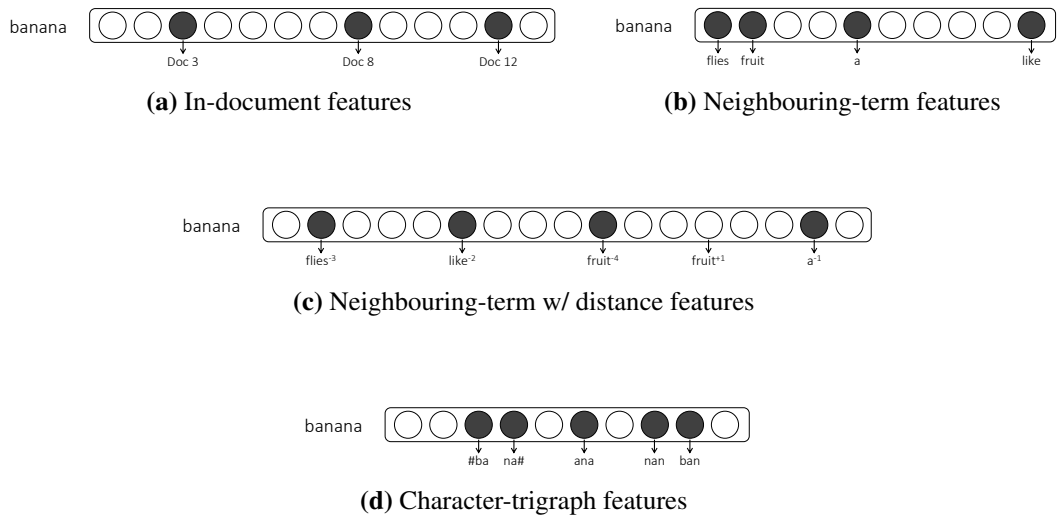

**Figure 3.4:** Examples of different feature-based distributed representations of the term "banana". The representations in (a), (b), and (c) are based on external contexts in which the term frequently occurs, while (d) is based on properties intrinsic to the term. The representation scheme in (a) depends on the documents containing the term while the scheme shown in (b) and (c) depends on other terms that appears in its neighbourhood. The scheme (b) ignores inter-term distances. Therefore, in the sentence "Time flies like an arrow; fruit flies like a banana", the feature "fruit" describes both the terms "banana" and "arrow". However, in the representation scheme of (c) the feature "fruit$^{-4}$" is positive for "banana", and the feature "fruit$^{+1}$" for "arrow".

**Distributed representations** Under distributed representations every term is represented by a vector $\vec{v} \in \mathbb{R}^{|k|}$. $\vec{v}$ can be a sparse or a dense vector—a vector of hand-crafted features or a latent representation in which the individual dimensions are not interpretable in isolation. The key underlying hypothesis for any distributed representation scheme, however, is that by representing a term by its attributes allows for defining some notion of similarity between the different terms based on the chosen properties. For example, in Figure 3.3b "banana" is more similar to "mango" than "dog" because they are both fruits, but yet different because of other properties that are not shared between the two, such as shape.

A key consideration in any feature based distributed representation is the choice of the features themselves. One approach involves representing terms by features that capture their distributional properties. This is motivated by the *distributional hypothesis* [168] that states that terms that are used (or occur) in similar context tend to be semantically similar. Firth [169] famously purported this idea of *distributional semantics*[1] by stating "*a word is characterized by the company it keeps*". However, the distribution of different types of context may model different semantics of a term. Figure 3.4 shows three different sparse vector representations of the term "banana" corresponding to different distributional feature spaces—documents containing the term (*e.g.*, LSA [170]), neighbouring terms in a window (*e.g.*, HAL [171], COALS [172], and [173]), and neighbouring terms with distance (*e.g.*, [174]). Finally, Figure 3.4d shows a vector representation of "banana" based on the character trigraphs in the term itself—instead of external contexts in which the term occurs. In Section 3.2.2 we will discuss how choosing different distributional features for term representation leads to different nuanced notions of semantic similarity between them. When the vectors are high-dimensional, sparse, and based on observable features we refer to them as *observed* (or *explicit*) vector representations [174]. When the vectors are dense, small ($k \ll |T|$), and learnt from data then we instead refer to them as *latent* vector spaces, or *embeddings*. In both observed and latent vector spaces, several distance metrics can be used to define the similarity between terms, although cosine similarity is commonly used.

$$sim(\vec{v}_i, \vec{v}_j) = cos(\vec{v}_i, \vec{v}_j) = \frac{\vec{v}_i^\intercal \vec{v}_j}{\|\vec{v}_i\| \|\vec{v}_j\|} \tag{3.15}$$

Most embeddings are learnt from observed features, and hence the discussions in Section 3.2.2 about different notions of similarity are also relevant to the embedding models. In Section 3.2.3 and Section 3.2.4 we discuss observed and latent space representations. In the context of neural models, distributed representations generally refer to learnt embeddings. The idea of 'local' and 'distributed' representations has a specific significance in the context of neural networks. Each concept, entity, or term can be represented within a neural network by the activation of a single neuron (local representation) or by the combined pattern of activations of several neurons (distributed representation) [175].

---

[1] Readers should take note that while many distributed representations take advantage of *distributional* properties, the two concepts are not synonymous. A term can have a distributed representation based on non-distributional features—*e.g.*, parts of speech classification and character trigraphs in the term.

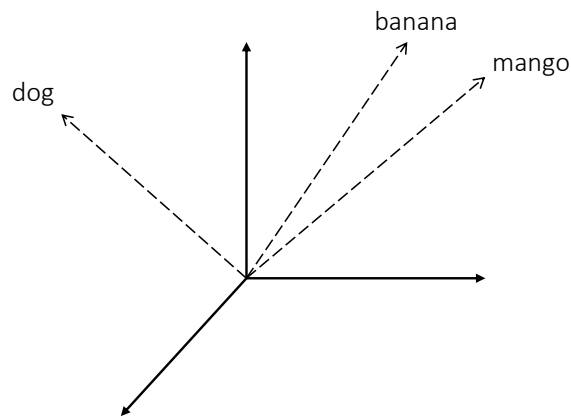

**Figure 3.5:** A vector space representation of terms puts "banana" closer to "mango" because they share more common attributes than "banana" and "dog".

Finally, with respect to *compositionality*, it is important to understand that distributed representations of items are often derived from local or distributed representation of its parts. For example, a document can be represented by the sum of the one-hot vectors or embeddings corresponding to the terms in the document. The resultant vector, in both cases, corresponds to a distributed bag-of-terms representation. Similarly, the character trigraph representation of terms in Figure 3.4d is simply an aggregation over the one-hot representations of the constituent trigraphs.

### 3.2.2 Notions of similarity

Any vector representation inherently defines some notion of relatedness between terms. Is "Seattle" closer to "Sydney" or to "Seahawks"? The answer depends on the type of relationship we are interested in. If we want terms of similar *type* to be closer, then "Sydney" is more similar to "Seattle" because they are both cities. However, if we are interested to find terms that co-occur in the same document or passage, then "Seahawks"—Seattle's football team—should be closer. The former represents a *typical*, or type-based notion of similarity while the latter exhibits a more *topical* sense of relatedness.

**Table 3.1:** A toy corpus of short documents that we consider for the discussion on different notions of similarity between terms under different distributed representations. The choice of the feature space that is used for generating the distributed representation determines which terms are closer in the vector space, as shown in Figure 3.6.

| Sample documents | | | |
| --- | --- | --- | --- |
| doc 01 | Seattle map | doc 09 | Denver map |
| doc 02 | Seattle weather | doc 10 | Denver weather |
| doc 03 | Seahawks jerseys | doc 11 | Broncos jerseys |
| doc 04 | Seahawks highlights | doc 12 | Broncos highlights |
| doc 05 | Seattle Seahawks Wilson | doc 13 | Denver Broncos Lynch |
| doc 06 | Seattle Seahawks Sherman | doc 14 | Denver Broncos Sanchez |
| doc 07 | Seattle Seahawks Browner | doc 15 | Denver Broncos Miller |
| doc 08 | Seattle Seahawks Ifedi | doc 16 | Denver Broncos Marshall |

If we want to compare "Seattle" with "Sydney" and "Seahawks based on their respective vector representations, then the underlying feature space needs to align with the notion of similarity that we are interested in. It is, therefore, important for the readers to build an intuition about the choice of features and the notion of similarity they encompass. This can be demonstrated by using a toy corpus, such as the one in Table 3.1. Figure 3.6a shows that the "in documents" features naturally lend to a topical sense of similarity between the terms, while the "neighbouring terms with distances" features in Figure 3.6c gives rise to a more typical notion of relatedness. Using "neighbouring terms" without the inter-term distances as features, however, produces a mixture of topical and typical relationships. This is because when the term distances (denoted as superscripts) are considered in the feature definition then the document "Seattle Seahawks Wilson" produces the bag-of-features $\{Seahawks^{+1}, Wilson^{+2}\}$ for "Seattle" which is non-overlapping with the bag-of-features $\{Seattle^{-1}, Wilson^{+1}\}$ for "Seahawks". However, when the feature definition ignores the term-distances then there is a partial overlap between the bag-of-features $\{Seahawks, Wilson\}$ and $\{Seattle, Wilson\}$ corresponding to "Seattle" and "Seahawks", respectively. The overlap increases when a larger window-size over the neighbouring terms is employed pushing the notion of similarity closer to a topical definition. This effect of the windows size on the latent vector space was reported by Levy and Goldberg [176] in the context of term embeddings.

Readers should note that the set of all inter-term relationships goes beyond the two notions of typical and topical that we discuss in this section. For example, vector representations could cluster terms closer based on linguistic styles—*e.g.*, terms that appear in thriller novels versus in children's rhymes, or in British versus American English. However, the notions of typical and topical similarities frequently come up in discussions in the context of many IR and NLP tasks—sometimes under different names such as *Paradigmatic* and *Syntagmatic* relations[2] [178–181]—and the idea itself goes back at least as far as Saussure [182–185].

### 3.2.3 Observed feature spaces

Observed feature space representations can be broadly categorized based on their choice of distributional features (*e.g.*, in documents, neighbouring terms with or without distances, *etc.*) and different weighting schemes (*e.g.*, TF-IDF, positive pointwise mutual information, *etc.*) applied over the raw counts. We direct the readers to [186, 187] which are good surveys of many existing observed vector representation schemes.

Levy et al. [174] demonstrated that explicit vector representations are amenable to the term analogy task using simple vector operations. A term analogy task involves answering questions of the form "*man* is to *woman* as *king* is to ____?"—the correct answer to which in this case happens to be "queen". In NLP, term analogies are typically performed by simple vector operations of the following form followed by a nearest-neighbour search,

$$\vec{v}_{Seahawks} - \vec{v}_{Seattle} + \vec{v}_{Denver} \approx \vec{v}_{Broncos} \tag{3.16}$$

---

[2]Interestingly, the notion of Paradigmatic (typical) and Syntagmatic (topical) relationships show up almost universally—not just in text. In vision, for example, the different images of "nose" are typically similar to each other, while sharing topical relationship with images of "eyes" and "ears". Curiously, Barthes [177] extended the analogy to garments. Paradigmatic relationships exist between items of the same type (*e.g.*, different style of boots) and the proper Syntagmatic juxtaposition of items from these different Paradigms—from hats to boots—forms a fashionable ensemble.

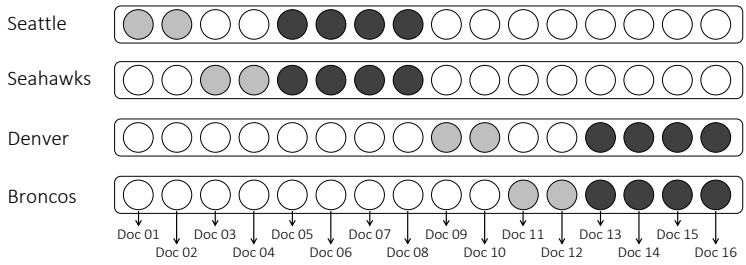

**(a)** "In-documents" features

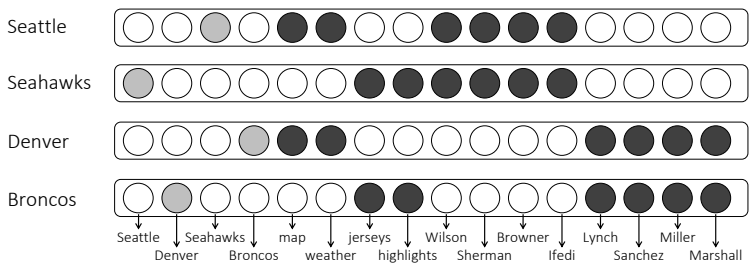

**(b)** "Neighbouring terms" features

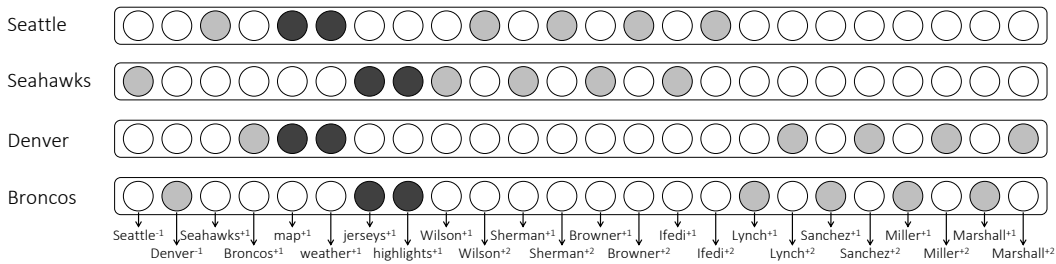

**(c)** "Neighbouring terms w/ distances" features

**Figure 3.6:** The figure shows different distributed representations for the four terms—"Seattle", "Seahawks", "Denver", and "Broncos"—based on the toy corpus in Table 3.1. Shaded circles indicate non-zero values in the vectors—the darker shade highlights the vector dimensions where more than one vector has a non-zero value. When the representation is based on the documents that the terms occur in then "Seattle" is more similar to "Seahawks" than to "Denver". The representation scheme in (a) is, therefore, more aligned with a topical notion of similarity. In contrast, in (c) each term is represented by a vector of neighbouring terms—where the distances between the terms are taken into consideration—which puts "Seattle" closer to "Denver" demonstrating a typical, or type-based, similarity. When the inter-term distances are ignored, as in (b), a mix of typical and topical similarities is observed. Finally, it is worth noting that neighbouring-terms based vector representations leads to similarities between terms that do not necessarily occur in the same document, and hence the term-term relationships are less sparse than when only in-document features are considered.

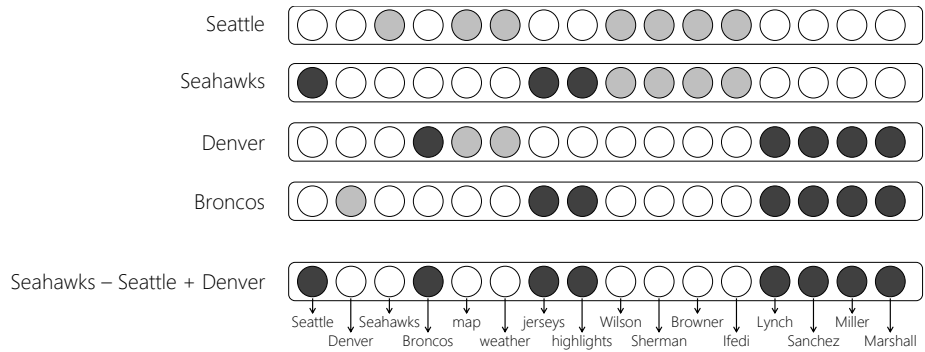

**Figure 3.7:** A visual demonstration of term analogies via simple vector algebra. The shaded circles denote non-zero values. Darker shade is used to highlight the non-zero values along the vector dimensions for which the output of $\vec{v}_{Seahawks} - \vec{v}_{Seattle} + \vec{v}_{Denver}$ is positive. The output vector is closest to $\vec{v}_{Broncos}$ as shown in this toy example.

It may be surprising to some readers that the vector obtained by the simple algebraic operations $\vec{v}_{Seahawks} - \vec{v}_{Seattle} + \vec{v}_{Denver}$ produces a vector close to the vector $\vec{v}_{Broncos}$. We present a visual intuition of why this works in practice in Figure 3.7, but we refer the readers to [174, 188] for a more rigorous mathematical handling of this subject.

### 3.2.4 Embeddings

While observed vector spaces based on distributional features can capture interesting relationships between terms, they have one big drawback—the resultant representations are highly sparse and high-dimensional. The number of dimensions, for example, may be the same as the vocabulary size, which is unwieldy for most practical tasks. An alternative is to learn lower dimensional representations that retains useful attributes from the observed feature spaces.

An *embedding* is a representation of items in a new space such that the properties of—and the relationships between—the items are preserved. Goodfellow et al. [189] articulate that the goal of an embedding is to generate a *simpler* representation—where simplification may mean a reduction in the number of dimensions, a decrease in the sparseness of the representation, disentangling the principle components of the vector space, or a combination of these goals. In the context of term embeddings, the explicit feature vectors—like those discussed in Section 3.2.3—constitutes the original representation. An embedding trained from these features assimilate the properties of the terms and the inter-term relationships observable in the original feature space.

Common approaches for learning embeddings include either factorizing the term-feature matrix (*e.g.* LSA [170]) or using gradient descent based methods that try to predict the features given the term (*e.g.*, [190, 191]). Baroni et al. [192] empirically demonstrate that these feature-predicting models that learn lower dimensional representations, in fact, also perform better than explicit counting based models on different tasks—possibly due to better generalization across terms—although some counter evidence the claim of better performances from embedding models have also been reported in the literature [193].

The sparse feature spaces of Section 3.2.3 are easier to visualize and leads to more intuitive explanations—while their latent counterparts may be more practically useful. Therefore, it may be useful to *think sparse, but act dense* in many scenarios. In the rest of this section, we will describe some of these neural and non-neural latent space models.

**Latent Semantic Analysis (LSA)** LSA [170] involves performing *singular value decomposition* (SVD) [194] on a term-document (or term-passage) matrix $X$ to obtain its low-rank approximation [195]. SVD on $X$ involves solving $X = U\Sigma V^T$, where $U$ and $V$ are orthogonal matrices and $\Sigma$ is a diagonal matrix.[3]

$$
\underset{(\vec{t}_i^\intercal)\rightarrow}{\overset{X\ (\vec{d}_j)\downarrow}{\begin{bmatrix} x_{1,1} & \cdots & x_{1,|D|} \\ \vdots & \ddots & \vdots \\ x_{|T|,1} & \cdots & x_{|T|,|D|} \end{bmatrix}}} = \underset{(\vec{t}_i^\intercal)\rightarrow}{\overset{U}{\begin{bmatrix} \begin{bmatrix} \\ \vec{u}_1 \\ \\ \end{bmatrix} \cdots \begin{bmatrix} \\ \vec{u}_l \\ \\ \end{bmatrix} \end{bmatrix}}} \cdot \overset{\Sigma}{\begin{bmatrix} \sigma_1 & \cdots & 0 \\ \vdots & \ddots & \vdots \\ 0 & \cdots & \sigma_l \end{bmatrix}} \cdot \overset{V^\intercal\ (\vec{d}_j)\downarrow}{\begin{bmatrix} [\ \vec{v}_1\ ] \\ \vdots \\ [\ \vec{v}_l\ ] \end{bmatrix}} \quad (3.17)
$$

---

[3] The matrix visualization is taken from https://en.wikipedia.org/wiki/Latent_semantic_analysis.

$\sigma_1, \ldots, \sigma_l, \vec{u}_1, \ldots, \vec{u}_l$, and $\vec{v}_1, \ldots, \vec{v}_l$ are the singular values, and the left and the right singular vectors, respectively. The $k$ largest singular values—and corresponding singular vectors from $U$ and $V$—is the rank $k$ approximation of $X$ ($X_k = U_k \Sigma_k V_k^T$) and $\Sigma_k \vec{t}_i$ is the embedding for the $i^{\text{th}}$ term.

While LSA operate on a term-document matrix, matrix factorization based approaches can also be applied to term-term matrices [172, 196, 197].

**Probabilistic Latent Semantic Analysis (PLSA)** PLSA [198] learns low-dimensional representations of terms and documents by modelling their co-occurrence $p(t, d)$ as follows,

$$p(t, d) = p(d) \sum_{c \in C} p(c|d) P(t|c) \qquad (3.18)$$

Where, $C$ is the set of latent topics—and the number of topics $|C|$ is a hyperparameter of the model. Both $p(c|d)$ and $P(t|c)$ are modelled as multinomial distributions and their parameters are typically learned using the EM algorithm [199]. After learning the parameters of the model, a term $t_i$ can be represented as a distribution over the latent topics $[p(c_0|t_i), \ldots, p(c_{|C|-1}|t_i)]$. In a related approach called Latent Dirichlet Allocation (LDA) [200], each document is represented by a Dirichlet prior instead of a fixed variable.

*Neural term embedding* models are typically trained by setting up a prediction task. Instead of factorizing the term-feature matrix—as in LSA—neural models are trained to predict the term from its features. The model learns dense low-dimensional representations in the process of minimizing the prediction error. These approaches are based on the *information bottleneck method* [201]—discussed more in Section 3.5.2—with the low-dimensional representations acting as the bottleneck. The training data may contain many instances of the same term-feature pair proportional to their frequency in the corpus (*e.g.*, word2vec [191]), or their counts can be pre-aggregated (*e.g.*, GloVe [202]).

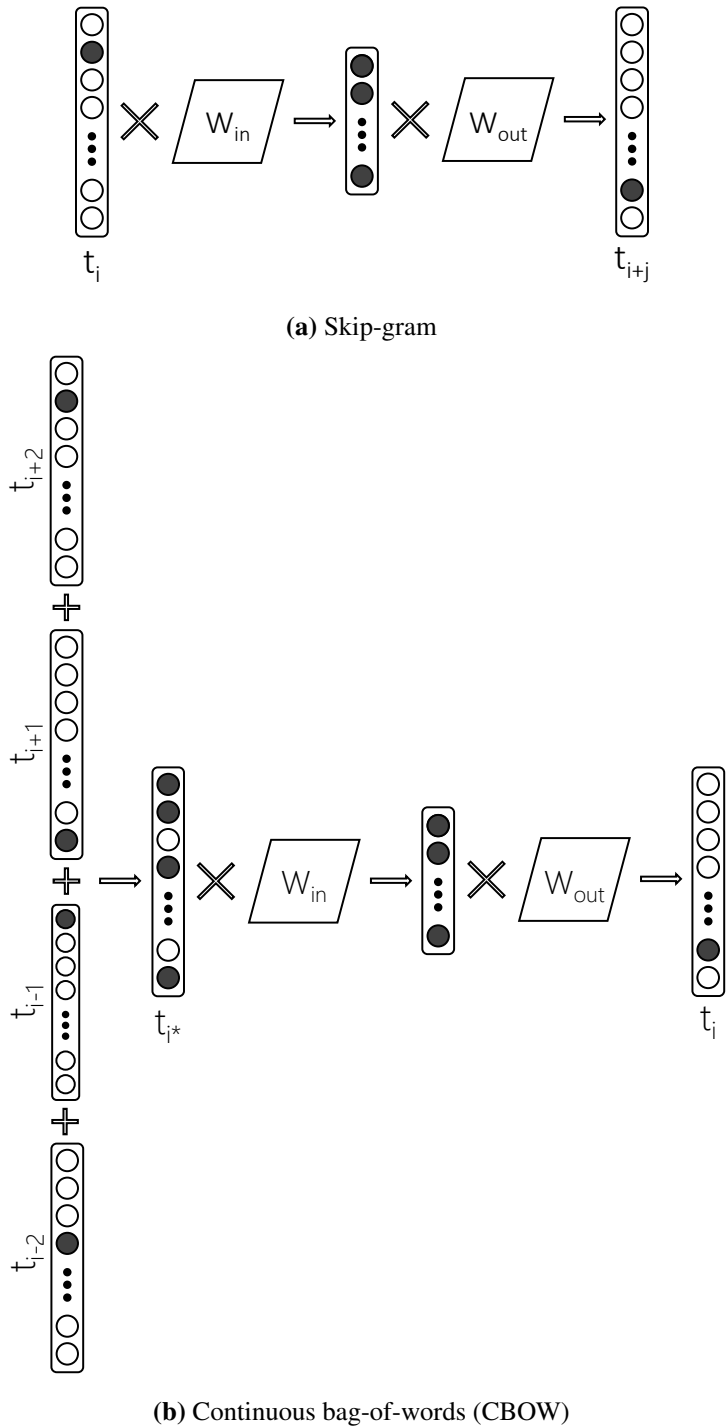

**Figure 3.8:** The (a) skip-gram and the (b) continuous bag-of-words (CBOW) architectures of word2vec. The architecture is a neural network with a single hidden layer whose size is much smaller than that of the input and the output layers. Both models use one-hot representations of terms in the input and the output. The learnable parameters of the model comprise of the two weight matrices $W_{in}$ and $W_{out}$ that corresponds to the embeddings the model learns for the input and the output terms, respectively. The skip-gram model trains by minimizing the error in predicting a term given one of its neighbours. The CBOW model, in contrast, predicts a term from a bag of its neighbouring terms.

**Word2vec** For word2vec [191, 203–206], the features for a term are made up of its neighbours within a fixed size window over the text. The *skip-gram* architecture (see Figure 3.8a) is a simple one hidden layer neural network. Both the input and the output of the model are one-hot vectors and the loss function is as follows,

$$\mathscr{L}_{skip-gram} = -\frac{1}{|S|} \sum_{i=1}^{|S|} \sum_{-c \leq j \leq +c, j \neq 0} log(p(t_{i+j}|t_i)) \quad (3.19)$$

$$\text{where,} \quad p(t_{i+j}|t_i) = \frac{\exp\left((W_{out}\vec{v}_{t_{i+j}})^\intercal (W_{in}\vec{v}_{t_i})\right)}{\sum_{k=1}^{|T|} \exp\left((W_{out}\vec{v}_{t_k})^\intercal (W_{in}\vec{v}_{t_i})\right)} \quad (3.20)$$

S is the set of all windows over the training text and c is the number of neighbours we want to predict on either side of the term $t_i$. The denominator for the softmax function for computing $p(t_{i+j}|t_i)$ sums over all the terms in the vocabulary. This is prohibitively costly and in practice either hierarchical-softmax [207] or negative sampling is employed, which we discuss more in Section 3.4.2. Note that the model has two different weight matrices $W_{in}$ and $W_{out}$ that constitute the learnable parameters of the models. $W_{in}$ gives us the IN embeddings corresponding to the input terms and $W_{out}$ corresponds to the OUT embeddings for the output terms. Generally, only $W_{in}$ is used and $W_{out}$ is discarded after training. We discuss an IR application that makes use of both the IN and the OUT embeddings in Section 3.3.1.

The *continuous bag-of-words* (CBOW) architecture (see Figure 3.8b) is similar to the skip-gram model, except that the task is to predict the middle term given all the neighbouring terms in the window. The CBOW model creates a single training sample with the sum of the one-hot vectors of the neighbouring terms as input and the one-hot vector $\vec{v}_{t_i}$—corresponding to the middle term—as the expected output. Contrast this with the skip-gram model that creates $2 \times c$ samples by individually pairing each neighbouring term with the middle term. During training, the skip-gram model trains slower than the CBOW model [191] because it creates more training samples from the same windows of text.

$$\mathcal{L}_{CBOW} = -\frac{1}{|S|}\sum_{i=1}^{|S|} log(p(t_i|t_{i-c},\ldots,t_{i-1},t_{i+1},\ldots,t_{i+c})) \quad (3.21)$$

Word2vec gained particular popularity for its ability to perform term analogies using simple vector algebra, similar to what we discussed in Section 3.2.3. For domains where the interpretability of the embeddings is important, Sun et al. [208] introduced an additional constraint in the loss function to encourage more sparseness in the learnt representations.

$$\mathcal{L}_{sparse-CBOW} = \mathcal{L}_{sparse-CBOW} - \lambda \sum_{t \in T} \|\vec{v}_t\|_1 \quad (3.22)$$

**GloVe** The skip-gram model trains on individual term-neighbour pairs. If we aggregate all the training samples such that $x_{ij}$ is the frequency of the pair $\langle t_i, t_j \rangle$ in the training data, then the loss function changes to,

$$\mathcal{L}_{skip-gram} = -\sum_{i=1}^{|T|}\sum_{j=1}^{|T|} x_{ij} log(p(t_j|t_i)) \quad (3.23)$$

$$= -\sum_{i=1}^{|T|} x_i \sum_{j=1}^{|T|} \frac{x_{ij}}{x_i} log(p(t_j|t_i)) \quad (3.24)$$

$$= -\sum_{i=1}^{|T|} x_i \sum_{j=1}^{|T|} \bar{p}(t_j|t_i) log(p(t_j|t_i)) \quad (3.25)$$

$$= \sum_{i=1}^{|T|} x_i H(\bar{p}(t_j|t_i), p(t_j|t_i)) \quad (3.26)$$

$H(\ldots)$ is the cross-entropy error between the actual co-occurrence probability $\bar{p}(t_j|t_i)$ and the one predicted by the model $p(t_j|t_i)$. This is similar to the loss function for GloVe [202] if we replace the cross-entropy error with a squared-error and apply a saturation function $f(\ldots)$ over the actual co-occurrence frequencies.

$$\mathcal{L}_{GloVe} = -\sum_{i=1}^{|T|}\sum_{j=1}^{|T|} f(x_{ij})(log(x_{ij} - \vec{v}_{w_i}^\intercal \vec{v}_{w_j}))^2 \tag{3.27}$$

$$\tag{3.28}$$

where,

$$f(x) = \begin{cases} (x/x_{max})^\alpha, & \text{if} x \leq x_{max} \\ 1, & \text{otherwise} \end{cases} \tag{3.29}$$

GloVe is trained using AdaGrad [209]. Similar to word2vec, GloVe also generates two different (IN and OUT) embeddings, but unlike word2vec it generally uses the sum of the IN and the OUT vectors as the embedding for each term in the vocabulary.

**Paragraph2vec** Following the popularity of word2vec [191, 203], similar neural architectures [181, 210–214] have been proposed that trains on term-document co-occurrences. The training typically involves predicting a term given the ID of a document or a passage that contains the term. In some variants, as shown in Figure 3.9, neighbouring terms are also provided as input. The key motivation for training on term-document pairs is to learn an embedding that is more aligned with a topical notion of term-term similarity—which is often more appropriate for IR tasks. The term-document relationship, however, tends to be more sparse [215]—including neighbouring term features may compensate for some of that sparsity. In the context of IR tasks, Ai et al. [213, 214] proposed a number of IR-motivated changes to the original Paragraph2vec [210] model training—including, document frequency based negative sampling and document length based regularization.

## 3.3 Term embeddings for IR

Traditional IR models use local representations of terms for query-document matching. The most straight-forward use case for term embeddings in IR is to enable

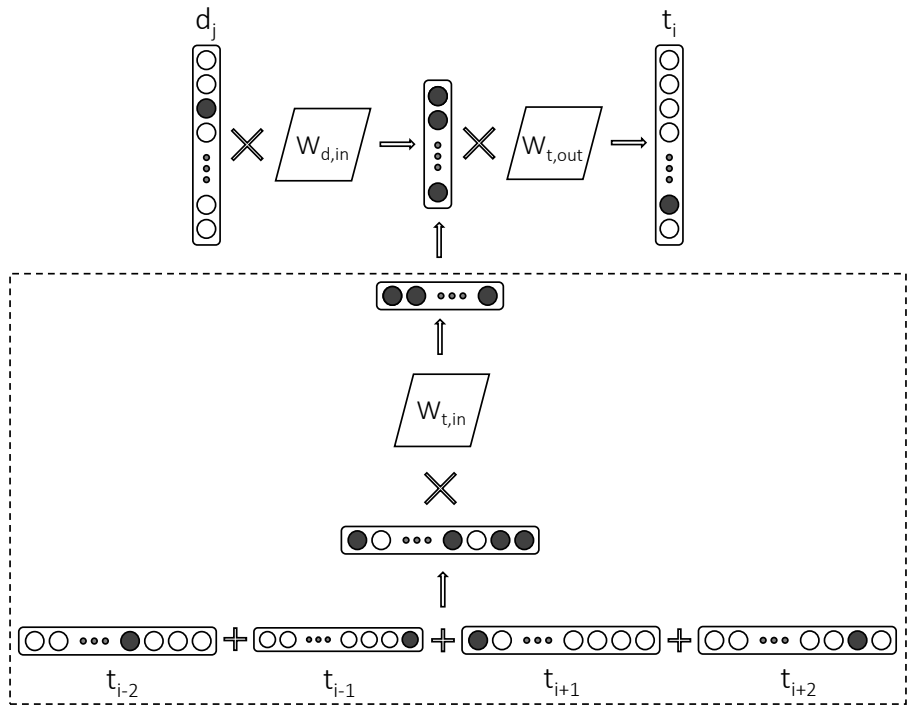

**Figure 3.9:** The paragraph2vec architecture as proposed by Le and Mikolov [210] trains by predicting a term given a document (or passage) ID containing the term. By trying to minimize the prediction error, the model learns an embedding for the term as well as for the document. In some variants of the architecture, optionally the neighbouring terms are also provided as input—as shown in the dotted box.

*inexact* matching in the embedding space. In Section 2.1, we argued the importance of inspecting non-query terms in the document for garnering evidence of relevance. For example, even from a shallow manual inspection, it is possible to conclude that the passage in Figure 3.10a is *about* Albuquerque because it contains "metropolitan", "population", and "area" among other informative terms. On the other hand, the passage in Figure 3.10b contains "simulator", "interpreter", and "Altair" which suggest that the passage is instead more likely related to computers and technology. In traditional term counting based IR approaches these signals are often ignored.

Unsupervised term embeddings can be incorporated into existing IR approaches for inexact matching. These approaches can be broadly categorized as those that compare the query with the document directly in the embedding space;

Albuquerque is the most populous city in the U.S. state of New Mexico. The high-altitude city serves as the county seat of Bernalillo County, and it is situated in the central part of the state, straddling the Rio Grande. The city population is 557,169 as of the July 1, 2014 population estimate from the United States Census Bureau, and ranks as the 32nd-largest city in the U.S. The Albuquerque metropolitan statistical area (or MSA) has a population of 907,301 according to the United States Census Bureau's most recently available estimate for 2015.

**(a)** About Albuquerque

Allen suggested that they could program a BASIC interpreter for the device; after a call from Gates claiming to have a working interpreter, MITS requested a demonstration. Since they didn't actually have one, Allen worked on a simulator for the Altair while Gates developed the interpreter. Although they developed the interpreter on a simulator and not the actual device, the interpreter worked flawlessly when they demonstrated the interpreter to MITS in Albuquerque, New Mexico in March 1975; MITS agreed to distribute it, marketing it as Altair BASIC.

**(b)** Not about Albuquerque

**Figure 3.10:** Two passages both containing exactly a single occurrence of the query term "Albuquerque". However, the passage in (a) contains other terms such as "population" and "area" that are relevant to a description of the city. In contrast, the terms in passage (b) suggest that it is unlikely to be about the city, and only mentions the city potentially in a different context.

and those that use embeddings to generate suitable query expansion candidates from a global vocabulary and then perform retrieval based on the expanded query. We discuss both these classes of approaches in the remainder of this section.

### 3.3.1 Query-document matching

One strategy for using term embeddings in IR involves deriving a dense vector representation for the query and the document from the embeddings of the individual terms in the corresponding texts. The term embeddings can be aggregated in different ways, although using the *average word (or term) embeddings* (AWE) is quite common [165, 210, 216–220]. Non-linear combinations of term vectors—such as using Fisher Kernel Framework [221]—have also been explored, as well as other families of aggregate functions of which AWE has been shown to be a special case [222].

The query and the document embeddings themselves can be compared using a variety of similarity metrics, such as cosine similarity or dot-product. For example,

$$sim(q,d) = cos(\vec{v}_q, \vec{v}_d) = \frac{\vec{v}_q^\intercal \vec{v}_d}{\|\vec{v}_q\|\|\vec{v}_d\|} \tag{3.30}$$

$$\text{where,} \quad \vec{v}_q = \frac{1}{|q|} \sum_{t_q \in q} \frac{\vec{v}_{t_q}}{\|\vec{v}_{t_q}\|} \tag{3.31}$$

$$\vec{v}_d = \frac{1}{|d|} \sum_{t_d \in d} \frac{\vec{v}_{t_d}}{\|\vec{v}_{t_d}\|} \tag{3.32}$$

An important consideration here is the choice of the term embeddings that is appropriate for the retrieval scenario. While, LSA [170], word2vec [203], and GloVe [202] are commonly used—it is important to understand how the notion of inter-term similarity modelled by a specific vector space may influence its performance on a retrieval task. In the example in Figure 3.10, we want to rank documents that contains related terms—such as "population" or "area"—higher. These terms are topically similar to the query term "Albuquerque". Intuitively, a document about "Tucson"—which is typically similar to "Albuquerque"—is unlikely to satisfy the user intent. The discussion in Section 3.2.2 on how input features influence the notion of similarity in the learnt vector space is relevant here.

Models, such as LSA [170] and Paragraph2vec [210], that consider term-document pairs generally capture topical similarities in the learnt vector space. On the other hand, word2vec [203] and GloVe [202] embeddings may incorporate a mixture of topical and typical notions of relatedness. The inter-term relationships modelled in these latent spaces may be closer to type-based similarities when trained with short window sizes or on short text, such as on keyword queries [165, 176].

In Section 3.2.4, we note that the word2vec model learns two different embeddings—IN and OUT—corresponding to the input and the output terms. In retrieval, if a query contains a term $t_i$ then—in addition to the frequency of occurrences of $t_i$ in the document—we may also consider the presence of a different term $t_j$ in the document to be a supporting evidence of relevance if the pair of terms $\langle t_i, t_j \rangle$ frequently co-occurs in the collection. As shown in Equation 3.19, in the skip-gram model this probability of co-occurrence $p(t_j|t_i)$ is proportional to $(W_{out}\vec{v}_{t_j})^\intercal(W_{in}\vec{v}_{t_i})$—i.e., the dot product between the IN embeddings of $t_i$ and the OUT embeddings of $t_j$. Therefore, Nalisnick et al. [218] point out that when using word2vec embeddings for estimating the relevance of a document to a query, it is more appropriate to compute the IN-OUT similarity between the query and the document terms. In other words, the query terms should be represented using the IN embeddings and the document terms using the OUT embeddings. Table 3.2 highlights the difference between IN-IN or IN-OUT similarities between terms.

**Table 3.2:** Different nearest neighbours in the word2vec embedding space based on whether we compute IN-IN, OUT-OUT, or IN-OUT similarities between the terms. The examples are from [165, 218] where the word2vec embeddings are trained on search queries. Training on short query text, however, makes the inter-term similarity more pronouncedly typical (where, "Yale" is closer to "Harvard" and "NYU") when both terms are represented using their IN vectors. In contrast, the IN-OUT similarity (where, "Yale" is closer to "faculty" and "alumni") mirrors more the topical notions of relatedness.

| yale | | | seahawks | | |
| --- | --- | --- | --- | --- | --- |
| IN-IN | OUT-OUT | IN-OUT | IN-IN | OUT-OUT | IN-OUT |
| yale | yale | yale | seahawks | seahawks | seahawks |
| harvard | uconn | faculty | 49ers | broncos | highlights |
| nyu | harvard | alumni | broncos | 49ers | jerseys |
| cornell | tulane | orientation | packers | nfl | tshirts |
| tulane | nyu | haven | nfl | packers | seattle |
| tufts | tufts | graduate | steelers | steelers | hats |

The proposed *Dual Embedding Space Model* (DESM)[4] [165, 218] estimates the query-document relevance as follows,

$$DESM_{in-out}(q,d) = \frac{1}{|q|}\sum_{t_q \in q} \frac{\vec{v}_{t_q,in}^\intercal \vec{v}_{d,out}}{\|\vec{v}_{t_q,in}\|\|\vec{v}_{d,out}\|} \quad (3.33)$$

$$\vec{v}_{d,out} = \frac{1}{|d|}\sum_{t_d \in d} \frac{\vec{v}_{t_d,out}}{\|\vec{v}_{t_d,out}\|} \quad (3.34)$$

---

[4]The dual term embeddings trained on Bing queries is available for download at https://www.microsoft.com/en-us/download/details.aspx?id=52597

An alternative to representing queries and documents as an aggregate of their term embeddings is to incorporate the term representations into existing IR models, such as the ones we discussed in Section 3.1. Zuccon et al. [154] proposed the *Neural Translation Language Model* (NTLM) that uses the similarity between term embeddings as a measure for term-term translation probability $p(t_q|t_d)$ in Equation 3.11.

$$p(t_q|t_d) = \frac{cos(\vec{v}_{t_q}, \vec{v}_{t_d})}{\sum_{t \in T} cos(\vec{v}_t, \vec{v}_{t_d})} \quad (3.35)$$

On similar lines, Ganguly et al. [223] proposed the *Generalized Language Model* (GLM) which extends the Language Model based approach in Equation 3.9 to,

$$\begin{aligned}p(d|q) = \prod_{t_q \in q} \Bigg( &\lambda \frac{tf(t_q, d)}{|d|} \\ &+ \alpha \frac{\sum_{t_d \in d} (sim(\vec{v}_{t_q}, \vec{v}_{t_d}) \cdot tf(t_d, d))}{\sum_{t_{d_1} \in d} \sum_{t_{d_2} \in d} sim(\vec{v}_{t_{d_1}}, \vec{v}_{t_{d_2}}) \cdot |d|^2} \\ &+ \beta \frac{\sum_{\bar{t} \in N_t} (sim(\vec{v}_{t_q}, \vec{v}_{\bar{t}}) \cdot \sum_{\bar{d} \in D} tf(\bar{t}, \bar{d}))}{\sum_{t_{d_1} \in N_t} \sum_{t_{d_2} \in N_t} sim(\vec{v}_{t_{d_1}}, \vec{v}_{t_{d_2}}) \cdot \sum_{\bar{d} \in D} |\bar{d}| \cdot |N_t|} \\ &+ (1 - \alpha - \beta - \lambda) \frac{\sum_{\bar{d} \in D} tf(t_q, \bar{d})}{\sum_{\bar{d} \in D} |\bar{d}|} \Bigg)\end{aligned} \quad (3.36)$$

Where, $N_t$ is the set of nearest-neighbours of term $t$. Ai et al. [214] incorporate paragraph vectors [210] into the query-likelihood model [90].

Another approach, based on the Earth Mover's Distance (EMD) [224], involves estimating similarity between pairs of documents by computing the minimum distance in the embedding space that each term in the first document needs to travel to reach the terms in the second document. This measure, commonly referred to as the *Word Mover's Distance* (WMD), was originally proposed by Wan et al. [225, 226], but used WordNet and topic categories instead of embeddings for defining the distance between terms. Term embeddings were later incorporated into the model by Kusner et al. [227, 228]. Finally, Guo et al. [229] incorporated similar notion of distance into the *Non-linear Word Transportation* (NWT) model that

3.3. Term embeddings for IR                                                                                     71estimates relevance between a a query and a document. The NWT model involves solving the following constrained optimization problem,

$$\max \sum_{t_q \in q} \log \left( \sum_{t_d \in u(d)} f(t_q, t_d) \cdot max\big(cos(\vec{v}_{t_q}, \vec{v}_{t_d}), 0\big)^{idf(t_q)+b} \right) \quad (3.37)$$

$$\text{subject to} \quad f(t_q, t_d) \geq 0, \quad \forall t_q \in q, t_d \in d \quad (3.38)$$

$$\text{and} \quad \sum_{t_q \in q} f(t_q, t_d) = \frac{tf(t_d) + \mu \frac{\sum_{\bar{d} \in D} tf(t_q, \bar{d})}{\sum_{\bar{d} \in D} |\bar{d}|}}{|d| + \mu}, \quad \forall t_d \in d \quad (3.39)$$

$$\text{where,} \quad idf(t) = \frac{|D| - df(t) + 0.5}{df(t) + 0.5} \quad (3.40)$$

$u(d)$ is the set of all unique terms in document $d$, and $b$ is a constant.

Another term-alignment based distance metric was proposed by Kenter and de Rijke [230] for computing short-text similarity. The design of the *saliency-weighted semantic network* (SWSN) is motivated by the BM25 [80] formulation.

$$swsn(s_l, s_s) = \sum_{t_l \in s_l} idf(t_l) \cdot \frac{sem(t_l, s_s) \cdot (k_1 + 1)}{sem(t_l, s_s) + k_1 \cdot \left(1 - b + b \cdot \frac{|s_s|}{avgsl}\right)} \quad (3.41)$$

$$\text{where,} \quad sem(t, s) = \max_{\bar{t} \in s} cos(\vec{v}_t, \vec{v}_{\bar{t}}) \quad (3.42)$$

Here $s_s$ is the shorter of the two sentences to be compared, and $s_l$ the longer sentence.

Figure 3.11 highlights the distinct strengths and weaknesses of matching using local and distributed representations of terms for retrieval. For the query "Cambridge", a local representation (or exact matching) based model can easily distinguish between the passage on Cambridge (Figure 3.11a) and the one on Oxford (Figure 3.11b). However, the model is easily duped by a non-relevant passage that has been artificially injected with the term "Cambridge" (Figure 3.11c). The embedding space based matching, on the other hand, can spot that the other terms in the passage pro-the city of cambridge is a university city and the county town of cambridgeshire , england . it lies in east anglia , on the river cam , about 50 miles ( 80 km ) north of london . according to the united kingdom census 2011 , its population was ( including students ) . this makes cambridge the second largest city in cambridgeshire after peterborough , and the 54th largest in the united kingdom . there is archaeological evidence of settlement in the area during the bronze age and roman times ; under viking rule cambridge became an important trading centre . the first town charters were granted in the 12th century , although city status was not conferred until 1951 .

**(a)** Passage about the city of Cambridge

oxford is a city in the south east region of england and the county town of oxfordshire . with a population of . it is the 52nd largest city in the united kingdom , and one of the fastest growing and most ethnically diverse . oxford has a broad economic base . its industries include motor manufacturing , education , publishing and a large number of information technology and businesses , some being academic offshoots . the city is known worldwide as the home of the university of oxford , the oldest university in the world . buildings in oxford demonstrate examples of every english architectural period since the arrival of the saxons , including the radcliffe camera . oxford is known as the city of dreaming spires , a term coined by poet matthew arnold .

**(b)** Passage about the city of Oxford

the cambridge ( giraffa camelopardalis ) is an african ungulate mammal , the tallest living terrestrial animal and the largest ruminant . its species name refers to its shape and its colouring . its chief distinguishing characteristics are its extremely long neck and legs , its , and its distinctive coat patterns . it is classified under the family , along with its closest extant relative , the okapi . the nine subspecies are distinguished by their coat patterns . the scattered range of giraffes extends from chad in the north to south africa in the south , and from niger in the west to somalia in the east . giraffes usually inhabit savannas , grasslands , and open woodlands .

**(c)** Passage about giraffes, but 'giraffe' is replaced by 'Cambridge'

**Figure 3.11:** A visualization of IN-OUT similarities between terms in different passages with the query term "Cambridge". The visualization reveals that, besides the term "Cambridge", many other terms in the passages about both Cambridge and Oxford have high similarity to the query term. The passage (c) is adapted from a passage on giraffes by replacing all the occurrences of the term "giraffe" with "cambridge". However, none of the other terms in (c) are found to be relevant to the query term. An embedding based approach may be able to determine that passage (c) is non-relevant to the query "Cambridge", but fail to realize that passage (b) is also non-relevant. A term counting-based model, on the other hand, can easily identify that passage (b) is non-relevant but may rank passage (c) incorrectly high.

<as>egment type="">3.3. Term embeddings for IR    73</as>egment>

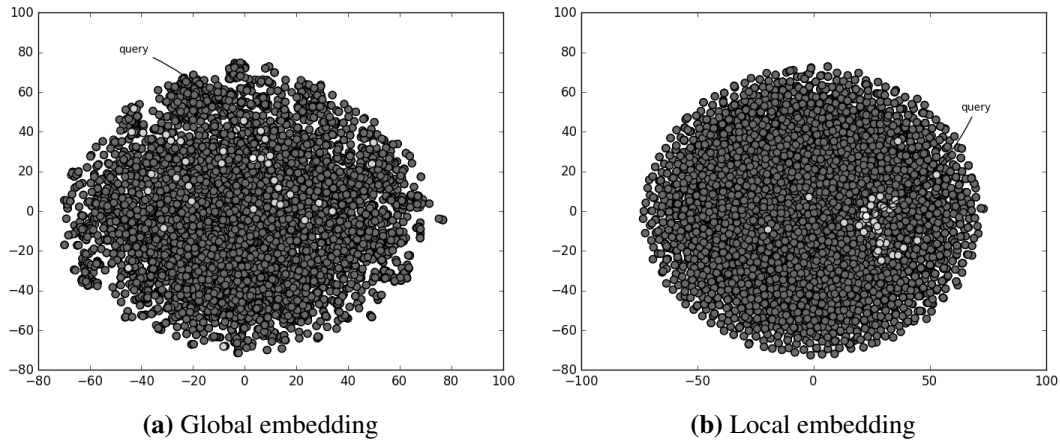

(a) Global embedding  (b) Local embedding

**Figure 3.12:** A two-dimensional visualization of term embeddings when the vector space is trained on a (a) global corpus and a (b) query-specific corpus, respectively. The grey circles represent individual terms in the vocabulary. The white circle represents the query "ocean remote sensing" as the centroid of the embeddings of the individual query terms, and the light grey circles correspond to good expansion terms for this query. When the representations are query-specific then the meaning of the terms are better disambiguated, and more likely to result in the selection of good expansion terms.

vide clear indication that the passage is not *about* a city, but fails to realize that the passage about Oxford (Figure 3.11b) is inappropriate for the same query.

Embedding based models often perform poorly when the retrieval is performed over the full document collection [165]. However, as seen in the example of Figure 3.11, the errors made by embedding based models and exact matching models may be different—and the combination of the two is often preffered [165, 198, 214, 223]. Another technique is to use the embedding based model to re-rank only a subset of the documents retrieved by a different—generally an exact matching based—IR model. The chaining of different IR models where each successive model re-ranks a smaller number of candidate documents is called *Telescoping* [111]. Telescoping evaluations are common in the neural IR literature [7, 163–165, 231] and the results are representative of performances of these models on re-ranking tasks. However, as Mitra et al. [165] demonstrate, good performances on re-ranking tasks may not be indicative how the model would perform if the retrieval involves larger document collections.

### 3.3.2 Query expansion

Instead of comparing the query and the document directly in the embedding space, an alternative approach is to use term embeddings to find good expansion candidates from a global vocabulary, and then retrieving documents using the expanded query. Different functions [139, 166, 232] have been proposed for estimating the relevance of candidate terms to the query—all of them involves comparing the candidate term individually to every query term using their vector representations, and then aggregating the scores. For example, [139, 166] estimate the relevance of candidate term $t_c$ as,

$$score(t_c, q) = \frac{1}{|q|} \sum_{t_q \in q} cos(\vec{v}_{t_c}, \vec{v}_{t_q}) \qquad (3.43)$$

Term embedding based query expansion on its own performs worse than pseudo-relevance feedback [166]. But like the models in the previous section, shows better performances when used in combination with PRF [232].

Diaz et al. [139] explored the idea of query-specific term embeddings and found that they are more effective in identifying good expansion terms than a global representation (see Figure 3.12). The local model proposed by Diaz et al. [139] incorporate relevance feedback in the process of learning the term embeddings—a set of documents is retrieved for the query and a query-specific term embedding model is trained. This *local* embedding model is then employed for identifying expansion candidates for the query for a second round of document retrieval.

Term embeddings have also been explored for re-weighting query terms [233] and finding relevant query re-writes [211], as well as in the context of other IR tasks such as cross-lingual retrieval [217] and entity retrieval [57, 58]. In Section 3.6, we will discuss neural network models with deeper architectures and their applications to retrieval.

## 3.4 Supervised learning to rank

Learning to rank (LTR) for IR uses training data $rel_q(d)$, such as human relevance labels and click data, to train towards an IR objective. Unlike traditional IR approaches, these models typically have large number of learnable parameters that require many training samples to be tuned [35]. LTR models represent a rankable item—*e.g.*, a query-document pair—as a feature vector $\vec{x} \in \mathbb{R}^n$. The ranking model $f : \vec{x} \to \mathbb{R}$ is trained to map the vector to a real-valued score such that for a given query more relevant documents are scored higher and some chosen rank-based metric is maximized. The model training is said to be end-to-end if the parameters of $f$ are learned all at once rather than in parts, and if the vector $\vec{x}$ contains simple features rather than models. Liu [39] categorizes the different LTR approaches based on their training objectives.

- In the *pointwise approach*, the relevance information $rel_q(d)$ is in the form of a numerical value associated with every query-document pair with input vector $\vec{x}_{q,d}$. The numerical relevance label can be derived from binary or graded relevance judgments or from implicit user feedback, such as a clickthrough rate. A regression model is typically trained on the data to predict the numerical value $rel_q(d)$ given $\vec{x}_{q,d}$.

- In the *pairwise approach*, the relevance information is in the form of preferences between pairs of documents with respect to individual queries (*e.g.*, $d_i \succ_q d_j$). The ranking problem in this case reduces to that of a binary classification to predict the more relevant document.

- Finally, the *listwise approach* involves directly optimizing for a rank-based metric such as NDCG—which is more challenging because these metrics are often not continuous (and hence not differentiable) with respect to the model parameters.

Many machine learning models—including support vector machines [234], neural networks [235], and boosted decision trees [236]—have been employed over the years for the LTR task, and a correspondingly large number of different loss functions have been explored.

### 3.4.1 Input features

Traditional LTR models employ hand-crafted features [39] for representing query-document pairs in $\vec{x}$. The design of these features typically encodes key IR insights and belong to one of the three categories.

- *Query-independent* or *static* features (*e.g.*, incoming link count and document length)

- *Query-dependent* or *dynamic* features (*e.g.*, BM25)

- *Query-level* features (*e.g.*, query length)

In contrast, in recently proposed neural LTR models the deep architecture is responsible for feature learning[5] from simple vector representations of the input which may resemble the schemes described in Section 3.5.1 (*e.g.*, [164]) or the interaction-based representations that we discuss later in Section 3.6.3 (*e.g.*, [7, 238]). These features, learnt from the query and document texts, can be combined with other features that may not be possible to infer from the content, such as document popularity [239].

### 3.4.2 Loss functions

In ad hoc retrieval, the LTR model needs to rank the documents in a collection $D$ in response to a query. When training a neural model for this task, the ideal ranking of documents for a query $q$ from the training dataset can be determined based on the relevance labels $rel_q(d)$ associated with each document $d \in D$. In the pointwise approach, the neural model is trained to directly estimate $rel_q(d)$, which can be a numeric value or a categorical label.

---

[5]In the literature, when the model is responsible for feature learning the task is sometimes categorized as "learning to match" [83, 237]. However, from a machine learning viewpoint, this distinction between whether $\vec{x}$ is a vector of hand-engineered features or a vector encoding of query-document text makes little difference to the LTR formulation described here. We, therefore, avoid making this distinction in favor of a more general definition.

**Regression loss** Given $\vec{x}_{q,d}$, the task of estimating the relevance label $rel_q(d)$ can be cast as a regression problem, and a standard loss function—such as the *square loss*—can be employed.

$$\mathscr{L}_{\text{squared}} = \|rel_q(d) - s(\vec{x}_{q,d})\|^2 \qquad (3.44)$$

Where, $s(\vec{x}_{q,d})$ is the score predicted by the model and $rel_q(d)$ can either be the value of the relevance label [240] or the one-hot representation when the label is categorical [241].

**Classification loss** When the relevance labels in the training data are categorical, it makes more sense to treat the label prediction problem as a multiclass classification. The neural model under this setting, estimates the probability of a label $y$ given $\vec{x}_{q,d}$. The probability of the correct label $y_{q,d}$ $(= rel_q(d))$ can be obtained by the softmax function,

$$p(y_{q,d}|q,d) = p(y_{q,d}|\vec{x}_{q,d}) = \frac{e^{\gamma \cdot s(\vec{x}_{q,d}, y_{q,d})}}{\sum_{y \in Y} e^{\gamma \cdot s(\vec{x}_{q,d}, y)}} \qquad (3.45)$$

The softmax function normalizes the score of the correct label against the set of all possible labels $Y$. The *cross-entropy* loss can then be applied [242] as follows,

$$\mathscr{L}_{\text{classification}} = -log\Big(p(y_{q,d}|q,d)\Big) = -log\Big(\frac{e^{\gamma \cdot s(\vec{x}_{q,d}, y_{q,d})}}{\sum_{y \in Y} e^{\gamma \cdot s(\vec{x}_{q,d}, y)}}\Big) \qquad (3.46)$$

However, a ranking model does not need to estimate the true relevance label accurately as long as it ranks the relevant documents $D^+$ over all the other candidates in $D$. Typically, only a few documents from $D$ are relevant to $q$. If we assume a binary notion of relevance, then the problem is similar to multi-label classification—or, multiclass classification if we assume a single relevant document $d^+$ per query— where the candidate documents are the classes. Next, we discuss loss functions for LTR models that tries to predict the relevant document by maximizing $p(d^+|q)$. Note that this is different from the classification loss in Equation 3.46 which maximizes $p(y_{q,d}|q,d)$.

**Contrastive loss** In representation learning models, a relevant document should be closer to the query representation than a non-relevant document. The contrastive loss [243, 244]—common in image retrieval—learns the model parameters by minimizing the distance between a relevant pair, while increasing the distance between dissimilar items.

$$\mathscr{L}_{\text{Contrastive}}(q,d,y_{q,d}) = (1 - y_{q,d}) \cdot \mathscr{L}_{pos}(\text{dist}_{q,d}) + y_{q,d} \cdot \mathscr{L}_{neg}(\text{dist}_{q,d}) \qquad (3.47)$$

Contrastive loss assumes that the relevance label $y_{q,d} \in \{0,1\}$ is binary. For each training sample, either $\mathscr{L}_{pos}$ or $\mathscr{L}_{neg}$ is applied over the distance $\text{dist}_{q,d}$ as predicted by the model. In particular, Hadsell et al. [244] use the following formulation of this loss function.

$$\mathscr{L}_{\text{Contrastive}}(q,d,y_{q,d}) = (1 - y_{q,d}) \cdot \frac{1}{2}\left(\max(0, m - \text{dist}_{q,d})\right)^2 \qquad (3.48)$$
$$+ y_{q,d} \cdot \frac{1}{2}(\text{dist}_{q,d})^2 \qquad (3.49)$$

Where, $m$ is a margin.

**Cross-Entropy loss over documents** The probability of ranking $d^+$ over all the other documents in the collection $D$ is given by the softmax function,

$$p(d^+|q) = \frac{e^{\gamma \cdot s(q,d^+)}}{\sum_{d \in D} e^{\gamma \cdot s(q,d)}} \qquad (3.50)$$

The cross-entropy (CE) loss then maximizes the difference between scores generated by the model for relevant and less relevant documents.

$$\mathcal{L}_{\text{CE}}(q, d^+, D) = -log\Big(p(d^+|q)\Big) \qquad (3.51)$$

$$= -log\Big(\frac{e^{\gamma \cdot s(q, d^+)}}{\sum_{d \in D} e^{\gamma \cdot s(q, d)}}\Big) \qquad (3.52)$$

However, when $D$ is the full collection then computing the softmax (*i.e.* the denominator in Equation 3.52) is prohibitively expensive. Coincidentally, the CE loss is also useful for non-IR tasks, such as language modelling [190, 191], where the model needs to predict a single term from a large vocabulary given its neighbours as input. Several different approaches have been proposed in the LM literature to address this computational complexity that is relevant to our discussion. We briefly describe some of these strategies here.

**Hierarchical softmax** Instead of computing $p(d^+|q)$ directly, Goodman [245] groups the candidates $D$ into a set of classes $C$, and then predicts the correct class $c^+$ given $q$ followed by predicting $d^+$ given $\langle c^+, q \rangle$.

$$p(d^+|q) = p(d^+|c^+, x) \cdot p(c^+|q) \qquad (3.53)$$

The computational cost in this modified approach is a function of $|C| + |c^+|$ which is typically much smaller than $|D|$. Further computational efficiency can be achieved by employing a hierarchy of such classes [207, 246]. The hierarchy of classes is typically based on either similarity between candidates [191, 247, 248], or frequency binning [249]. Zweig and Makarychev [250] and Grave et al. [251] have explored strategies for building the hierarchy that directly minimizes the computational complexity.

**Importance sampling (IS)** An alternative to computing the exact softmax, is to approximately estimate it using sampling based approaches. Note, that we can rewrite Equation 3.52 as follows,

$$\mathscr{L}_{\text{CE}}(q,d^+,D) = -log\left(\frac{e^{\gamma \cdot s(q,d^+)}}{\sum_{d \in D} e^{\gamma \cdot s(q,d)}}\right) \tag{3.54}$$

$$= -\gamma \cdot s(q,d^+) + log \sum_{d \in D} e^{\gamma \cdot s(q,d)} \tag{3.55}$$

To train a neural model using back-propagation, we need to compute the gradient $\nabla_\theta$ of the loss $\mathscr{L}_{\text{CE}}$ with respect to the model parameters $\theta$,

$$\nabla_\theta \mathscr{L}_{\text{CE}}(q,d^+,Y) = -\gamma \nabla_\theta \cdot s(q,d^+) + \nabla_\theta log \sum_{d \in D} e^{\gamma \cdot s(q,d)} \tag{3.56}$$

$$= -\gamma \nabla_\theta \cdot s(q,d^+) + \frac{\nabla_\theta \sum_{d \in D} e^{\gamma \cdot s(q,d)}}{\sum_{d \in D} e^{\gamma \cdot s(q,d)}} \tag{3.57}$$

$$= -\gamma \nabla_\theta \cdot s(q,d^+) + \frac{\sum_{d \in D} \nabla_\theta e^{\gamma \cdot s(q,d)}}{\sum_{d \in D} e^{\gamma \cdot s(q,d)}} \tag{3.58}$$

$$= -\gamma \nabla_\theta \cdot s(q,d^+) + \frac{\sum_{d \in D} \gamma \cdot e^{\gamma \cdot s(q,d)} \nabla_\theta s(q,d)}{\sum_{d \in D} e^{\gamma \cdot s(q,d)}} \tag{3.59}$$

$$= -\gamma \nabla_\theta \cdot s(q,d^+) + \gamma \sum_{d \in D} \frac{e^{\gamma \cdot s(q,d)}}{\sum_{d \in D} e^{\gamma \cdot s(q,d)}} \nabla_\theta s(q,d) \tag{3.60}$$

$$= -\gamma \nabla_\theta \cdot s(q,d^+) + \gamma \sum_{d \in D} p(d|q) \nabla_\theta s(q,d) \tag{3.61}$$

As Senécal and Bengio [252] point out, the first component of the gradient $\gamma \nabla_\theta s(q,d^+)$ is the positive reinforcement to the model for the correct candidate $d^+$ and the second component $\gamma \sum_{d \in D} p(d|q) \nabla_\theta s(q,d)$ is the negative reinforcement corresponding to all the other (incorrect) candidates. The key idea behind sampling based approaches is to estimate the second component without computing the costly sum over the whole candidate set. In IS [253–256], Monte-Carlo method is used to estimate the second component.

**Noise Contrastive Estimation (NCE)** In NCE [257–259], the task is modified to that of a binary classification. The model is trained to distinguish a sample drawn from a true distribution $p(d|q)$ from a sample drawn from a noisy distribution $\tilde{p}(d)$. The training data contains $k$ noisy samples for every true sample. Let, $\mathscr{E}$ and $\bar{\mathscr{E}}$ indicate that a sample is drawn from the true and the noisy distributions, respectively. Then,

$$p(\mathscr{E}|q,d) = \frac{p(d|q)}{p(d|q) + k \times \tilde{p}(d)} \tag{3.62}$$

$$p(\bar{\mathscr{E}}|q,d) = \frac{k \times \tilde{p}(d)}{p(d|q) + k \times \tilde{p}(d)} \tag{3.63}$$

We want our model to learn the true distribution $p(d|q)$. Remember, that according to our model,

$$p(d|q) = \frac{e^{\gamma \cdot s(q,d)}}{\sum_{\bar{d} \in D} e^{\gamma \cdot s(q,\bar{d})}} \tag{3.64}$$

$$= \frac{e^{\gamma \cdot s(q,d)}}{z(q)} \tag{3.65}$$

A key efficiency trick involves setting $z(q)$ to 1 [258–260]. Therefore,

$$p(d|q) = e^{\gamma \cdot s(q,d)} \tag{3.66}$$

Putting Equation 3.66 back in Equation 3.62 and 3.63.

$$p(\mathscr{E}|q,d) = \frac{e^{\gamma \cdot s(q,d)}}{e^{\gamma \cdot s(q,d)} + k \times \tilde{p}(d)} \tag{3.67}$$

$$p(\bar{\mathscr{E}}|q,d) = \frac{k \times \tilde{p}(d)}{e^{\gamma \cdot s(q,d)} + k \times \tilde{p}(d)} \tag{3.68}$$

Finally, the NCE loss is given by,

$$\mathscr{L}_{\text{NCE}} = - \sum_{\langle x,d^+ \rangle} \left( log\ p(\mathscr{E}|x,d^+) + \sum_{i=1}^{k} log\ p(\bar{\mathscr{E}}|x,y_i^-) \right) \quad (3.69)$$

$$= - \sum_{\langle x,d^+ \rangle} \left( log\ \frac{e^{\gamma \cdot s(q,d^+)}}{e^{\gamma \cdot s(q,d^+)} + k \times \tilde{p}(d^+)} + \sum_{i=1}^{k} log\ \frac{k \times \tilde{p}(y_i^-)}{e^{\gamma \cdot s(q,d_i^-)} + k \times \tilde{p}(y_i^-)} \right) \quad (3.70)$$

Note, that the outer summation iterates over all the positive $\langle x,d^+ \rangle$ pairs in the training data.

**Negative sampling (NEG)** Mikolov et al. [203] modify the NCE loss by replacing $k \times \tilde{p}(d)$ with 1 in Equation 3.67 and 3.68.

$$p(\mathscr{E}|q,d) = \frac{e^{\gamma \cdot s(q,d)}}{e^{\gamma \cdot s(q,d)} + 1} \quad (3.71)$$

$$= \frac{1}{1 + e^{-\gamma \cdot s(q,d)}} \quad (3.72)$$

$$p(\bar{\mathscr{E}}|q,d) = \frac{1}{1 + e^{\gamma \cdot s(q,d)}} \quad (3.73)$$

which changes the NCE loss to the NEG loss.

$$\mathscr{L}_{\text{NEG}} = - \sum_{\langle x,d^+ \rangle} \left( log\ \frac{1}{1 + e^{-\gamma \cdot s(q,d^+)}} + \sum_{i=1}^{k} log\ \frac{1}{1 + e^{\gamma \cdot s(q,d_i^-)}} \right) \quad (3.74)$$

**BlackOut** Related to both IS and NCE, is BlackOut [261]. It is an extension of the DropOut [262] method that is often employed to avoid over-fitting in neural models with large number of parameters. DropOut is typically applied to the input or hidden layers of the network and involves randomly dropping a subset of the neural units and their corresponding connections. BlackOut applies the same idea to the output layer of the network for efficiently computing the loss. We refer readers to [261] for more rigorous discussions on the relationship between IS, NCE, and DropOut.

For document retrieval Huang et al. [164] approximate the cross-entropy loss of Equation 3.52 by replacing $D$ with $D'$—where, $D' = \{d^+\} \cup D^-$ and $D^-$ is a fixed number of randomly sampled candidates. Mitra et al. [7] use a similar loss function but focus on the document re-ranking task where the neural model needs to distinguish the relevant documents from less relevant (but likely not completely non-relevant) candidates. Therefore, in their work the re-ranking model is trained with negative examples which comprise of documents retrieved by an existing IR system but manually judged as less relevant, instead of being sampled uniformly from the collection. IS, NCE, NEG, and these other sampling based approaches approximate the comparison with the full collection based on a sampled subset. For additional notes on these approaches, we refer the readers to [263–265].

In a typical retrieval scenario, however, multiple documents may be relevant to the same query $q$, and the notion of relevance among this set of documents $D^+$ may be further graded. Some LTR approaches consider pairs of documents for the same query and minimize the average number of inversions in ranking—i.e., $d_i \succ_q d_j$ but $d_j$ is ranked higher than $d_i$. The pairwise loss employed in these approaches has the following form [266],

$$\mathcal{L}_{pairwise} = \phi(s_i - s_j) \quad (3.75)$$

where, some possible choices for $\phi$ include,

- Hinge function $\phi(z) = \max(0, 1-z)$ [267, 268]

- Exponential function $\phi(z) = e^{-z}$ [269]

- Logistic function $\phi(z) = \log(1 + e^{-z})$ [235]

**RankNet loss** RankNet [235] is a *pairwise* loss function that has been a common choice for training neural LTR models and was also for many years an industry favourite, such as at the commercial Web search engine Bing.[6] Under the RankNet loss, the model is trained on triples $\langle q, d_i, d_j \rangle$ consisting of a query $q$ and a pair of documents $d_i$ and $d_j$ with different relevance labels—such that $d_i$ is more relevant than $d_j$ (i.e., $d_i \succ_q d_j$)—and corresponding feature vectors $\langle \vec{x}_i, \vec{x}_j \rangle$. The model $f :$ $\mathbb{R}^n \to \mathbb{R}$, typically a neural network but can also be any other machine learning model whose output is differentiable with respect to its parameters, computes the scores $s_i = f(\vec{x}_i)$ and $s_j = f(\vec{x}_j)$, where ideally $s_i > s_j$. Given the scores $\langle s_i, s_j \rangle$, the probability that $d_i$ would be ranked higher than $d_j$ is given by,

$$p_{ij} \equiv p(s_i > s_j) \equiv \frac{1}{1 + e^{-\sigma(s_i - s_j)}} \tag{3.76}$$

---

[6]https://www.microsoft.com/en-us/research/blog/ranknet-a-ranking-retrospective/

Where, the constant $\sigma$ determines the shape of the sigmoid. During training, the probability of ranking $d_i$ higher than $d_j$ for $q$ is maximized. Let $S_{ij} \in \{-1, 0, +1\}$ be the true preference label between $d_i$ and $d_j$ for the training sample— denoting $d_i$ is less, equal, or more relevant than $d_j$, respectively. Then the desired probability of ranking $d_i$ over $d_j$ is given by $\bar{p}_{ij} = \frac{1}{2}(1 + S_{ij})$. The cross-entropy loss $\mathscr{L}$ between the desired probability $\bar{p}_{ij}$ and the predicted probability $p_{ij}$ is given by,

$$\mathscr{L} = -\bar{p}_{ij} log(p_{ij}) - (1 - \bar{p}_{ij}) log(1 - p_{ij}) \tag{3.77}$$

$$= \frac{1}{2}(1 - S_{ij})\sigma(s_i - s_j) + log(1 + e^{-\sigma(s_i - s_j)}) \tag{3.78}$$

$$= log(1 + e^{-\sigma(s_i - s_j)}) \quad \text{if, } d_i \succ_q d_j (S_{ij} = 1) \tag{3.79}$$

Note that $\mathscr{L}$ is differentiable with respect to the model output $s_i$ and hence the model can be trained using gradient descent. We direct the interested reader to [270] for more detailed derivations for computing the gradients for RankNet.

Readers should note the obvious connection between the CE loss described previously and the RankNet loss. If in the denominator of Equation 3.52, we only sum over a pair of relevant and non-relevant documents then it reduces to the logistic-loss function of RankNet described in Equation 3.79. So, at the level of a single training sample, the key distinction between the two is whether we compare the relevant document to a single less relevant candidate or the full collection. However, in case of RankNet, it is important to consider how the pairs are sampled as the training is influenced by their distribution.

The key limitation of pairwise objective functions is that the rank inversion of any pair of documents is considered equally harmful. This is, however, generally untrue for most IR metrics where a significantly large penalty is associated with inversions at the top rank positions. For example, consider two different result lists for the same query—result list A ranks two relevant documents at position one and 50, while result list B ranks the same two relevant documents at positions three and 40. While the result set A has more rank inversions compared to result set B (48 vs. 40), it would fare better on typical IR metrics, such as NDCG. Therefore, to optimize for a rank-based metric we need to incorporate listwise objectives—that are sensitive to these differences—in our model training. However, the rank-based metrics are generally non-continuous and non-differentiable, which makes them difficult to incorporate in the loss function.

**LambdaRank loss** Burges et al. [271] make two key observations: (i) the gradient should be bigger for pairs of documents that produce a bigger impact in NDCG by swapping positions, and (ii) to train a model we don't need the costs themselves, only the gradients (of the costs w.r.t model scores). This leads to the LambdaRank loss which weights the gradients from the RankNet loss by the NDCG delta that would result from swapping the rank position of the pair of documents.

$$\lambda_{LambdaRank} = \lambda_{RankNet} \cdot |\Delta NDCG| \quad (3.80)$$

This formulation of LambdaRank can optimize directly for NDCG [272, 273], and any other IR measure by incorporating the corresponding delta change in Equation 3.80.

**ListNet and ListMLE loss** The probability of observing a particular rank order can be estimated from the individual document scores using different models [274–276]. For example, according to the Luce model [274], given four items $\{d_1, d_2, d_3, d_4\}$ the probability of observing a particular rank-order, say $[d_2, d_1, d_4, d_3]$, is given by:

$$p(\pi|s) = \frac{\phi(s_2)}{\phi(s_1) + \phi(s_2) + \phi(s_3) + \phi(s_4)} \times \frac{\phi(s_1)}{\phi(s_1) + \phi(s_3) + \phi(s_4)} \times \frac{\phi(s_4)}{\phi(s_3) + \phi(s_4)} \quad (3.81)$$

Where, $\pi$ is a particular permutation and $\phi$ is a transformation (e.g., linear, exponential, or sigmoid) over the score $s_i$ corresponding to item $d_i$. Using this model, we can compute the probability distribution over all possible permutations based on the model scores and the ground truth labels. The K-L divergence between these two distributions gives us the ListNet loss [277].

However, computing the probability distribution over all possible permutations is computationally expensive, even when restricted to only the top-K items. The ListMLE loss [278] instead computes the probability of the ideal permutation based on the ground truth. However, with categorical labels more than one ideal permutation may be possible which should be handled appropriately.

Many of the challenges discussed in this section are common to both retrieval tasks as well as multiclass and multilabel classification with extremely large number of classes—often referred to as *extreme classification* [279–281]. Ad hoc retrieval can be posed as an extreme classification task under a binary notion of relevance and a fixed collection constraint. New loss functions (*e.g.* the spherical loss family [282–284]) have been explored for these large scale classification tasks which may be relevant for neural retrieval research. The problem of learning from sparse biased labels [285, 286] is also an important challenge in these frameworks. Finally, deep neural models for LTR with large number of parameters may require large training data for supervised learning. Alternative training schemes—*e.g.*, using weak supervision signals [287, 288] or adversarial learning [141, 289]—are emerging.

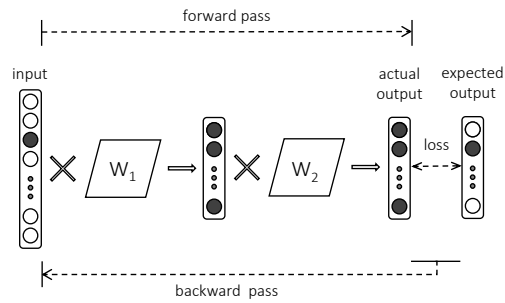

**(a)** A neural network with a single hidden layer.

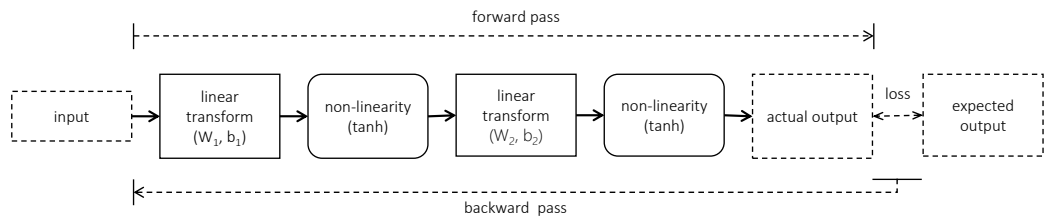

**(b)** The same neural network viewed as a chain of computational steps.

**Figure 3.13:** Two different visualizations of a feed-forward neural network with a single hidden layer. In (a), the addition of the bias vector and the non-linearity function is implicit. Figure (b) shows the same network but as a sequence of computational nodes. Most neural network toolkits implement a set of standard computational nodes that can be connected to build more sophisticated neural architectures.

## 3.5 Deep neural networks

Deep neural network models consist of chains of tensor operations. The tensor operation can range from parameterized linear transformations (*e.g.*, multiplication with a weight matrix, or the addition of a bias vector) to elementwise application of non-linear functions, such as *tanh* or *rectified linear units* (ReLU) [290–292]. Figure 3.13 shows a simple *feed-forward* neural network with *fully-connected* layers. For an input vector $\vec{x}$, the model produces the output $\vec{y}$ as follows,

$$\vec{y} = tanh(W_2 \cdot tanh(W_1 \cdot \vec{x} + \vec{b}_1) + \vec{b}_2) \qquad (3.82)$$

The model training involves tuning the parameters $W_1$, $\vec{b}_1$, $W_2$, and $\vec{b}_2$ to minimize the loss between the expected output and the output predicted by the final layer.

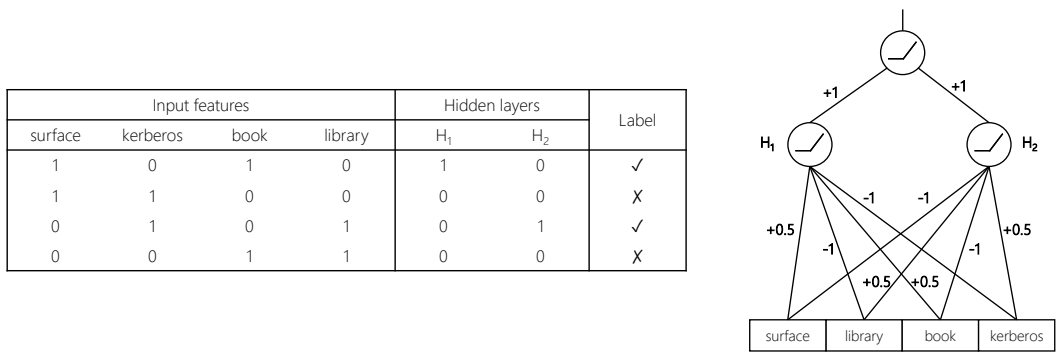

**Figure 3.14:** Consider a toy binary classification task on a corpus of four short texts—"surface book", "kerberos library", "library book", and "kerberos surface"—where the model needs to predict if the text is related to computers. The first two texts—"Surface Book" and "kerberos library"—are positive under this classification, and the latter two negative. The input feature space consists of four binary features that indicate whether each of the four terms from the vocabulary is present in the text. The table shows that the specified classes are not linearly separable with respect to the input feature space. However, if we add couple of hidden nodes, as shown in the diagram, then the classes can be linearly separated with respect to the output of the hidden layer.

The parameters are usually trained discriminatively using backpropagation [293–295]. During forward-pass each layer generates an output conditioned on its input, and during backward pass each layer computes the error gradient with respect to its parameters and its inputs.

The design of a DNN typically involves many choices of architectures and hyper-parameters. Neural networks with as few a single hidden layer—but with sufficient number of hidden nodes—can theoretically approximate any function [296]. In practice, however, deeper architectures—sometimes with as many as 1000 layers [297]—have been shown to perform significantly better than shallower networks. For readers who are less familiar with neural network models, we present a simple example in Figure 3.14 to illustrate how hidden layers enable these models to capture non-linear relationships. We direct readers to [298] for further discussions on how additional hidden layers help.

The rest of this section is dedicated to the discussion of input representations and standard architectures for deep neural models.

### 3.5.1 Input text representations

Neural models that learn representations of text take raw text as input. A key consideration is how the text should be represented at the input layer of the model. Figure 3.15 shows some of the common input representations of text.

Some neural models [256, 299–301] operate at the character-level. In these models, each character is typically represented by a one-hot vector. The vector dimensions—referred to as *channels*—in this case equals the number of allowed characters in the vocabulary. These models incorporate the least amount of prior knowledge about the language in the input representation—for example, these models are often required to learn about tokenization from scratch by treating space as just another character in the vocabulary. The representation of longer texts, such as sentences, can be derived by concatenating or summing the character-level vectors as shown in Figure 3.15a.

The input text can also be pre-tokenized into terms—where each term is represented by either a sparse vector or using pre-trained term embeddings (Figure 3.15d). Terms may have a one-hot (or local) representation where each term has an unique ID (Figure 3.15b), or the term vector can be derived by aggregating one-hot vectors of its constituting characters (or character *n*-graphs) as shown in Figure 3.15c. If pre-trained embeddings are used for term representation, then the embedding vectors can be further tuned during training or kept fixed.

Similar to character-level models, the term vectors are further aggregated (by concatenation or sum) to obtain the representation of longer chunks of text, such as sentences. While one-hot representations of terms (Figure 3.15b) are common in many NLP tasks, historically pre-trained embeddings (*e.g.*, [302, 303]) and character *n*-graph based representations (*e.g.*, [7, 164]) are more commonplace in IR.

### 3.5.2 Architectures

In this section, we describe few standard neural architectures commonly used in IR. For broader overview of neural architectures and design patterns please refer to [21, 189, 293].

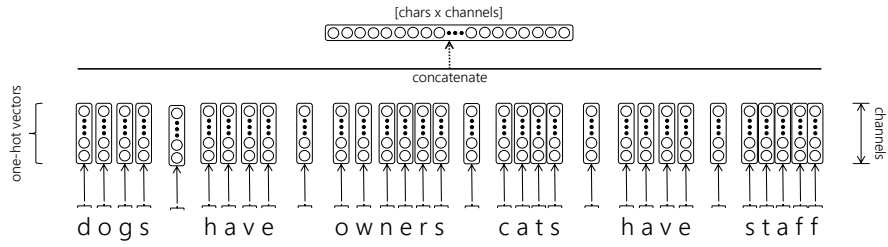

**(a)** Character-level input

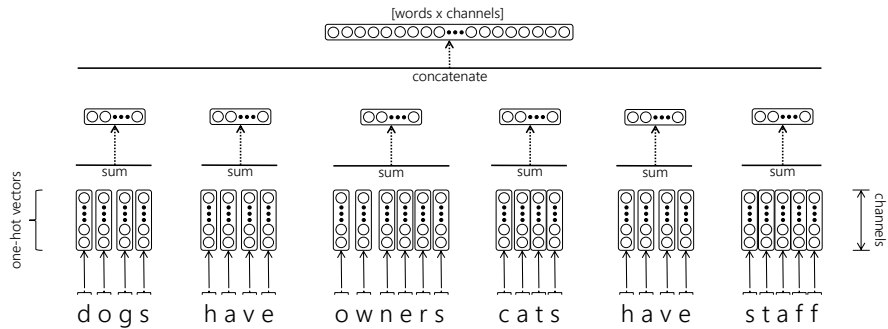

**(b)** Term-level input w/ bag-of-characters per term

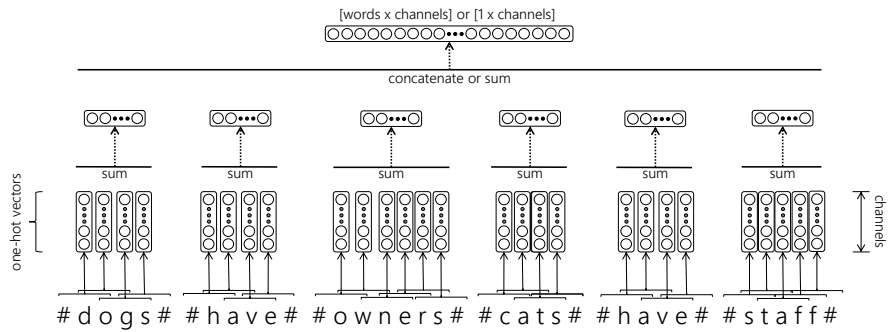

**(c)** Term-level input w/ bag-of-trigraphs per term

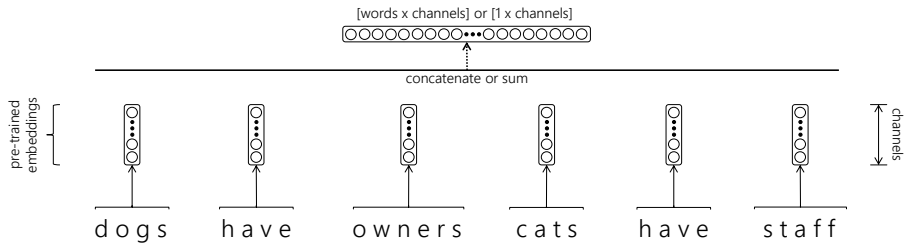

**(d)** Term-level input w/ pre-trained term embeddings

**Figure 3.15:** Examples of different representation strategies for text input to deep neural network models. The smallest granularity of representation can be a character or a term. The vector can be a sparse local representation, or a pre-trained embedding.

**Shift-invariant neural operations** Convolutional [20, 292, 304, 305] and recurrent [306–309] architectures are commonplace in many deep learning applications. These neural operations are part of a broader family of shift-invariant architectures. The key intuition behind these architectures stem from the natural regularities observable in most inputs. In vision, for example, the task of detecting a face should be invariant to whether the image is shifted, rotated, or scaled. Similarly, the meaning of an English sentence should, in most cases, stay consistent independent of which part of the document it appears in. Therefore, intuitively a neural model for object recognition or text understanding should not learn an independent logic for the same action applied to different parts of the input space. All shift-invariant neural operations fundamentally employ a window-based approach. A fixed size window moves over the input space with fixed stride in each step. A (typically parameterized) function—referred to as a *kernel*, or a *filter*, or a *cell*—is applied over each instance of the window. The parameters of the cell are shared across all the instances of the input window. The shared parameters not only imply a smaller number of total parameters in the model, but also more supervision per parameter per training sample due to the repeated application.

Figure 3.16a shows an example of a cell being applied on a sequence of terms—with a window size of three terms—in each step. A common cell implementation involves multiplying with a weight matrix—in which case the architecture in Figure 3.16a is referred as *convolutional*. An example of a cell without any parameters is *pooling*—which consists of aggregating (*e.g.*, by computing the max or the average per channel) over all the terms in the window. Note, that the length of the input sequence can be variable in both cases and the length of the output of a convolutional (or pooling) layer is a function of the input length. Figure 3.16b shows an example of *global pooling*—where the window spans over the whole input—being applied on top of a convolutional layer. The global pooling strategy is common for generating a fixed size output from a variable length input.[7]

In convolution or pooling, each window is applied independently. In con-trast, in the *recurrent* architecture of Figure 3.16e the cell not only considers the input window but also the output of the previous instance of the cell as its input. Many different cell architectures have been explored for recurrent neural networks (RNN)—although Elman network [310], Long Short-Term Memory (LSTM) [309], and Gated Recurrent Unit (GRU) [311, 312] are commonly used. RNNs are typically applied to sequences but can also be useful for two (and higher) dimensional inputs [313].

---

[7]It may be obvious, but worth pointing out, that a *global convolutional* layer is exactly the same as a fully-connected layer.

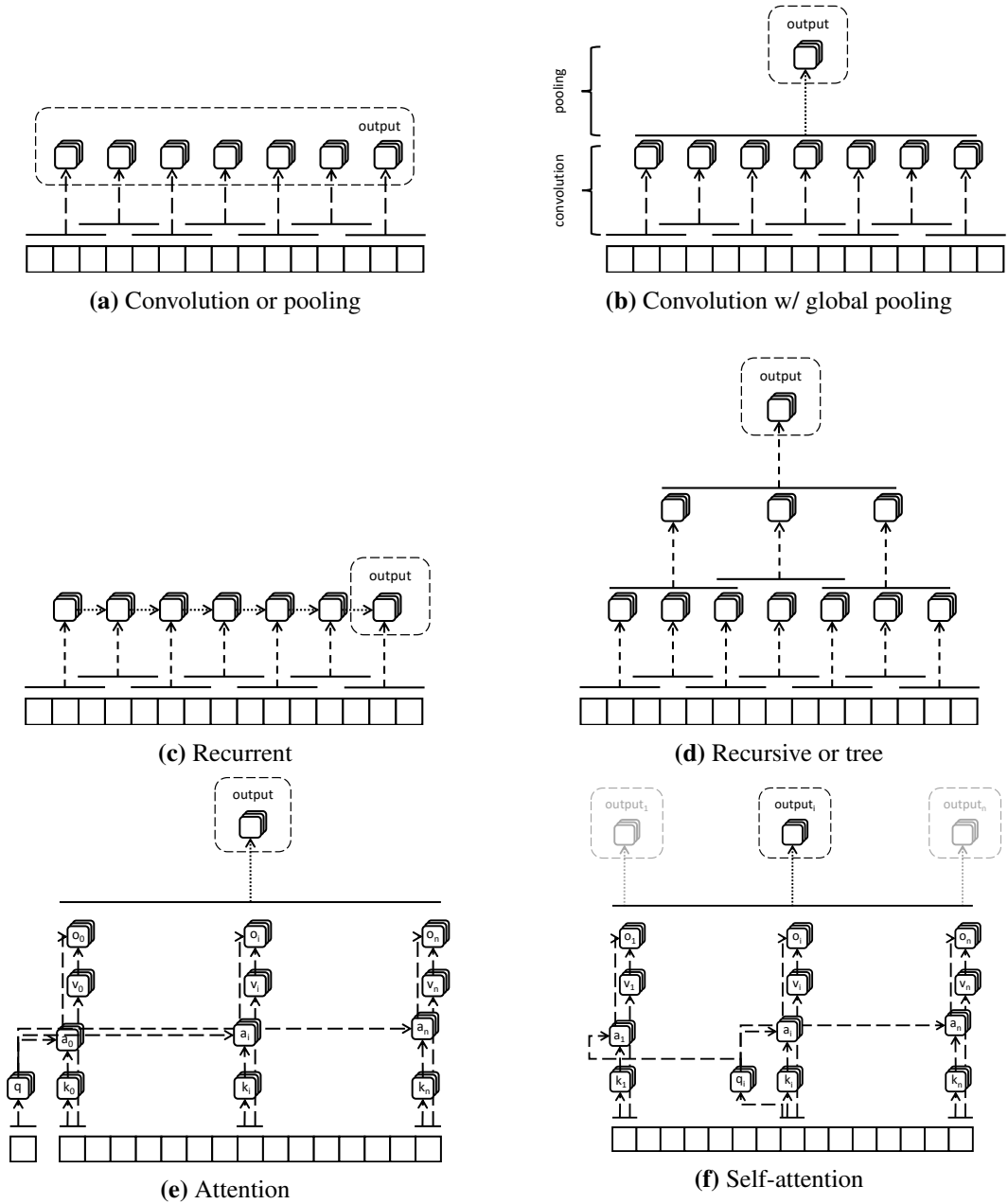

**Figure 3.16:** Standard shift-invariant neural architectures including convolutional neural networks (CNN), recurrent neural networks (RNN), pooling layers, tree-structured neural networks, attention layer, and self-attention layer.

One consideration when using convolutional or recurrent layers is how the window outputs are aggregated. Convolutional layers are typically followed by pooling or fully-connected layers that perform a global aggregation over all the window instances. While a fully-connected layer is aware of each window position, a global pooling layer is typically agnostic to it. However, unlike a fully-connected layer, a global max-pooling operation can be applied to a variable size input. Where a global aggregation strategy may be less appropriate (*e.g.*, long sequences), recurrent networks with memory [314–316] and/or attention [44, 317–320] may be useful. Figure 3.16e shows *tree-structured* (or *recursive*) neural networks [321–325] where the same cell is applied at multiple levels in a tree-like hierarchical fashion resulting in a recursive aggregation strategy.

Finally, attention mechanisms—in particular, self-attention [326]—have demonstrated remarkable usefulness for many NLP and IR tasks. In a typical attention setting, we have a set of *n* items that we can attend over and an input context, and we produce a probability distribution $\{a_1, \ldots, a_i, \ldots, a_n\}$ of attending to each item as a function of similarity between a learned representation $q$ of the context and learned representations $k_i$ of the items. The final output $o$ is the aggregate of learned value $v_i$ corresponding to each item weighted by their attention probabilities.

$$o = \sum_i^n \frac{\phi(q, k_i)}{\sum_j^n \phi(q, k_j)} \times v_i \quad (3.83)$$

In self-attention, we repeat the above process *n* times treating one of the *n* items themselves as the context in each case. Self-attention layers have been operationalized in Transformer-based [326] architectures, *e.g.*, BERT [327].

**Auto-encoders** The autoencoder architecture [294, 328, 329] is based on the *information bottleneck method* [201]. The goal is to learn a compressed representation $\vec{x} \in \mathbb{R}^k$ of items from their higher-dimensional vector representations $\vec{v} \in \mathbb{R}^K$, such that $k \ll K$. The model has an hour-glass shape as shown in Figure 3.17a and is trained by feeding in the high-dimensional vector inputs and trying to re-construct the same representation at the output layer. The lower-dimensional middle layer forces the encoder part of the model to extract the *minimal sufficient statistics* of $\vec{v}$ into $\vec{x}$, such that the decoder part of the network can reconstruct the original input back from $\vec{x}$. The model is trained by minimizing the reconstruction error between the input $\vec{v}$ and the actual output of the decoder $\vec{v'}$. The squared-loss is commonly employed.

$$\mathcal{L}_{autoencoder}(\vec{v}, \vec{v'}) = \|\vec{v} - \vec{v'}\|^2 \tag{3.84}$$

**Siamese networks** Siamese networks were originally proposed for comparing fingerprints [330] and signatures [331]. Yih et al. [332] later adapted the same architecture for comparing short texts. The siamese network, as seen in Figure 3.17b, resembles the autoencoder architecture (if you squint hard enough)—but unlike the latter is trained on pairs of inputs $\langle input_1, input_2 \rangle$. The architecture consists of two models ($model_1$ and $model_2$) that project $input_1$ and $input_2$, respectively, to $\vec{v}_1$ and $\vec{v}_2$ in a common latent space. A pre-defined metric (*e.g.*, cosine similarity) is used to then compute the similarity between $\vec{v}_1$ and $\vec{v}_2$. The model parameters are optimized such that $\vec{v}_1$ and $\vec{v}_2$ are closer when the two inputs are expected to be similar, and further away otherwise.

One possible loss function is the logistic loss. If each training sample consist of a triple $\langle \vec{v}_q, \vec{v}_{d1}, \vec{v}_{d2} \rangle$, such that $sim(\vec{v}_q, \vec{v}_{d1})$ should be greater than $sim(\vec{v}_q, \vec{v}_{d2})$, then we minimize,

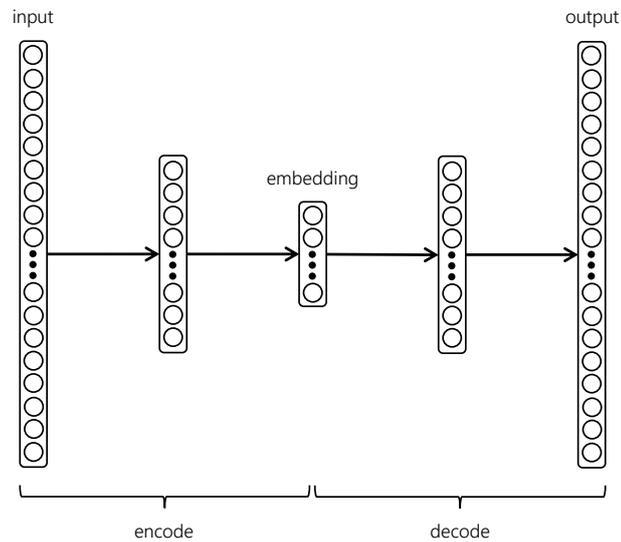

**(a)** Autoencoder

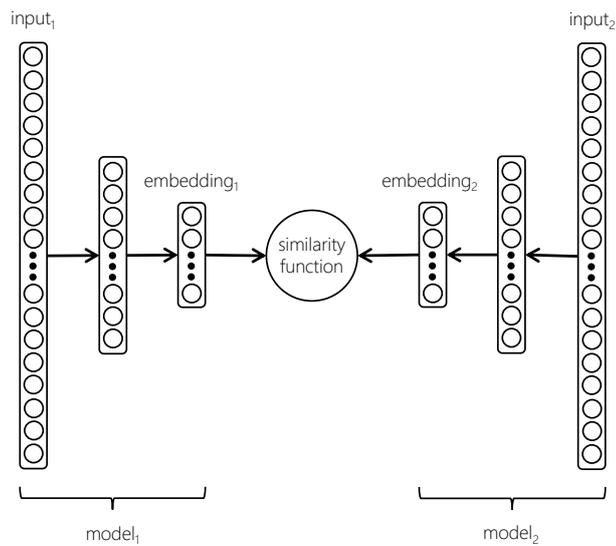

**(b)** Siamese network

**Figure 3.17:** Both (a) the autoencoder and (b) the Siamese network architectures are designed to learn compressed representations of inputs. In an autoencoder the embeddings are learnt by minimizing the self-reconstruction error, whereas a Siamese network focuses on retaining the information that is necessary for determining the similarity between a pair of items (say, a query and a document).

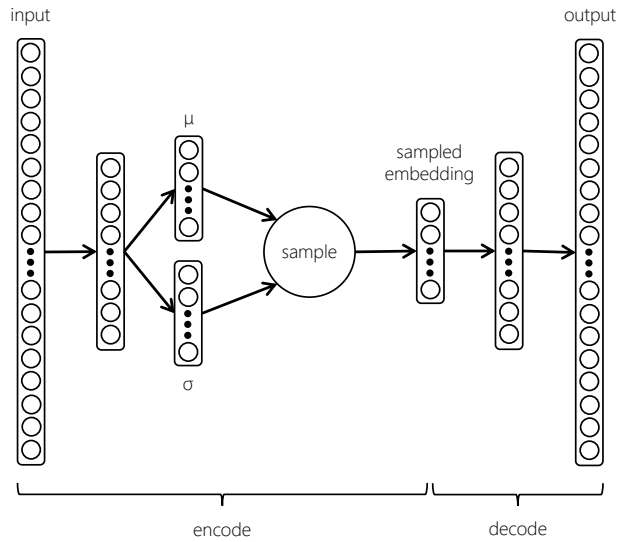

**Figure 3.18:** Instead of directly generating an encoded representation, variational autoencoders sample the latent vector from the generated vector of means $\mu$ and standard deviations $\sigma$. This local variation forces the model to learn a smoother and more continuous latent space.

$$\mathcal{L}_{siamese}(\vec{v}_q, \vec{v}_{d1}, \vec{v}_{d2}) = log\left(1 + e^{-\gamma(sim(\vec{v}_q, \vec{v}_{d1}) - sim(\vec{v}_q, \vec{v}_{d2}))}\right) \quad (3.85)$$

Where, $\gamma$ is a constant that is often set to 10. Typically, both the models—$model_1$ and $model_2$—share identical architectures, but can also choose to share the same parameters. In image retrieval, the contrastive loss [243, 244] is also used for training Siamese networks.

It is important to note that, unlike the autoencoder, the minimal sufficient statistics retained by a Siamese network is dictated by which information it deems important for determining the similarity between the paired items.

**Variational autoencoders (VAE)** In Variational autoencoders [333, 334], the encoder part of the network generates two separate vectors—the vector of means $\mu$ and the vector of standard deviations $\sigma$. The latent representation $\vec{x}$ of the input is then generated by sampling a random variable $x_i$ with mean $\mu_i$ and standard deviation $\sigma_i$ along each of the $k$ latent dimensions.

$$\vec{x} = [x_0 \sim N(\mu_0, \sigma_0^2), \ldots, x_i \sim N(\mu_i, \sigma_i^2), \ldots, x_{k-1} \sim N(\mu_{k-1}, \sigma_{k-1}^2)] \quad (3.86)$$

By sampling the latent representation, we expose the decoder to a certain degree of local variations in its input that should force the model to learn a smoother continuous latent space. The VAE is trained by jointly minimizing the reconstruction loss—similar to vanilla autoencoders—and an additional component to the loss function which is the KL-divergence between the latent variable $x_i$ and a unit gaussian.

$$\mathscr{L}_{\text{VAE}} = \mathscr{L}_{reconstruction} + \mathscr{L}_{KL-divergence} \quad (3.87)$$

$$= \|\vec{v} - \vec{v'}\|^2 + \sum_i^k \sigma_i^2 + \mu_i^2 - \log(\sigma_i) - 1 \quad (3.88)$$

Without the $\mathscr{L}_{KL-divergence}$ component the model can learn very different $\mu$ for different classes of inputs and minimize the $\lambda$ to be arbitrarily small such that the learnt latent space is no longer smooth or continuous. Readers should note that the sampling step is non-differentiable, but the model can be trained using the "reparameterization trick" proposed by Kingma and Welling [333].

An important application of VAE is for the synthesis of new items (*e.g.*, images [335] or text [336]) not observed in the training collection. Another class of techniques for synthesis includes the Generative Adversarial Networks.

**Generative Adversarial Networks (GAN)** Goodfellow et al. [337] proposed a framework for training generative models under an adversarial setting. GANs typically consist of two separate neural networks—a *generator* network and a *discriminator* network. The goal of the generator network is to synthesize new (fake) items that mimic similar distributions as items that exist in the training collection. The goal of the discriminator network is to correctly distinguish between a true item and an item produced by the generator. The generator is trained to maximize the probability of the discriminator wrongly classifying the true and the generated item— which corresponds to a minimax two-player game.

### 3.5.3 Neural toolkits

In recent years, the advent of numerous flexible toolkits [338–345] has had a catalytic influence on the area of neural networks. Most of the toolkits define a set of common neural operations that—like Lego[8] blocks—can be composed to build complex network architectures. Each instance of these neural operations or *computation nodes* can have associated learnable parameters that are updated during training, and these parameters can be shared between different parts of the network if necessary. Every computation node under this framework must implement the appropriate logic for,

- computing the output of the node given the input (forward-pass)

- computing the gradient of the loss with respect to the inputs, given the gradient of the loss with respect to the output (backward-pass)

- computing the gradient of the loss with respect to its parameters, given the gradient of the loss with respect to the output (backward-pass)

A deep neural network, such as the one in Figure 3.13 or ones with much more complex architectures (*e.g.*, [297, 346, 347]), can then be specified by chaining instances of these available computation nodes, and trained end-to-end on large datasets using backpropagation over GPUs or CPUs. In IR, various application interfaces [348, 349] bind these neural toolkits with existing retrieval/indexing frameworks, such as Indri [156]. Refer to [350] for a comparison of different neural toolkits based on their speed of training using standard performance benchmarks.

## 3.6 Deep neural models for IR

Traditionally, deep neural network models have much larger number of learnable parameters than their shallower counterparts. A DNN with a large set of parameters can easily overfit to smaller training datasets [351]. Therefore, during model design

---

[8]`https://en.wikipedia.org/wiki/Lego`

it is typical to strike a balance between the number of model parameters and the size of the data available for training. Data for ad hoc retrieval mainly consists of,

- Corpus of search queries

- Corpus of candidate documents

- Ground truth—in the form of either explicit human relevance judgments or implicit labels (*e.g.*, from clicks)—for query-document pairs

While both large scale corpora of search queries [85, 352] and documents [353–355] are publicly available for IR research, the amount of relevance judgments that can be associated with them are often limited outside of large industrial research labs—mostly due to user privacy concerns. We note that we are interested in datasets where the raw text of the query and the document is available. Therefore, this excludes large scale public labelled datasets for learning-to-rank (*e.g.*, [356]) that don't contain the textual contents.

The proportion of labelled and unlabelled data that is available influences the *level of supervision* that can be employed for training these deep models. Most of the models we covered in Section 3.3 operate under the data regime where large corpus of documents or queries is available, but limited (or no) labelled data. Under such settings where no direct supervision or relevance judgments is provided, typically an *unsupervised* approach is employed (*e.g.*, using auto-encoding [357] or masked language modeling [327]). The unlabelled document (or query) corpus is used to learn good text representations, and then these learnt representations are incorporated into an existing retrieval model or a query-document similarity metric. If small amounts of labelled data are available, then that can be leveraged to train a retrieval model with few parameters that in turn uses text representations that is pre-trained on larger unlabelled corpus. Examples of such *semi-supervised* training includes models such as [163, 238, 302]. In contrast, *fully-supervised* models—*e.g.*, [7, 8, 164, 358, 359]—optimize directly for the target task by training on large number of labelled query-document pairs.

It is also useful to distinguish between deep neural models that focus on ranking long documents, from those that rank short texts (*e.g.*, for the question-answering task, or for document ranking where the document representation is based on a short text field like title). The challenges in short text ranking are somewhat distinct from those involved in the ad hoc retrieval task [360]. When computing similarity between pairs of short-texts, vocabulary mismatches are more likely than when the retrieved items contain long text descriptions [95]. Neural models that perform matching in a latent space tend to be more robust towards the vocabulary mismatch problem compared to lexical term-based matching models. On the other hand, documents with long body texts may contain mixture of many topics and the query matches may be spread over the whole document. A neural document ranking model must effectively aggregate the relevant matches from different parts of a long document. In the rest of this section, we discuss several neural architectures and approaches to document ranking.

### 3.6.1 Document auto-encoders

Salakhutdinov and Hinton [357] proposed *Semantic Hashing*—one of the earliest deep neural models for ad hoc retrieval. The model is a deep autoencoder trained under unsupervised setting on unlabelled document collection. The model considers each document as a bag-of-terms and uses one-hot vector representation for the terms—considering only top two thousand most frequent terms in the corpus after removing stopwords. Salakhutdinov and Hinton [357] first pre-train the model layer-by-layer, and then train it further end-to-end for additional tuning. After fine tuning the output of the model are thresholded to generate binary vector encoding of the documents. Given a search query, a corresponding hash is generated, and the relevant candidate documents quickly retrieved that match the same hash vector. A standard IR model can then be employed to rank between the selected documents.

Semantic hashing is an example of a document encoder based approach to IR. Variational autoencoders have also been explored [361] on similar lines. While vocabulary sizes of few thousand distinct terms may be too small for most practical IR tasks, a larger vocabulary or a different term representation strategy—such as the character trigraph based representation of Figure 3.15c—may be considered in practice. Another shortcoming of the autoencoder architecture is that it minimizes the document reconstruction error which may not align well with the goal of the target IR task. A better alternative may be to train on query-document paired data where the choice of what constitutes as the minimal sufficient statistics of the document is influenced by what is important for determining relevance of the document to likely search queries. In line with this intuition, we next discuss the Siamese architecture based models.

### 3.6.2 Siamese networks

In recent years, several deep neural models based on the Siamese architecture have been explored especially for short text matching. The *Deep Semantic Similarity Model* (DSSM) [164] is one such architecture that trains on query and document title pairs where both the pieces of texts are represented as bags-of-character-trigraphs. The DSSM architecture consists of two deep models—for the query and the document—with all fully-connected layers and cosine distance as the choice of similarity function in the middle. Huang et al. [164] proposed to train the model on clickthrough data where each training sample consists of a query $q$, a positive document $d^+$ (a document that was clicked by a user on the SERP for that query), and a set of negative documents $D^-$ randomly sampled with uniform probability from the full collection. The model is trained my minimizing the cross-entropy loss,

$$\mathscr{L}_{dssm}(q, d^+, D^-) = -log\Big(\frac{e^{\gamma \cdot cos(\vec{q}, \vec{d^+})}}{\sum_{d \in D} e^{\gamma \cdot cos(\vec{q}, \vec{d})}}\Big) \tag{3.89}$$

$$\text{where,} \quad D = \{d^+\} \cup D^- \tag{3.90}$$

While, DSSM [164] employs deep fully-connected architecture for the query and the document models, more sophisticated architectures involving convolutional layers [231, 303, 362, 363], recurrent layers [364, 365], and tree-structured networks [324] have also been explored. The similarity function can also be parameter-ized and implemented as additional layers of the neural network as in [358]. Most of these models have been evaluated on the short text matching task, but Mitra et al. [7] recently reported meaningful performances on the long document ranking task from models like DSSM [164] and CDSSM [231] under telescoping evaluation. Mitra et al. [7] also show that sampling the negative documents uniformly from the collection is less effective to using documents that are closer to the query intent but judged as non-relevant by human annotators in similar evaluation settings.

**Table 3.3:** Comparing the nearest neighbours for "seattle" and "taylor swift" in the CDSSM embedding spaces when the model is trained on query-document pairs vs. query prefix-suffix pairs. The former resembles a topical notion of similarity between terms while the latter is more typical in the definition of inter-term similarities.

| seattle | | taylor swift | |
| --- | --- | --- | --- |
| Query-Document | Prefix-Suffix | Query-Document | Prefix-Suffix |
| weather seattle | chicago | taylor swift.com | lady gaga |
| seattle weather | san antonio | taylor swift lyrics | meghan trainor |
| seattle washington | denver | how old is taylor swift | megan trainor |
| ikea seattle | salt lake city | taylor swift twitter | nicki minaj |
| west seattle blog | seattle wa | taylor swift new song | anna kendrick |

**Notions of similarity** It is important to emphasize that our earlier discussion in Section 3.2.2 on different notions of similarity between terms that can be learnt by shallow embedding models is also relevant in the context of these deeper architectures. In the case of Siamese networks, such as the convolutional-DSSM (CDSSM) [231], the notion of similarity being modelled depends on the choice of the paired data that the model is trained on. When the CDSSM is trained on query and document title pairs [231] then the notion of similarity is more *topical* in nature. Mitra and Craswell [19] trained the same CDSSM architecture on query prefix-suffix pairs which, in contrast, captures a more *typical* notion of similarity, as shown in Table 7.2. In a related work, Mitra [18] demonstrated that the CDSSM model when trained on session-query pairs is amenable to vector-based text analogies.

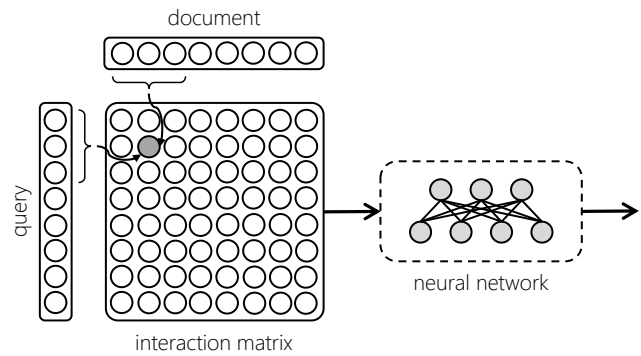

**Figure 3.19:** Schematic view of an interaction matrix generated by comparing windows of text from the query and the document. A deep neural network—such as a CNN—operates over the interaction matrix to find patterns of matches that suggest relevance of the document to the query.

$$\vec{v}_{\text{things to do in london}} - \vec{v}_{\text{london}} + \vec{v}_{\text{new york}} \approx \vec{v}_{\text{new york tourist attractions}} \quad (3.91)$$

$$\vec{v}_{\text{university of washington}} - \vec{v}_{\text{seattle}} + \vec{v}_{\text{denver}} \approx \vec{v}_{\text{university of colorado}} \quad (3.92)$$

$$\vec{v}_{\text{new york}} + \vec{v}_{\text{newspaper}} \approx \vec{v}_{\text{new york times}} \quad (3.93)$$

By modelling different notions of similarity these deep neural models tend to be more suitable for other IR tasks, such as query auto-completion [19] or session-based personalization [18].

### 3.6.3 Interaction-based networks

Siamese networks represent both the query and the document using single embedding vectors. Alternatively, we can individually compare different parts of the query with different parts of the document, and then aggregate these partial evidences of relevance. Especially, when dealing with long documents—that may contain a mixture of many topics—such a strategy may be more effective than trying to represent the full document as a single low-dimensional vector. Typically, in these approaches a sliding window is moved over both the query and the document text and each instance of the window over the query is compared (or "interacts") against The President of the United States of America (POTUS) is the elected head of state and head of government of the United States. The president leads the executive branch of the federal government and is the commander in chief of the United States Armed Forces. Barack Hussein Obama II (born August 4, 1961) is an American politician who is the 44th and current President of the United States. He is the first African American to hold the office and the first president born outside the continental United States.

<div align="center">(a) Lexical model</div>

The President of the United States of America (POTUS) is the elected head of state and head of government of the United States. The president leads the executive branch of the federal government and is the commander in chief of the United States Armed Forces. Barack Hussein Obama II (born August 4, 1961) is an American politician who is the 44th and current President of the United States. He is the first African American to hold the office and the first president born outside the continental United States.

<div align="center">(b) Semantic model</div>

**Figure 3.20:** Analysis of term importance for estimating the relevance of a passage to the query "United States President" by a lexical and a semantic deep neural network model. The lexical model only considers the matches of the query terms in the document but gives more emphasis to earlier occurrences. The semantic model is able to extract evidence of relevance from related terms such as "Obama" and "federal".

each instance of the window over the document text (see Figure 3.19). The terms within each window can be represented in different ways including, one-hot vectors, pre-trained embeddings, or embeddings that are updated during the model training. A neural model (typically convolutional) operates over the generated interaction matrix and aggregates the evidence across all the pairs of windows compared.

The interaction matrix based approach have been explored both for short text matching [302, 303, 366–369], as well as for ranking long documents [7, 238, 370, 371].

### 3.6.4 Lexical matching networks

Much of the explorations in neural IR models have focused on learning good representations of text. However, these representation learning models tend to perform poorly when dealing with rare terms and search intents. In Section 2.1.2, we highlighted the importance of modelling rare terms in IR. Based on similar motivaions, Guo et al. [163] emphasized the importance of modelling lexical matches using deep neural networks, and proposed to use histogram-based features in their DNN model to capture lexical notion of relevance. Neural models that focus on lexical matching typically have fewer parameters, and can be trained under small data

regimes—unlike their counterparts that focus on learning representations of text.

### 3.6.5 BERT

BERT-based [327] architectures have recently demonstrated significant performance improvements on retrieval tasks [15, 16]. The model architecture comprises of stacked Transformer [326] layers. The query and document are concatenated and then tokenized as a single sequence of subword terms for input. The relevance estimation task is cast as a binary classification problem—*i.e.*, given a query-document pair predict if they are relevant or nonrelevant—although other training objectives have also been explored [372].

## 3.7 Conclusion

We surveyed a large body work in this section. We introduced the fundamentals of traditional IR models and representation learning with neural networks. We presented some of the recent (shallow and deep) neural approaches for document ranking and question-answer matching. Readers should note that this is an active area for research, and new architectures and learning methods are continuously emerging. So, it is likely that by the time this thesis is published, many of the methods described here may have already been superseded by more recent and advanced methods. In the subsequent chapters of this thesis, we will cover our contributions in the form of new neural models and approaches for some of these IR tasks.

# Chapter 4

# Learning to rank with Duet networks

In traditional Web search, the query consists of only few terms but the body text of the documents may typically have tens or hundreds of sentences. In the absence of click information, such as for newly-published or infrequently-visited documents, the body text can be a useful signal to determine the relevance of the document for the query. Therefore, extending existing neural text representation learning approaches to long body text for document ranking is an important challenge in IR. However, as was noted previously [373], despite the recent surge in interests towards applying deep neural networks (DNN) for retrieval, their success on ad hoc retrieval tasks has been rather limited. Some papers [166, 238] report worse performance of neural embedding models when compared to traditional term-based approaches, such as BM25 [80].

Traditional IR approaches consider terms as discrete entities. The relevance of the document to the query is estimated based on, amongst other factors, the number of matches of query terms in the document, the parts of the document in which the matches occur, and the proximity between the matches. In contrast, latent semantic analysis (LSA) [170], probabilistic latent semantic analysis (PLSA) [198] and latent Dirichlet allocation (LDA) [200, 374] learn low-dimensional vector representations of terms, and match the query against the document in the latent semantic space. In Section 2.1, we emphasized the importance of both lexical and latent matching in IR. Lexical matching can be particularly important when the query terms are new or rare. On the other hand, matches between learned latent representations of query

The President of the United States of America (POTUS) is the elected head of state and head of government of the United States. The president leads the executive branch of the federal government and is the commander in chief of the United States Armed Forces. Barack Hussein Obama II (born August 4, 1961) is an American politician who is the 44th and current President of the United States. He is the first African American to hold the office and the first president born outside the continental United States.

The President of the United States of America (POTUS) is the elected head of state and head of government of the United States. The president leads the executive branch of the federal government and is the commander in chief of the United States Armed Forces. Barack Hussein Obama II (born August 4, 1961) is an American politician who is the 44th and current President of the United States. He is the first African American to hold the office and the first president born outside the continental United States.

**(a)** Local subnetwork

**(b)** Distributed subnetwork

**Figure 4.1:** Visualizing the drop in the local and the distributed subnetwork's retrieval score by individually removing each of the passage terms for the query "united states president". Darker green signifies a bigger drop. The local subnetwork uses only exact term matches. The distributed subnetwork uses matches based on a learned representation.

and document are important for addressing the vocabulary mismatch problem.

Retrieval models can be classified based on what representations of text they employ at the point of matching the query against the document. At the point of match, if each term is represented by a unique identifier (*local* representation [175]) then the query-document relevance is a function of the pattern of occurrences of the exact query terms in the document. However, if the query and the document text is first projected into a continuous latent space, then it is their distributed representations that are compared. Along these lines, Guo et al. [163] classify recent DNNs for short-text matching as either *interaction*-focused [302, 303, 366] or *representation*-focused [164, 231, 303, 358, 362]. They claim that IR tasks are different from NLP tasks, and that it is more important to focus on exact matching for the former and on learning text embeddings for the latter. Mitra et al. [165], on the other hand, claim that models that compare the query and the document in the latent semantic space capture a different sense of relevance than models that focus on exact term matches, and therefore the combination of the two is more favourable. Our work is motivated by the latter intuition that it is important to match the query and the document using both local and distributed representations of text. We propose a novel ranking model comprised of two separate DNNs that model query-document relevance using local and distributed representations, respectively. The two DNNs, referred to henceforth as the *local subnetwork* and the *distributed subnetwork*, are jointly trained as part of a single model, that we name as the *Duet* network because the two subnetworks co-operate to achieve a common goal. Figure 4.1 demonstrates how each subnetwork models the same document given a fixed query. While the local subnetwork captures properties like exact match position and proximity, the distributed subnetwork detects synonyms (*e.g.*, 'Obama'), related terms (*e.g.*, 'federal'), and even well-formedness of content (*e.g.*, 'the', 'of').[1]

In this chapter, we show that the combination of the two DNNs not only outperforms the individual subnetworks, but also demonstrates large improvements over traditional baselines and other previously proposed models based on DNNs on the document ranking task. Unlike other previous work [166, 238], our model significantly outperforms classic IR approaches by using a DNN to learn text representation.

Deep neural network models are known to benefit from large training data, achieving state-of-the-art performance in areas where large scale training corpora are available [21, 256]. Some of the lack of positive results from neural models in ad hoc retrieval is likely due to the scarce public availability of large quantity of training data necessary to learn effective representations of text. In Section 4.5, we will present some analysis on the effect of training data on the performance of these DNN models. In particular, we found that–unsurprisingly–the performance of the distributed model improves drastically in the presence of more data. Unlike some previous work [164, 231, 362] that train on clickthrough data with randomly sampled documents as negative examples, we train our model on human-judged labels. Our candidate set for every query consists of documents that were retrieved by the commercial search engine Bing, and then labelled by crowdsourced judges. We found that training with the documents that were rated non-relevant by the human judges as the negative examples is more effective than randomly sampling negative

---

[1] While surprising, this last property is important for detecting quality web content [375].

examples from the corpus.

In Section 4.4 we present additional improvements to the Duet network benchmarked on the MS MARCO passage ranking task [52] and TREC 2019 Deep Learning track [15].

To summarize, the key contributions of this chapter are:

1. We propose a novel Duet network that jointly learns two deep neural networks that match query and document based on their lexical similarity and similarity in their learned latent representations, respectively.

2. We demonstrate that Duet out-performs previous state-of-the-art neural and traditional non-neural baselines.

3. We demonstrate that training with documents judged as non-relevant as the negative examples is more effective than randomly sampling them from corpus.

4. We report additional improvement to the original Duet network evaluated on two recently released public benchmarks with sufficiently large training data.

## 4.1 The Duet network

Figure 4.2 provides a detailed schematic view of the Duet network. The distributed subnetwork projects the query and the document text into an embedding space before matching, while the local subnetwork operates over an interaction matrix comparing every query term to every document term. The final score under the Duet setting is the sum of scores from the local and the distributed subnetworks,

$$\text{Duet}(q,d) = \text{Duet}_{\text{local}}(q,d) + \text{Duet}_{\text{distrib}}(q,d) \tag{4.1}$$

Where both the query and the document are considered as ordered list of terms, $q = [t_{q_1}, \ldots, t_{q_{|q|}}]$ and $d = [t_{d_1}, \ldots, t_{d_{|d|}}]$. Each query term $t_q$ and document term $t_d$ is represented by a $m \times 1$ vector where $m$ is the input representation of the text (*e.g.*, the number of terms in the vocabulary for the local subnetwork). The query $q$ and the document $d$ is, in turn, represented by the matrices $X_q$ and $X_d$, respectively.

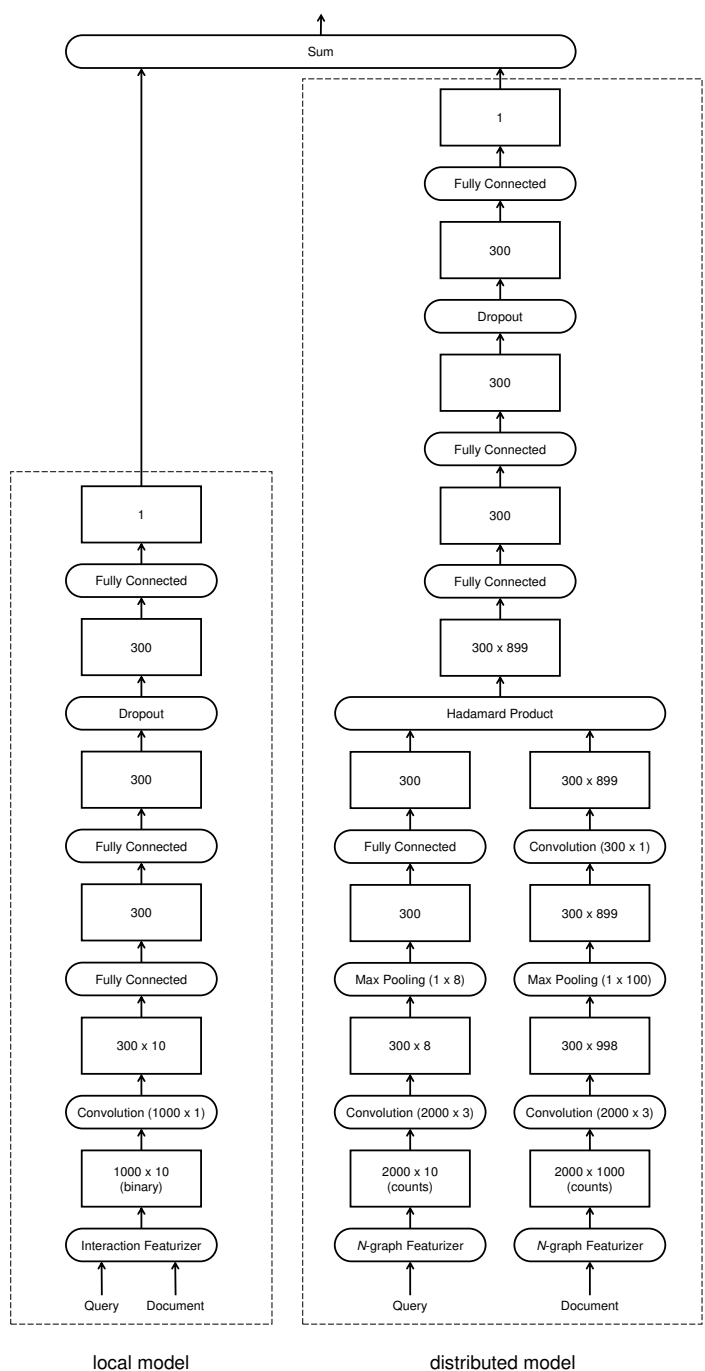

**Figure 4.2:** The Duet network is composed of the local subnetwork (left) and the distributed subnetwork (right). The local subnetwork takes an interaction matrix of query and document terms as input, whereas the distributed subnetwork learns embeddings of the query and the document text before matching. The parameters of both subnetworks are optimized jointly during training. Hyperparameters such as $n_{hidden}$ and $n_{pool}$ shown as in the document ranking task.

$$X_q = \begin{bmatrix} \begin{bmatrix} \vec{v}_{t_{q_1}} \end{bmatrix} \cdots \begin{bmatrix} \vec{v}_{t_{q_{|q|}}} \end{bmatrix} \end{bmatrix}, \qquad X_d = \begin{bmatrix} \begin{bmatrix} \vec{v}_{t_{d_1}} \end{bmatrix} \cdots \begin{bmatrix} \vec{v}_{t_{d_{|d|}}} \end{bmatrix} \end{bmatrix} \qquad (4.2)$$

We fix the length of the inputs across all the queries and the documents such that we consider only the first $n_q$ terms in the query and the first $n_d$ terms in the document. If either the query or the document is shorter than these target dimensions, then the input vectors are padded with zeros. The truncation of the document body text to the first $n_d$ terms is performed only for our subnetwork and its variants, but not for the baseline models. For all the neural and the non-neural baseline models we consider the full body text.

### 4.1.1 Local subnetwork

Match positions of the query terms in the document not only reflect where potentially the relevant parts of the document are localized (*e.g.*, title, first paragraph, closing paragraph) but also how clustered the individual query term matches are with each other. Figure 4.3 shows the position of matches on two different queries and a sample of relevant and non-relevant documents. In the first query, we see that the query term matches in the relevant document are much more clustered than in the non-relevant documents. We observe this behaviour also in the second query but in addition notice that the clustered matches are localized near the beginning of the relevant document. Match proximity serves as a foundation for traditional methods such as sequential dependence models [155].

The local subnetwork estimates document relevance based on patterns of exact matches of query terms in the document. To this end, each term is represented by its one-hot encoding in a $m_{local}$-dimensional space, where $m_{local}$ is the size of the vocabulary. The subnetwork then generates the $n_d \times n_q$ binary matrix $X = X_d^\top X_q$, capturing every exact match (and position) of query terms in the document. This interaction matrix is similar to the visual representation of term matches in Figure 4.3, and captures both the exact term matches and the match positions. It is also similar to the indicator matching matrix proposed previously by Pang et al. [302]. While the interaction matrix $X$ perfectly captures every query term match in the document, it does not retain any information about the actual terms themselves. Therefore, the local subnetwork cannot learn term-specific properties from the training corpus, nor model interactions between dissimilar terms.

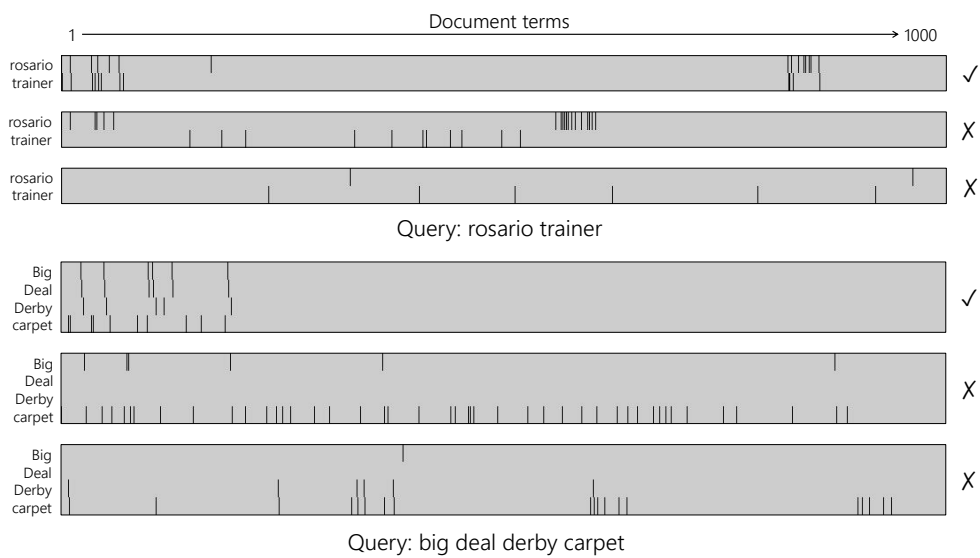

**Figure 4.3:** Visualizing patterns of query term matches in documents. Query terms are laid out along the vertical axis, and the document terms along the horizontal axis. The short vertical lines correspond to exact matches between pairs of query and document terms. For both queries, the first document was rated relevant by a human judge and the following two as non-relevant. The query term matches in the relevant documents are observed to be more clustered, and more localized near the beginning of the document.

The interaction matrix $X$ is first passed through a convolutional layer with $n_{hidden}$ filters, a kernel size of $n_d \times 1$, and a stride of 1. The output $Z_i$ corresponding to the $i^{th}$ convolutional window over $X$ is a function of the match between the $t_{q_i}$ term against all the terms in the document,

$$Z_i = \tanh\left(X_i^\top \cdot W\right) \qquad (4.3)$$

Where $X_i$ is the row $i$ of $X$, tanh is performed elementwise, and the $n_d \times n_{hidden}$ matrix $W$ contains the learnable parameters of the convolutional layer. The output $Z$ of the convolutional layer is a matrix of dimension $n_{hidden} \times n_q$. The output of the convolutional layer is then passed through two fully-connected layers, a dropout layer, and a final fully-connected layer that produces a single real-valued output. All the nodes in the local subnetwork uses the hyperbolic tangent function for non-linearity.

### 4.1.2  Distributed subnetwork

The distributed subnetwork learns dense lower-dimensional vector representations of the query and the document text, and then computes the positional similarity between them in the learnt embedding space. Instead of one-hot encoding of terms, as in the local subnetwork, we use a character *n*-graph based representation of each term in the query and document. Our *n*-graph based input encoding is motivated by the trigraph encoding proposed by Huang et al. [164], but unlike their approach we don't limit our input representation to *n*-graphs of a fixed length. For each term, we count all the *n*-graphs present for $1 \leq n \leq n_{maxgraph}$. We then use this *n*-graph frequency vector of length $m_{distrib}$ to represent the term.

Instead of directly computing the interaction between the $m_{distrib} \times n_q$ matrix $X_q$ and the $m_{distrib} \times n_d$ matrix $X_d$, we first learn a series of nonlinear transformations to the character-based input. For both the query and the document, the first step is convolution. The $m_{distrib} \times n_{window}$ convolution window has filter size of $n_{hidden}$. It projects $n_{window}$ consecutive terms to a $n_{hidden}$-dimensional vector, then takes a stride by 1 position, and projects the next $n_{window}$ terms, and so on. For the query, the convolution step generates a tensor of dimensions $n_{hidden} \times (n_q - n_{window} + 1)$. For the document, it generates one of dimensions $n_{hidden} \times (n_d - n_{window} + 1)$.

Following this, we conduct a max-pooling step. For the query the pooling kernel dimensions are $1 \times (n_q - n_{window} + 1)$. For the document, it is $1 \times n_{pool}$. Thus, we get one $n_{hidden}$-dimensional embedding $\vec{v}_q$ for the query and a $n_{hidden} \times (n_d - n_{window} - n_{pool} + 2)$ matrix $\tilde{X}_d$ for the document. The document matrix $\tilde{X}_d$ can be interpreted as $(n_d - n_{window} - n_{pool} + 2)$ separate embeddings, each corresponding to different equal-sized spans of text within the document. Our choice of a window-based max-pooling strategy, instead of global max-pooling as employed by CDSSM [362], is motivated by the fact that the window-based approach allows the model to distinguish between matches in different parts of the document. As posited in the previous section, a model that is aware of match positions may be more suitable when dealing with long documents, especially those containing mixture of many different topics.

The output of the max-pooling layer for the query is then passed through a fully-connected layer. For the document, the $n_{hidden} \times (n_d - n_{window} - n_{pool} + 2)$ dimensional matrix output is operated on by another convolutional layer with filter size of $n_{hidden}$, kernel dimensions of $n_{hidden} \times 1$, and a stride of 1. The combination of these convolutional and max-pooling layers enable the distributed subnetwork to learn suitable representations of text for effective inexact matching.

To perform the matching, we conduct the element-wise or Hadamard product between the embedded document matrix and the extended or broadcasted query embedding,

$$\tilde{X} = (\underbrace{\vec{v}_q \ldots \ldots \ldots \ldots \ldots \vec{v}_q}_{(n_d - n_{window} - n_{pool} + 2) \text{ times}}) \circ \tilde{X}_d \quad (4.4)$$

After this, we pass the matrix through fully connected layers, and a dropout layer until we arrive at a single score. Like the local subnetwork, we use hyperbolic tangent function here for non-linearity.

### 4.1.3 Optimization

Each training sample consists of a query $q$, a relevant document $d^+$ and a set of non-relevant documents $D^- = \{d_0, \ldots, d_{n_{neg}}\}$. We use a softmax function to compute the posterior probability of the positive document given a query based on the score.

$$p(d^+|q) = \frac{e^{\text{Duet}(q,d^+)}}{\sum_{d \in D} e^{\text{Duet}(q,d)}} \quad (4.5)$$

$$\text{where,} \quad D = \{d^+\} \cup D^- \quad (4.6)$$

We maximize the log likelihood $\log p(d^+|q)$ using stochastic gradient descent.

## 4.2 Experiments

We conduct three experiments on a document ranking task to test: (1) the effectiveness of the Duet network compared to the local and distributed subnetworks separately, (2) the effectiveness of the Duet network compared to existing baselines for content-based web ranking, and (3) the effectiveness of training with judged negative documents compared to random negative documents.

In addition, we also evaluate the effectiveness of the Duet model on the TREC Complex Answer Retrieval (TREC CAR) task [376]. In this section, we detail both the experiment setup and the corresponding baseline implementations.

### 4.2.1 Data

**Document ranking task** The training dataset consist of 199,753 instances in the format described in Section 4.2.2. The queries in the training dataset are randomly sampled from Bing's search logs from a period between January 2012 and September 2014. Human judges rate the documents on a five-point scale (*perfect*, *excellent*, *good*, *fair*, and *bad*). The document body text is retrieved from Bing's Web document index. We use proprietary parsers for extracting the body text from raw HTML content. All query and document text are normalized by down-casing and removing all non-alphanumeric characters.

We consider two different test sets, both sampled from Bing search logs. The *weighted* set consist of queries sampled per their frequency in the search logs. Thus, frequent queries are well-represented in this dataset. Queries are sampled between October 2014 and December 2014. The *unweighted* set consist of queries sampled uniformly from the entire population of unique queries. The queries in this samples remove the bias toward popular queries found in the weighted set. The unweighted queries are sampled between January 2015 and June 2015.

**Table 4.1:** Statistics of the three test sets randomly sampled from Bing's search logs for the document ranking task. The candidate documents are generated by querying Bing and then rated using human judges.

|  | queries | documents | docs per query |
|---|---|---|---|
| training | 199,753 | 998,765 | 5 |
| weighted test | 7,741 | 171,302 | 24.9 |
| unweighted test | 6,808 | 71,722 | 10.6 |

Because all of our datasets are derived from sampling real query logs and because queries naturally repeat, there is some overlap in queries between the training and testing sets. Specifically, 14% of the testing queries in the weighted set occurr in the training set, whereas only 0.04% of the testing queries in the unweighted set occurr in the training set. We present both results for those who may be in environments with repeated queries (as is common in production search engines) and for those who may be more interested in cold start situations or tail queries. Table 4.1 summarizes statistics for the two test sets.

**TREC Complex Answer Retrieval task** The goal of TREC CAR task is to, given a document title and a section heading from the same document as a query, retrieve and rank passages from a provided collection. In order to support this task, the TREC CAR organizers present a large training set derived from English Wikipedia. The mediawiki format of articles is parsed to extract the title, the section headings, and the corresponding passages. The collection is filtered to exclude pages which belong to frequent categories, such as people and events, and articles with less than five sections are discarded.

For each heading, we construct a set that includes all passages from the page (in random order) as well as the same amount of passages randomly drawn from other pages. On average this process yields a mean of 35 passages per section which includes: (1) passages from the correct section, (2) passages from the same page, but from different sections, or (3) passages from other pages. The retrieval involves ranking the correct passages (1) higher than the passages from the wrong section or article (2 and 3). We split the dataset for training and testing at a 4:1 ratio.

### 4.2.2 Training

Besides the architecture (Figure 4.2), our model has the following free parameters: (1) the maximum order of the character-based representation for the distributed subnetwork $n_{maxgraph}$, (2) the maximum number of query terms $n_q$ and document terms $n_d$ considered by the model, (3) the convolutional filter size $n_{hidden}$ and window size $n_{window}$, (4) the windows size for max-pooling on the document input for the distributed subnetwork $n_{pool}$, (5) the number of negative documents to sample at training time $n_{neg}$, (6) the dropout rate, and (7) the learning rate..

We use a maximum order of five for our character *n*-graphs in the distributed subnetwork. Instead of using the full 62,193,780-dimensional vector, we only consider the top 2,000 most frequent *n*-graphs, resulting in 36 unigraphs (a-z and 0-9), 689 bigraphs, 1149 trigraphs, 118 4-graphs, and eight 5-graphs.

For both the document ranking and the TREC CAR tasks we limit the maximum number of query terms $n_q$ to 10, and fix the window size of the convolution $n_{window}$ to 3. The dropout rate is also set to 0.20 for both.

For the document ranking task, we consider the first 1000 terms in the document. Correspondingly, the max-pooling window size $n_{pool}$ is fixed at 100, and $n_{hidden}$ is set to 300. When training our model, we sample four negative documents for every relevant document. More precisely, for each query we generated a maximum of one training sample of each form, (1) One *excellent* document with four *fair* documents (2) One *excellent* document with four *bad* documents (3) One *good* document with four *bad* documents.

Pilot experiments showed that treating documents judged as *fair* or *bad* as the negative examples result in significantly better performance, than when the model is trained with randomly sampled negatives. For training, we discard all documents rated as *perfect* because a large portion of them fall under the navigational intent, which can be better satisfied by historical click based ranking signals. When dealing with long documents, it is necessary to use a small minibatch size of 8 to fit the whole data in GPU memory.

For TREC CAR, the average size of passages is significantly smaller than the documents in the previous ranking task. So we consider the first 100 terms in every passage and set $n_{pool}$ and $n_{hidden}$ to 10 and 64, respectively. Because of the (1) smaller size of the input, (2) the smaller number of model parameters, as well as (3) the use of single negative documents, we increase the minibatch size to 1024.

Finally, we choose 0.01 and 0.001 as the learning rates for the two tasks, respectively, based on corresponding validation sets. We implement our model using CNTK [339] and train the model with stochastic gradient descent based optimization (with automatic differentiation) on a single GPU.[2]

### 4.2.3 Baselines

**Document ranking task** Exact term matching is effectively performed by many classic information retrieval models. We used the Okapi BM25 [80] and query likelihood (QL) [90] models as representative of this class of model. We use Indri[3] for indexing and retrieval.

Match positions are handled by substantially fewer models. Metzler's dependence model (DM) [155] provides an inference network approach to modeling term proximity. We use the Indri implementation for our experiments.

Inexact term matching received both historic and modern treatments in the literature. Deerwester et al. [170] originally presented latent semantic analysis (LSA) as a method for addressing vocabulary mismatch by projecting terms and documents into a lower-dimension latent space. The dual embedding space model (DESM) [165, 218] computes a document relevance score by comparing every term in the document with every query term using pre-trained term embeddings. We used the same pre-trained term embeddings dataset that the authors made publicly available online for download[4]. These embeddings, for approximately 2.8M terms, were previously trained on a corpus of Bing queries. In particular, we use the DESM$_{\text{IN-OUT}}$ model, which was reported to have the best performance on the retrieval task, as a baseline here.

---

[2] A CNTK implementation of Duet is available at https://github.com/bmitra-msft/NDRM/blob/master/notebooks/Duet.ipynb under the MIT license.
[3] http://www.lemurproject.org/indri/
[4] https://www.microsoft.com/en-us/download/details.aspx?id=52597

Both the deep structured semantic model (DSSM) [164] and its convolutional variant CDSSM [362] consider only the document title for matching with the query. While some negative results have been reported for title-based DSSM and CDSSM on the *ad hoc* document retrieval tasks [163, 238], we include document-based variants appropriately retrained on the same set of positive query and document pairs as our model. As with the original implementation we choose the non-relevant documents for training by randomly sampling from the document corpus. For the CDSSM model, we concatenate the trigraph hash vectors of the first *n* terms of the body text followed by a vector that is a sum of the trigraph hash vectors for the remaining terms. The choice of *n* is constrained by memory requirements, and we pick 499 for our experiments.

The DRMM model [163] uses a DNN to perform term matching, with few hundred parameters, over histogram-based features. The histogram features, computed using exact term matching and pre-trained term embeddings based cosine similarities, ignoring the actual position of matches. We implemented the DRMM$_{\text{LCH} \times \text{IDF}}$ variant of the model on CNTK [339] using term embeddings trained on a corpus of 341,787,174 distinct sentences randomly sampled from Bing's Web index, with a corresponding vocabulary of 5,108,278 terms. Every training sample for our model is turned into four corresponding training samples for DRMM, comprised of the query, the positive document, and each one of the negative documents. This guarantees that both models observed the exact same pairs of positive and negative documents during training. We adopted the same loss function as proposed by Guo et al. [163].

**TREC Complex Answer Retrieval task** We rank results using Okapi BM25 [80] with $k_1$=1.2 and *b*=0.75. Porter stemming is applied to a Lucene 6.4.1 index and the query.

In addition, we experiment with three different query expansion approaches (terms, entities, and passages) and three vector space representations of queries and documents (tf-idf, GloVe embeddings, and RDF2Vec embeddings). Each of the possible combinations (*e.g.*, term-expansion + tf-idf vectors, or passage-expansion + term embedding vectors) defines a query representation. Results are ranked according to the cosine similarity between the vector representations of the query and the document.

We experiment with three different query expansion approaches:

- *Expansion terms (RM).* Feedback terms are derived using pseudo relevance feedback and the relevance model [158]. We use Galago's implementation[5] which is based on a Dirichlet smoothed language model for the feedback run. We achieve the best performance by expanding the query with top 10 terms extracted from the top 10 feedback documents.
- *Expansion entities.* We also expand the query using supporting entities retrieved by a search for the query. Best performance is achieved using 10 entities for expansion.
- *Passage Rocchio.* Inspired by the work of Banerjee and Mitra [377], we retrieve other passages, which have identical section heading to the heading part of our query, from the portion of the dataset reserved for training. For example, given a query such as "United States demographic", with respect to the entity United States, we collect supporting passages from the pages of other entities (*e.g.*,, United Kingdom), that fall under the section titled "Demographics". Headings are processed with tokenisation, stopword and digit removal, and stemming. We are able to retrieve at least one supporting passage for one-third of our queries. We obtain best performance from expanding the query with 5 passages.

We investigate three vector representation schemes for the query and the passage:

- *Local representation.* Each term in the vocabulary is represented by a one-hot vector. Queries and passages are represented as bag of terms, where the term frequencies are weighted by TF-IDF and are logarithmic L2-normalised.

---
[5]lemurproject.org/galago.php

- *Term Embeddings.* Under this scheme, each term is represented by their corresponding pre-trained GloVE [202] embeddings. The query and passage vectors are obtained by averaging the term embeddings—with TF-IDF weighting.

$$\vec{v}_q = \frac{1}{|q|} \sum_{t_q \in u(q)} \text{tf-idf}(t_q) \cdot \vec{v}_{t_q}$$

where, $u(d)$ is the set of unique words in query $q$.

- *Entity Embeddings.* Queries and documents are represented as their mentioned DBpedia entities, using the entity linker TagMe [378]—with default parameters. We obtain latent vector representations $\vec{v}_e$ of each linked entity $e$ using pre-computed **RDF2Vec** entity embeddings [379]. Query and passage representation is obtained from weighted average of these entity vectors. Entity vectors are weighted based on inlink and outlink statistics from the 2015-04 DBpedia Data Set [380].

$$\vec{v}_q = \frac{1}{|\{e \in ent(q)\}|} \sum_{e \in ent(q)} \text{link}(e) \cdot \vec{v}_e$$

where, $ent(d)$ is the set of entities in query $q$.

Additionally, we combine the ranking-score of these different baselines with supervised machine learning [381]. We train a linear model using RankLib [6] optimized for MAP, trained with coordinate ascent.

### 4.2.4  Evaluation

For the document ranking task, we report the normalized discounted cumulative gain (NDCG) metric computed at positions one and ten. All performance metrics are averaged over queries for each run. Whenever testing for significant differences in performance, we used the paired *t*-test with a Bonferroni correction. For the TREC CAR task, we report MRR, R-Prec, and MAP numbers for all the models.

---

[6]lemurproject.org/ranklib.php

**Table 4.2:** Performance on the document ranking task. All Duet runs significantly outperformed our local and distributed model ($p < 0.05$). All Duet runs also outperformed non-neural and neural baselines. The difference between the Duet model and the best performing baseline per dataset and position (italics) is statistically significant ($p < 0.05$). The best NDCG performance on each dataset and position is highlighted in bold.

|  | Weighted | | Unweighted | |
| --- | --- | --- | --- | --- |
|  | NDCG@1 | NDCG@10 | NDCG@1 | NDCG@10 |
| **Non-neural baselines** | | | | |
| LSA | 22.4 | 44.2 | 31.9 | 62.7 |
| BM25 | 24.2 | 45.5 | 34.9 | 63.3 |
| DM | 24.7 | 46.2 | 35.0 | 63.4 |
| QL | 24.6 | 46.3 | 34.9 | 63.4 |
| **Neural baselines** | | | | |
| DRMM | 24.3 | 45.2 | *35.6* | *65.1* |
| DSSM | 25.8 | 48.2 | 34.3 | 64.4 |
| CDSSM | *27.3* | 48.2 | 34.3 | 64.0 |
| DESM | 25.4 | *48.3* | 35.0 | 64.7 |
| **Our models** | | | | |
| Local model | 24.6 | 45.1 | 35.0 | 64.4 |
| Distributed model | 28.6 | 50.5 | 35.2 | 64.9 |
| Duet model | **32.2** | **53.0** | **37.8** | **66.4** |

## 4.3 Results

**Document ranking task** Table 4.2 reports NDCG based evaluation results on two test datasets for our model and all the baseline models. Our main observation is that Duet performs significantly better than the individual local and distributed models. This supports our underlying hypothesis that matching in a latent semantic space can complement exact term matches in a document ranking task, and hence a combination of the two is more appropriate. Note that the NDCG numbers for the local and the distributed subnetworks correspond to when these DNNs are trained individually, but for Duet the two DNNs are trained together as part of a single neural network.

Among the baseline models, including both traditional and neural network based models, CDSSM and DESM achieve the highest NDCG at position one and ten, respectively, on the weighted test set. On the unweighted test set DRMM is our best baseline model at both rank positions. Duet demonstrates significant improvements over all these baseline models on both test sets and at both NDCG positions.

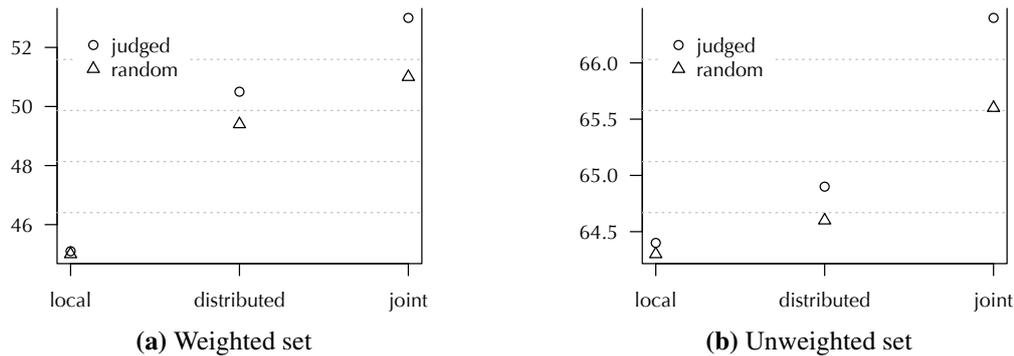

**Figure 4.4:** Duet demonstrates significantly better NDCG performance ($p < 0.05$) on both test sets when trained with judged non-relevant documents as the negative examples, instead of randomly sampling them from the document corpus. The distributed subnetwork also shows statistically significant NDCG gain ($p < 0.05$) on the weighted set, and a non-statistically significant NDCG gain on the unweighted set.

We also test our independent local and distributed models against their conceptually closest baselines. Because our local model captures both matching and proximity, we compared performance to dependence models (DM). While the performance in terms of NDCG@1 is statistically indistinguishable, both NDCG@10 results are statistically significant ($p < 0.05$). We compared our distributed model to the best neural model for each test set and metric. We found no statistically significant difference except for NDCG@10 for the weighted set.

We were interested in testing our hypotheses that training with labeled negative documents is superior to training with randomly sampled documents presumed to be negative. We conducted an experiment training with negative documents following each of the two protocols. Figure 4.4 shows the results of these experiments. We found that, across all our models, using judged nonrelevant documents was more effective than randomly sampling documents from the corpus and considering them as negative examples. Very recently, Xiong et al. [382] have presented similar evidence on the importance of sampling negative documents that are closer in relevance to the query than documents sampled from the collection at uniform probability, and operationalized the idea in the form of active metric learning [383–385].

**TREC Complex Answer Retrieval task** Results are presented in table 4.3. Not all query expansion approaches and vector space representations methods improve over the BM25 baseline. The most promising results—among the baseline methods which employ cosine similarity as a ranking function—are obtained when the query is expanded with supporting textual paragraphs. This is an interesting finding that reconfirms the results of previous work on the automatic generation of Wikipedia articles based on its structural information [377, 386]. The learning to rank model is our best performing baseline. Duet yields a substantial improvement over all presented approaches, including a 47% improvement in MAP over the BM25 baseline and a 14% improvement over the learning to rank model.

**Table 4.3:** Duet outperforms (statistically significant at $p < 0.05$) the best baseline model (italics) on the TREC Complex Answer Retrieval task. The best performances are highlighted in bold for each metric.

|  | MRR | R-Prec | MAP |
| --- | --- | --- | --- |
| **bm25** | | | |
| query only | 0.409 | 0.232 | 0.320 |
| **tf-idf (cs)** | | | |
| query only | 0.383 | 0.212 | 0.350 |
| query + RM1 | 0.384 | 0.205 | 0.324 |
| query + Rocchio | 0.466 | 0.286 | 0.400 |
| **GloVe (cs)** | | | |
| query only | 0.387 | 0.210 | 0.329 |
| query + RM1 | 0.339 | 0.177 | 0.289 |
| query + Rocchio | 0.410 | 0.236 | 0.349 |
| **RDF2Vec (cs)** | | | |
| entity-query only | 0.369 | 0.200 | 0.313 |
| ent-query + ent-RM1 | 0.377 | 0.208 | 0.320 |
| ent-query + ent-Rocchio | 0.375 | 0.206 | 0.316 |
| **Learning to Rank** | | | |
| all (cs) scores | *0.475* | *0.290* | *0.412* |
| **Duet** | | | |
| query only | **0.553** | **0.359** | **0.470** |

## 4.4 Further improvements

In follow up work, we explore several additional modifications to the original Duet architecture and demonstrate through an ablation study that incorporating these changes results in significant improvements on passage ranking. We evaluate the modified Duet model on the MS MARCO passage ranking task [52] and the TREC

Deep Learning track [15]. In the context of the document ranking task at TREC, we further modify the architecture to incorporate multiple-field representation of documents.

### 4.4.1 Duet on MS MARCO

In this section, we briefly describe several modifications to the Duet architecture in the context of passage ranking. A public implementation of the updated Duet model using PyTorch [387] is available online[7].

1. **Word embeddings.** We replace the character level *n*-graph encoding in the input of the distributed subnetwork with word embeddings. We see significant reduction in training time given a fixed number of minibatches and a fixed minibatch size. This change primarily helps us to train on a significantly larger amount of data under fixed training time constraints. We initialize the word embeddings using pre-trained GloVe [202] embeddings before training Duet.

2. **Inverse document frequency weighting.** In contrast to some of the other datasets on which Duet has been previously evaluated [7, 8], the MS MARCO dataset contains a relatively larger percentage of natural language queries and the queries are considerably longer on average. In traditional IR models, the inverse document frequency (IDF) [91] of a query term provides an effective mechanism for weighting the query terms by their discriminative power. In the original Duet architecture, the input to the local subnetwork corresponding to a query $q$ and a document $d$ is a binary interaction matrix $X \in \mathbb{R}^{|q| \times |d|}$ defined as follows:

$$X_{ij} = \begin{cases} 1, & \text{if } q_i = d_j \\ 0, & \text{otherwise} \end{cases} \tag{4.7}$$

---

[7]https://github.com/dfcf93/MSMARCO/blob/master/Ranking/Baselines/Duet.ipynb

We incorporate IDF in Duet by weighting the interaction matrix by the IDF of the matched terms. We adopt the Robertson-Walker definition of IDF [388] normalized to the range $[0, 1]$.

$$X'_{ij} = \begin{cases} \text{IDF}(q_i), & \text{if } q_i = d_j \\ 0, & \text{otherwise} \end{cases} \quad (4.8)$$

$$\text{IDF}(t) = \frac{\log(N/n_t)}{\log(N)} \quad (4.9)$$

Where, $N$ is the total number of passages in the collection and $n_t$ is the number of passages in which the term $t$ appears at least once.

3. **Non-linear combination of local and distributed subnetworks.** Zamani et al. [389] show that when combining different subnetworks in a neural ranking model, it is more effective if each subnetwork produce a vector output that are further combined by additional multi-layer perceptrons (MLP). In the original Duet, the local and the distributed subnetwork produce a single score that are linearly combined. In our updated architecture, both subnetworks produce a vector that are further combined by an MLP—with two hidden layers—to generate the estimated relevance score.

4. **Rectifier Linear Units (ReLU).** We replace the Tanh non-linearities in the original Duet with ReLU [390] activations.

5. **Bagging.** We observe some additional improvements from combining multiple Duet models—trained with different random seeds and on different random sample of the training data—using bagging [391].

**Experiments** We evaluate the proposed modifications to Duet on the recently released MS MARCO passage ranking task [52]. The task requires a model to rank approximately thousand passages for each query. The queries are sampled from Bing's search logs, and then manually annotated to restrict them to questions with specific answers. A BM25 [80] model is employed to retrieve the top thousand candidate passages for each query from the collection. For each query, zero or more candidate passages are deemed relevant based on manual annotations. The ranking model is evaluated on this passage re-ranking task using the mean reciprocal rank (MRR) metric [75]. Participants are required to submit the ranked list of passages per query for a development (dev) set and a heldout (eval) set. The ground truth annotations for the development set are available publicly, while the corresponding annotations for the evaluation set are heldout to avoid overfitting. A public leaderboard[8] presents all submitted runs from different participants on this task.

The MS MARCO task provides a pre-processed training dataset—called "triples.train.full.tsv"—where each training sample consists of a triple $\langle q, p_+, p_- \rangle$, where $q$ is a query and $p_+$ and $p_-$ are a pair of passages, with $p_+$ being more relevant to $q$ than $p_-$. Similar to the original Duet, we employ the cross-entropy with softmax loss to learn the parameters of our network $\mathcal{M}$:

$$\mathcal{L} = \mathbb{E}_{q,p_+,p_- \sim \theta}[\ell(\mathcal{M}_{q,p_+} - \mathcal{M}_{q,p_-})] \tag{4.10}$$

$$\text{where, } \ell(\Delta) = \log(1 + e^{-\sigma \cdot \Delta}) \tag{4.11}$$

Where, $\mathcal{M}_{q,p}$ is the relevance score for the pair $\langle q, p \rangle$ as estimated by the model $\mathcal{M}$. Note, that by considering a single negative passage per sample, our loss is equivalent to the RankNet loss [235].

We use the Adam optimizer with default parameters and a learning rate of 0.001. We set $\sigma$ in Equation 5.8 to 0.1 and dropout rate for the model to 0.5. We trim all queries and passages to their first 20 and 200 words, respectively. We restrict our input vocabulary to the $71,486$ most frequent terms in the collection and set the size of all hidden layers to 300. We use minibatches of size 1024 and train the model for 1024 minibatches. Finally, for bagging we train eight different Duet networks with different random seeds and on different samples of the training data. We train and evaluate our models using a Tesla K40 GPU—on which it takes a total of only 1.5 hours to train each single Duet model and to evaluate it on both dev and eval sets.

---

[8]http://www.msmarco.org/leaders.aspx

Table 4.4: Comparison of the different Duet variants and other state-of-the-art approaches from the public MS MARCO leaderboard. The update Duet benefits significantly from the modifications proposed in this paper.

| Model | MRR@10 Dev | MRR@10 Eval |
|---|---|---|
| **Other approaches** | | |
| BM25 | 0.165 | 0.167 |
| Single CKNRM [392] model | 0.247 | 0.247 |
| Ensemble of 8 CKNRM [392] models | 0.290 | 0.271 |
| IRNet (a proprietary deep neural model) | 0.278 | 0.281 |
| BERT [167] | 0.365 | 0.359 |
| **Duet variants** | | |
| Single Duet w/o IDF weighting for interaction matrix | 0.163 | - |
| Single Duet w/ Tanh non-linearity (instead of ReLU) | 0.179 | - |
| Single Duet w/o MLP to combine local and distributed scores | 0.208 | - |
| Single Duet | 0.243 | 0.245 |
| Ensemble of 8 Duet networks | 0.252 | 0.253 |

**Results** Table 4.4 presents the MRR@10 corresponding to all the Duet variants we evaluated on the dev set. The updated Duet with all the modifications described in Section 4.4.1 achieves an MRR@10 of 0.243. We perform an ablation study by leaving out one of the three modifications—(i) IDF weighting for interaction matrix, (ii) ReLU non-linearity instead of Tanh, and (iii) LP to combine local and distributed scores,—out at a time. We observe a 33% degradation in MRR by not incorporating the IDF weighting alone. It is interesting to note that the Github implementations[9] of the KNRM [393] and CKNRM [392] models also indicate that their MS MARCO submissions incorporated IDF term-weighting—potentially indicating the value of IDF weighting across multiple architectures. Similarly, we also observe a 26% degradation in MRR by using Tanh non-linearity instead of ReLU. Using a linear combination of scores from the local and the distributed subnetwork instead of combining their vector outputs using an MLP results in 14% degradation in MRR. Finally, we observe a 3% improvement in MRR by ensembling eight Duet networks using bagging. We also submit the individual Duet model and the ensemble of eight Duets for evaluation on the heldout set and observe similar numbers.

---
[9]https://github.com/thunlp/Kernel-Based-Neural-Ranking-Models

We include the MRR numbers for other non-Duet based approaches that are available on the public leaderboard in Table 4.4. As of writing this paper, BERT [327] based approaches—*e.g.*, [167]—are outperforming other approaches by a significant margin. Among the non-BERT based approaches, a proprietary deep neural network—called IRNet—currently demonstrates the best performance on the heldout evaluation set. This is followed, among others, by an ensemble of CKNRM [392] models and the single CKNRM model. The single Duet model achieves comparable MRR to the single CKNRM model on the eval set. The ensemble of Duets, however, performs slightly worse than the ensemble of the CKNRM models on the same set.

### 4.4.2   Duet on TREC Deep Learning track

The deep learning track at TREC 2019 makes large training datasets—suitable for traininig deep models with large number of learnable parameters—publicly available in the context of a document ranking and a passage ranking tasks. We benchmark Duet on both tasks.

In the context of the document ranking task, we adapt Duet to ingest a "multiple field" view of documents, based on findings from Zamani et al. [389]. We refer to this new architecture as Duet with Multiple Fields (DuetMF). We also combine the relevance estimates from DuetMF with several other traditional and neural retrieval methods in a learning-to-rank (LTR) [39] framework.

For the passage ranking task, we submit a single run based on an ensemble of eight Duet models. The architecture and the training scheme resembles that described in Section 4.4.1.

**TREC 2019 deep learning track**  The TREC 2019 deep learning track introduces: (i) a document retrieval task and (ii) a passage retrieval task. For both tasks, participants are provided a set of candidates—100 documents and 1000 passages,

respectively—per query that should be ranked. Participants can choose to either rerank provided candidates or retrieve from the full collection.

For the passage retrieval task, the track reuses the set of 500K+ manually-assessed binary training labels released as part of the Microsoft Machine Reading COmprehension (MS MARCO) challenge [52]. For the document retrieval task, the passage-level labels are transferred to their corresponding source documents—producing a training dataset of size close to 400K labels.

For evaluation, a shared test set of 200 queries is provided for both tasks, of which two different overlapping set of 43 queries were later selected for manual NIST assessments corresponding to the two tasks.

Full details of all datasets is available on the track website[10] and in the track overview paper [15].

**Duet with Multiple Fields (DuetMF).** Zamani et al. [389] study neural ranking models in the context of documents with multiple fields. In particular, they make the following observations:

Obs. 1: It is more effective to summarize the match between query and individual document fields by a vector—as opposed to a single score—before aggregating to estimate full document relevance to the query.

Obs. 2: It is better to learn different query representations corresponding to each document field under consideration.

Obs. 3: Structured dropout (*e.g.*, field-level dropout) is effective for regularization during training.

We incorporate all of these ideas in the updated Duet network as shown in Fig. 4.5.

Documents in the deep learning track dataset contains three text fields: (i) URL, (ii) title, and (iii) body. We employ Duet to match the query against each individual document fields. In line with Obs. 1 from [389], the field-specific Duet outputs a vector instead of a single score. We do not share the parameters of Duet between the field-specific instances based on Obs. 2. Following Obs. 3, we introduce structured dropouts at different stages of the model. We randomly dropout each of the local subnetworks for 50% of the training samples. Similarly, we also dropout different combinations of field-level subnetworks uniformly at random—taking care that at least one field-level model is always retained.

---
[10]https://microsoft.github.io/TREC-2019-Deep-Learning/

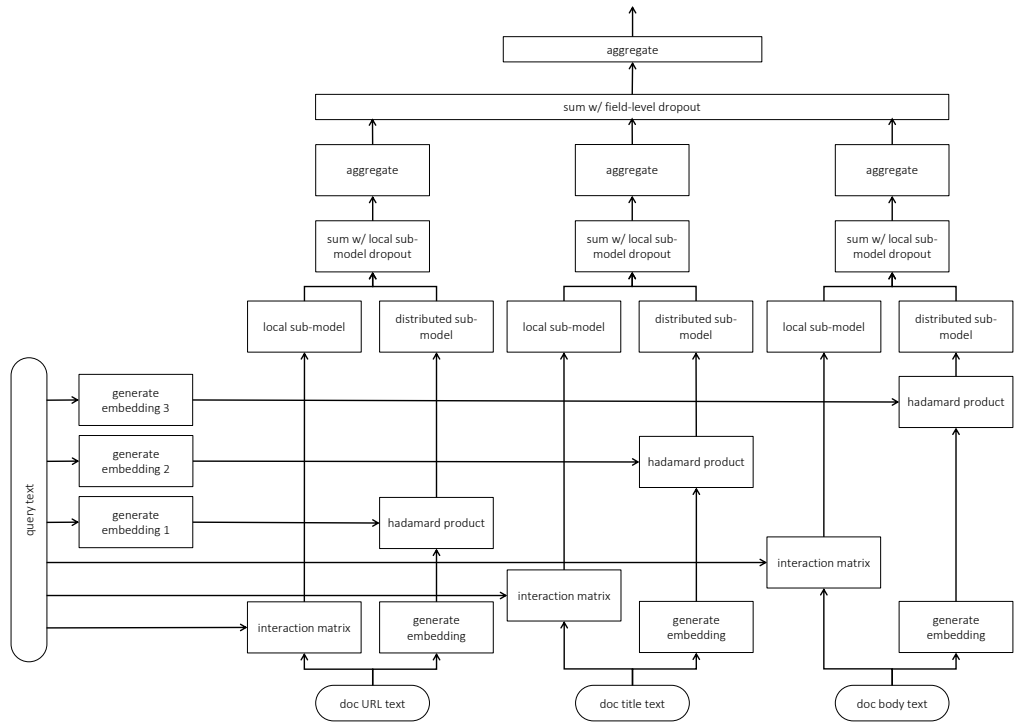

**Figure 4.5:** The modified Duet (DuetMF) that considers multiple document fields.

We consider the first 20 terms for queries and for document URLs and titles. For document body text, we consider the first 2000 terms. Similar to Section 4.4.1, we employ pretrained word embeddings as the input text representation for the distributed subnetworks. We train the word embeddings using a standard word2vec [203] implementation in FastText [394] on a combination of the MS MARCO document corpus and training queries.

The query and document field embeddings are learned by deep convolutional-pooling layers. We set the hidden layer size at all stages of the model to 300 and dropout rate for different layers to 0.5. For training, we employ the RankNet loss [235] over $<q, d_{\text{pos}}, d_{\text{neg}}>$ triples and the Adam optimizer [395]—with a mini-batch size of 128 and a learning rate of 0.0001 for training. We sample $d_{\text{neg}}$ uni-formly at random from the top 100 candidates provided that are not positively labeled. When employing structured dropout, the same sub-models are masked for both $d_{\text{pos}}$ and $d_{\text{neg}}$.

In light of the recent success of large pretrained language models—*e.g.*, [167]—we also experiment with an unsupervised pretraining scheme using the MS MARCO document collection. The pretraining is performed over $< q_{\text{pseudo}}, d_{\text{pos}}, d_{\text{neg}} >$—where $d_{\text{pos}}$ and $d_{\text{neg}}$ are randomly sampled from the collection and a pseudo-query $q_{\text{pseudo}}$ is generated by picking the URL or the title of $d_{\text{pos}}$ randomly (with equal probability) and masking the corresponding field on the document side for both $d_{\text{pos}}$ and $d_{\text{neg}}$. We see faster convergence during supervised training when the DuetMF model is pretrained in this fashion on the MS MARCO document collection. We posit that a more formal study should be performed in the future on pretraining Duet networks on large collections, such as Wikipedia and the BookCorpus [396].

In addition to the independent Duet model, we train a neural LTR model with two hidden layers—each with 1024 hidden nodes. The LTR run reranks a set of 100 document candidates retrieved by query likelihood (QL) [90] with Dirichlet smoothing ($\mu = 1250$) [150]. Several ranking algorithms based on neural and inference networks act as features: (i) DuetMF, (ii) Sequential Dependence Model (SDM) [155], and (iii) Pseudo-Relevance Feedback (PRF) [157, 158], (iv) BM25, [80], and (v) Dual Embedding Space Model (DESM) [165, 218].

We employ SDM with an order of 3, combine weight of 0.90, ordered window weight of 0.034, and an unordered window weight of 0.066 as our base candidate scoring function. We use these parameters to retrieve from the target corpus as well as auxiliary corpora of English language Wikipedia (`enwiki-20180901-pages-articles-multistream.xml.bz2`), LDC Gigaword (`LDC2011T07`). For PRF, initial retrievals—from either of the target, wikipedia, or gigaword corpora—adopted the SDM parameters above, however are used to rank 75-word passages with a 25-word overlap. These passages are then interpolated using the top *m* passages and standard relevance modeling techniques, from which we select the top 50 words to use as an expanded query for the final ranking of the target candidates. We do not explicitly adopt RM3 [160] because our LTR model implicitly combines our initial retrieval score and score from the expanded query. All code for the SDM and PRF feature computation is available at https://github.com/diazf/indri.

**Table 4.5:** Official TREC 2019 Deep Learning track results. The recall metric is computed at position 100 for the document retrieval task and at position 1000 for the passage retrieval task.

| Run description | Subtask | MRR | NDCG@10 | MAP | Recall |
|---|---|---|---|---|---|
| **Document retrieval task** | | | | | |
| LTR w/ DuetMF | fullrank | 0.876 | 0.578 | 0.237 | 0.368 |
| DuetMF model | rerank | 0.810 | 0.533 | 0.229 | 0.387 |
| **Passage retrieval task** | | | | | |
| Ensemble of 8 Duets | rerank | 0.806 | 0.614 | 0.348 | 0.694 |

We evaluate two different BM25 models with hyperparameters $<k_1 = 0.9, b = 0.4>$ and $<k_1 = 3.44, b = 0.87>$.

Corresponding to each of the DuetMF, SDM, PRF, and BM25 runs we generate two features based on the score and the rank that the model predicts for a document *w.r.t.* the target query.

We generate eight features by comparing the query against two different document fields (title and body) and using different DESM similarity estimates (INxIN, INxOUT, OUTxIN, OUTxOUT).

We add couple of features based on query length and domain quality—where the latter is defined simply as a ratio between how often documents from a given domain appear in the positively labeled training data and in the overall document collection.

Finally, for the passage ranking task, we adopt the exact same model and training procedure from Section 4.4.1. Our final submission is an ensemble of eight Duet networks.

Table 4.5 summarizes the official evaluation results for all three runs.

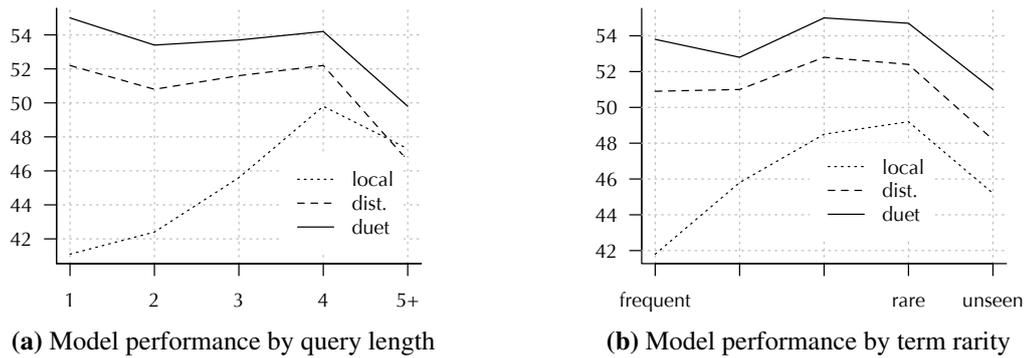

(a) Model performance by query length

(b) Model performance by term rarity

**Figure 4.6:** NDCG performance of different models by length of query and how rare the rarest query term is in the training data. For the rare term analysis, we place all query terms into one of five categories based on their occurrence counts in the training data. Then we then categorize each query in the test dataset based on the frequency of the rarest term belongs in the query. We include a category for queries with at least one term which has no occurrences in the training data.

## 4.5 Discussion

Our results demonstrated that our joint optimization of local and distributed subnetworks provides substantial improvement over several state-of-the-art baselines. Although the independent models were competitive with existing baselines, the combination provided a significant boost.

We also confirm that using judged negative documents should be used when available. We speculate that training with topically-similar (but non-relevant) documents allows the model to better discriminate between the documents provided by an earlier retrieval stage that are closer to each other *w.r.t.* relevance. This sort of staged ranking, first proposed by Cambazoglu et al. [397], is now a common web search engine architecture.

In Section 4.2.3 we described our baseline models according to which of the properties of effective retrieval systems they incorporate. It is reasonable to expect that models with certain properties are better suited to deal with certain segments of queries. For example, the relevant Web page for the query "what channel are the seahawks on today" may contain the name of the actual channel (*e.g.*,, "ESPN" or "FOX") and the actual date for the game, instead of the terms "channel" or "today". A retrieval model that only counts repetitions of query terms is likely to retrieve less relevant documents for this query – compared to a model that con-siders "ESPN" and "FOX" to be relevant document terms. In contrast, the query "pekarovic land company", which may be considered as a tail navigational intent, is likely to be better served by a retrieval model that simply retrieves documents containing many matches for the term "pekarovic". A representation learning model is unlikely to have a good representation for this rare term, and therefore may be less equipped to retrieve the correct documents. These anecdotal examples agree with the results in in Table 4.2 that show that on the weighted test set all the neural models whose main focus is on learning distributed representations of text (Duet model, distributed model, DESM, DSSM, and CDSSM) perform better than the models that only look at patterns of term matches (local model and DRMM). We believe that this is because the DNNs can learn better representations for more frequent queries, and perform particularly well on this segment. Figure 4.6 provides further evidence towards this hypothesis by demonstrating that the distributed model has a larger NDCG gap with the local model for queries containing more frequent terms, and when the number of terms in the query is small. The Duet model , however, is found to perform better than both the local and the distributed models across all these segments.

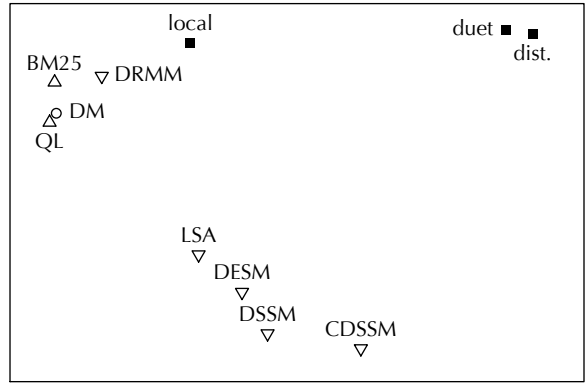

**Figure 4.7:** Principal component analysis of models based on retrieval performance across testing queries. Models using exact term matches (△), proximity (○), and inexact matches (▽) are presented. Our models are presented as black squares.

To better understand the relationship of our models to existing baselines, we compared the per-query performance amongst all models. We conjecture that similar models should perform similarly for the same queries. We represented a retrieval model as a vector where each position of the vector contains the performance of the model on a different query. We randomly sample two thousand queries from our weighted test set and represent all ranking models as vectors of their NDCG values against these two thousand queries. We visualized the similarity between models by projecting using principal component analysis on the set of performance vectors. The two-dimensional projection of this analysis is presented in Figure 4.7. The figure largely confirms our intuitions about properties of retrieval models. Models that use only local representation of terms are closer together in the projection, and further away from models that learn distributed representations of text. Interestingly, the plot does not distinguish between whether the underlying model is based on a neural network based or not – with neural networks of different retrieval properties appearing in each of the three clusters.

Another interesting distinction between deep neural models and traditional approaches is the effect of the training data size on model performance. BM25 has very few parameters and can be applied to new corpus or task with almost no training. On the other hand, DNNs like ours demonstrate significant improvements when trained with larger datasets. Figure 4.8 shows that the effect of training data size particularly pronounced for Duet and the distributed subnetwork that learns representations of text. The trends in these plots indicate that training on even larger datasets may result in further improvements in model performance over what is reported here. We believe this should be a promising direction for future work.

A last consideration when comparing these models is runtime efficiency. Web search engines receive tens of thousands of queries per second. Running a deep neural model on raw body text at that scale is a hard problem. The local subnetwork of our model operates on the term interaction matrix that should be reasonable to generate using an inverted index. For the distributed model, it is important to note that the $300 \times 899$ dimensional matrix representation of the document, that is used to compute the Hadamard product with the query, can be pre-computed and stored as part of the document cache. At runtime, only the Hadamard product and the subsequent part of the network needs to be executed. Such caching strategies, if

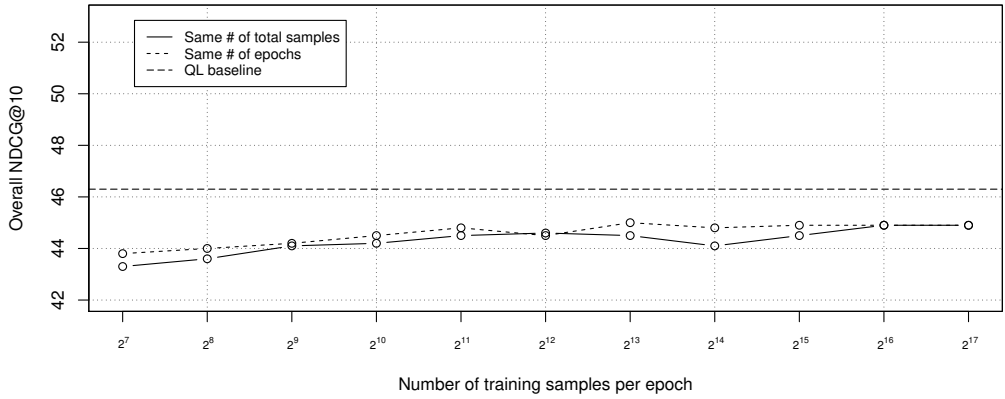

**(a)** Local subnetwork

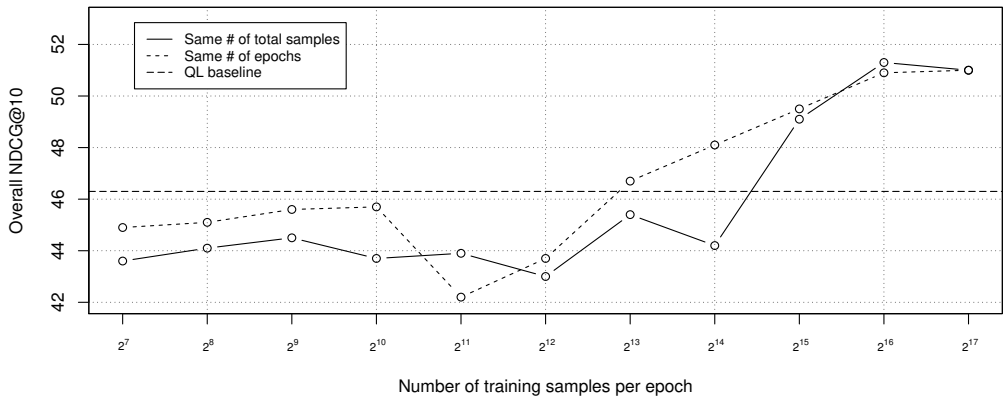

**(b)** Distributed subnetwork

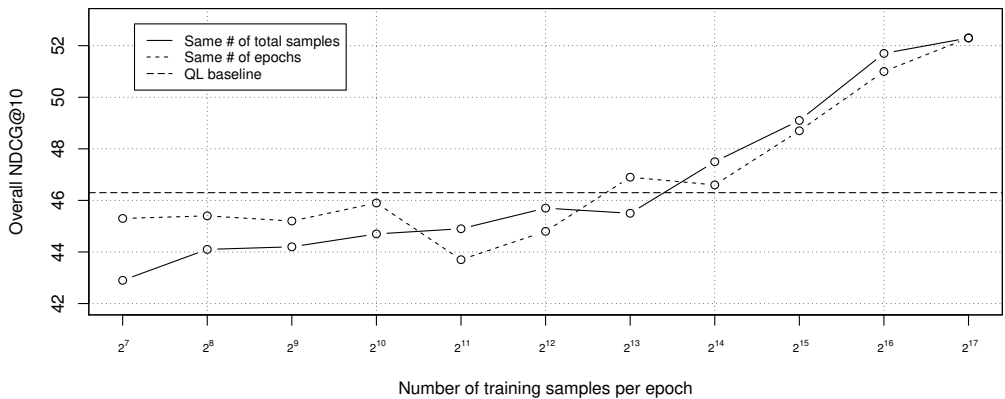

**(c)** Duet

**Figure 4.8:** We study the performance of our model variants when trained with different size datasets. For every, dataset size we train two models – one for exactly one epoch and another one with multiple epochs such that the total number of training samples seen by the model during training is 131,072.

employed effectively, can mitigate large part of the runtime cost of running a DNN based document ranking model at scale. In Chapter 5, we will revisit the question of runtime efficiency, but in the context of a family of neural IR models.

## 4.6 Conclusion

We propose a novel ranking model composed of two separate deep subnetworks, one that matches using local representation of text, and another that learns distributed representation before matching. The Duet of these two subnetworks achieve better performance compared to the sub-models individually on the document ranking and passage ranking tasks—as well as significant improvements over other neural and traditional non-neural baselines. Our analysis indicate that the improvements over traditional methods are more substantial in the presence of larger training datasets.

# Chapter 5

# Retrieve, not just rerank, using deep neural networks

In response to short text queries, search engines attempt to retrieve the top few relevant results by searching through collections containing billions of documents [398], often under a second [399]. Response time is a key consideration in web search. Even a 100ms latency has been shown to invoke negative user reactions [108, 109]. To achieve such short response times, these systems typically distribute the collection over multiple machines that can be searched in parallel [400]. Specialized data structures—such as inverted indexes [401, 402]—are used to dramatically cut down the number of documents required to be evaluated for any specific query. The index organization and query evaluation strategies, in particular, trade-off retrieval effectiveness and efficiency during the candidate generation stage. However, unlike in late stage re-ranking where machine learning (ML) models are commonplace [39, 403], the candidate generation frequently employs traditional retrieval models with few learnable parameters.

Query evaluation using state-of-the-art deep neural ranking models require time and resource intensive computations. Typically these models also require both query and document as input to inspect the interactions between query and document terms. The study of these neural ranking methods have, therefore, been largely limited to late stage re-ranking. Efficient retrieval using these complex machine learned relevance estimators is an important challenge [404].

Recently, a few different attempts [405–407] have been made to leverage neural methods for retrieval over large collections. All of these studies focus on neural methods that compute the latent representations of documents independent of the query. This allows the document embeddings to be precomputed. At query evaluation time, only the query embedding is computed by evaluating the corresponding portion of the deep neural model. This is followed by an approximate nearest-neighbour search over the collection—using the precomputed document embeddings. This approaches typically achieve significantly poorer retrieval performance compared to traditional IR methods—and generally need to be combined with classical IR functions [405, 406].

In this chapter, we describe a different approach—that assumes query term independence—to leverage state-of-the-art neural ranking models for retrieval over the full collection. Based on our initial study, we posit there the is significant opportunity to use neural methods in combination with impact-ordered inverted index [408–410]. These data structures employ score quantization for efficient retrieval. In the second half of this chapter, we propose a method to learn appropriate quantization schemes that optimize for retrieval effectiveness.

## 5.1   Query term independence assumption

Many traditional IR ranking functions [80, 90, 147, 148, 151] and early word embedding based IR methods [218, 223, 230] manifest the *query-term independence* (QTI) property—*i.e.*, the documents can be scored independently *w.r.t.* each query term, and then the scores accumulated. Given a document collection, these term-document scores can be precomputed [409]. Specialized IR data structures, such as inverted indexes [401, 402], in combination with clever organization strategies (*e.g.*, impact-ordering [408–410]) can take advantage of the simplicity of the accumulation function (typically a linear sum) to aggressively prune the set of documents that need to be assessed per query. This dramatically speeds up query evaluations enabling fast retrieval from large collections, containing billions of documents.

Recent deep neural architectures—such as BERT [167], Duet (see Chapter 4),

and CKNRM [392]—have demonstrated state-of-the-art performance on several IR tasks [1, 15]. However, the superior retrieval effectiveness comes at the cost of evaluating deep models with tens of millions to hundreds of millions of parameters at query evaluation time. In practice, this limits the scope of these models to late stage re-ranking.

Like traditional IR models, we can incorporate the QTI assumption into the design of the deep neural model—which would allow offline precomputation of all term-document scores. The query evaluation then involves only their linear combination—alleviating the need to run the computation intensive deep model at query evaluation time. We can further combine these precomputed machine-learned relevance estimates with an inverted index, to retrieve from the full collection. This significantly increases the scope of potential impact of neural methods in the retrieval process. We study this approach in this work. Of course, by operating independently per query term, the ranking model has access to less information compared to if it has the context of the full query. Therefore, we expect the ranking model to show some loss in retrieval effectiveness under this assumption. However, we trade this off with the expected gains in efficiency of query evaluations and the ability to retrieve, and not just re-rank, using deep models.

The efficiency benefits of our proposed approach is two-fold. First and foremost, incorporating the QTI assumption allows for the deep model evaluations to be performed at document indexing time, instead of at query evaluation time. While query evaluation has strict response time constraints [108, 109, 399], IR systems generally have more leeway dealing with heavy computation during the offline indexing process. Furthermore, the offline evaluation provides additional flexibility to group samples into large batches and can take advantage of large-scale parallelization by distributing the workload over large clusters of machines. Secondly, the computation complexity involved in exhaustively evaluating every document in a collection $D$ with respect to a set of queries $Q$ for a typical deep ranking models, that operate over individual query-document pairs, is $O(|D| \times |Q|)$. For models that incorporate the QTI assumption, the compute complexity changes to $O(|D| \times |T|)$, where $T$ is the vocabulary of all indexed terms. While this may not look like an obvious improvement over the $O(|D| \times |Q|)$ complexity, we note that rarely do we need to evaluate the document exhaustively with respect to every term in the vocabulary. In fact, we can rewrite the complexity for query term independent ranking models as $O(|D| \times k)$, where $k$ is the maximum number of terms that are practically important to evaluate for any given document. We posit that $k \ll |T|$ and that we can employ efficient methods, including simple heuristics, to preselect candidate terms for a given document. The compute complexity can be further improved if, say, the costliest part of the model—*e.g.*, the document encoder—needs to be evaluated only once per document and then only a small overhead is incurred for each of the $k$ candidate terms. In that case, the compute complexity may be closer to $O(|D|)$. A similar motivation has recently been operationalized in the Conformer-Kernel [11, 12] and DeepCT [411] architectures that incorporate the QTI assumption.

In this study, we incorporate the QTI assumption into three state-of-the-art neural ranking models—BERT, Duet, and CKNRM—and evaluate their effectiveness on the MS MARCO passage ranking task [52]. We surprisingly find that the two of the models suffer no statistically significant adverse affect *w.r.t.* ranking effectiveness on this task under the query term independence assumption. While the performance of BERT degrades under the strong query term independence assumption—the drop in MRR is reasonably small and the model maintains a significant performance gap compared to other non-BERT based approaches. We conclude that at least for a certain class of existing neural IR models, incorporating query term independence assumption may result in significant efficiency gains in query evaluation at minimal (or no) cost to retrieval effectiveness.

## 5.2 Related work

Several neural IR methods—*e.g.*, [163, 218, 223, 230]—already operate under query term independence assumption. However, recent performance breakthroughs on many IR tasks have been achieved by neural models [7, 167, 302, 303, 392] that learn latent representations of the query or inspect interaction patterns between query and document terms. In this work, we demonstrate the potential to incorporate query term independence assumption in these recent representation learning and interaction focused models.

Some neural IR models [164, 357, 412] learn (dense low-dimensional or sparse high-dimensional) vector representations of document that can be computed independently of the query. The query-document relevance is then estimated as simple similarity functions (*e.g.*, cosine or dot-product) of the learned representations. These models are also amenable to precomputation of document representations and fast retrieval using approximate nearest neighbor search [413]—or even traditional IR data structures [407]. However, these approaches do not work when the model architecture incorporates early interactions between query and document representations—*e.g.*, [7, 167, 238, 392]. The approach proposed in this study allows for interactions between individual query terms and documents.

## 5.3 Model

IR functions that assume QTI observe the following general form:

$$S_{q,d} = \sum_{t \in q} s_{t,d} \tag{5.1}$$

Where, $s \in \mathbb{R}_{\geq 0}^{|V| \times |C|}$ is the set of positive real-valued scores as estimated by the relevance model corresponding to documents $d \in C$ in collection $C$ w.r.t. to terms $t \in V$ in vocabulary $V$—and $S_{q,d}$ denotes the aggregated score of document $d$ w.r.t. to query $q$. For example, in case of BM25 [80]:

$$s_{t,d} = \text{idf}_t \cdot \frac{\text{tf}_{td} \cdot (k_1 + 1)}{\text{tf}_{td} + k_1 \cdot \left(1 - b + b \cdot \frac{|d|}{avgdl}\right)} \tag{5.2}$$

Where, tf and idf denote term-frequency and inverse document frequency, respectively—and $k_1$ and $b$ are the free parameters of the BM25 model.

Deep neural models for ranking, in contrast, do not typically assume QTI. Instead, they learn complex matching functions to compare the candidate document to the full query. The parameters of such a model $\phi$ is typically learned discriminatively by minimizing a loss function of the following form:

$$\mathscr{L} = \mathbb{E}_{q \sim \theta_q,\, d_+ \sim \theta_{d_+},\, d_- \sim \theta_{d_-}} [\ell(\Delta_{q,d_+,d_-})] \tag{5.3}$$

$$\text{where,} \quad \Delta_{q,d_+,d_-} = \phi_{q,d_+} - \phi_{q,d_-} \tag{5.4}$$

We use $d_+$ and $d_-$ to denote a pair of relevant and non-relevant documents, respectively, *w.r.t.* query $q$. The instance loss $\ell$ in Equation 5.8 can take different forms—*e.g.*, ranknet [235] or hinge [267].

$$\ell_{\text{ranknet}}(\Delta_{q,d_+,d_-}) = \log(1 + e^{-\sigma \cdot \Delta_{q,d_+,d_-}}) \tag{5.5}$$

$$\ell_{\text{hinge}}(\Delta_{q,d_+,d_-}) = \max\{0, \varepsilon - \Delta_{q,d_+,d_-}\} \tag{5.6}$$

Given a neural ranking model $\phi$, we define $\Phi$—the corresponding model under the QTI assumption—as:

$$\Phi_{q,d} = \sum_{t \in q} \phi_{t,d} \tag{5.7}$$

The new model $\Phi$ preserves the same architecture as $\phi$ but estimates the relevance of a document independently *w.r.t.* each query term, as shown in Figure 5.1.

The parameters of $\Phi$ are learned using the modified loss:

$$\mathscr{L} = \mathbb{E}_{q \sim \theta_q,\, d_+ \sim \theta_{d_+},\, d_- \sim \theta_{d_-}} [\ell(\delta_{q,d_+,d_-})] \tag{5.8}$$

$$\text{where,} \quad \delta_{q,d_+,d_-} = \sum_{t \in q} \phi_{t,d_+} - \phi_{t,d_-} \tag{5.9}$$

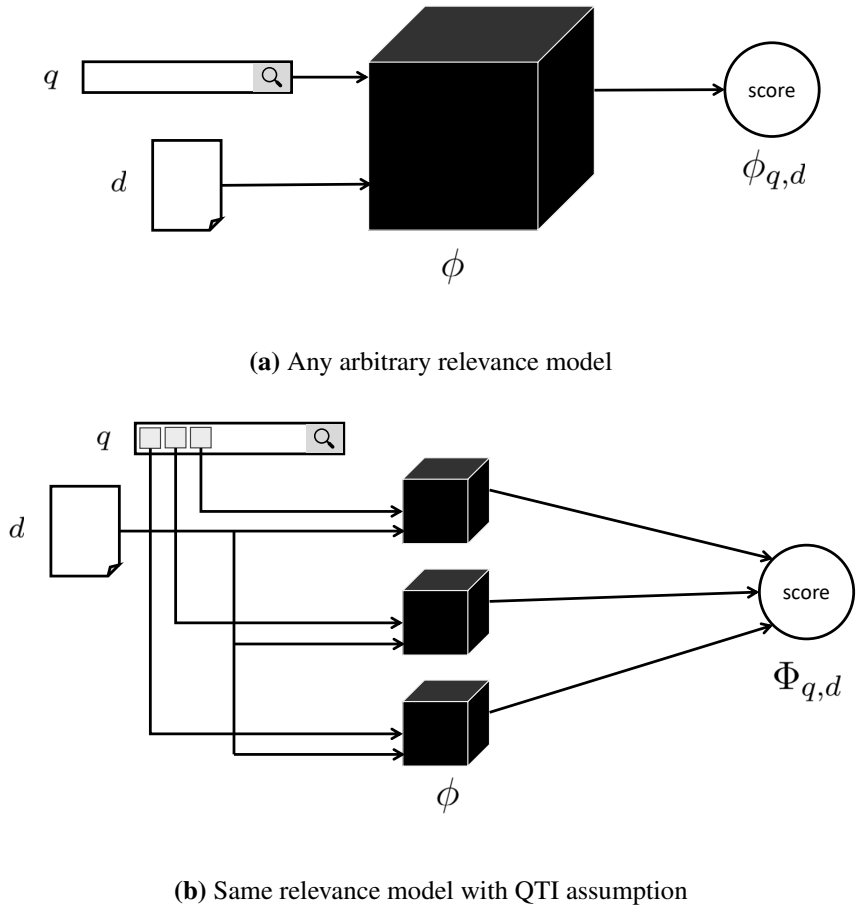

(a) Any arbitrary relevance model

(b) Same relevance model with QTI assumption

**Figure 5.1:** A visual representation of incorporating QTI assumption into any relevance model. We treat the model in (a) as a black-box and re-visualize the same model under QTI assumption in (b).

Given collection $C$ and vocabulary $V$, we precompute $\phi_{t,d}$ for all $t \in V$ and $d \in C$. In practice, the total number of combinations of $t$ and $d$ may be large but we can enforce additional constraints on which $\langle t, d \rangle$ pairs to evaluate, and assume no contributions from remaining pairs. During query evaluation, we can lookup the precomputed score $\phi_{t,d}$ without dedicating any additional time and resource to evaluate the deep ranking model. We employ an inverted index, in combination with the precomputed scores, to perform retrieval from the full collection using the learned relevance function $\Phi$. We note that several IR data structures assume that $\phi_{t,d}$ be always positive which may not hold for any arbitrary neural architecture. But this can be rectified[1] by applying a rectified linear unit [390] activation on the model's output.

---
[1] Pun intended.

## 5.4 Experiments

### 5.4.1 Task description

We study the effect of the QTI assumption on deep neural IR models in the context of the MS MARCO passage ranking task [52]. We find this ranking task to be suitable for this study for several reasons. Firstly, with one million question queries sampled from Bing's search logs, 8.8 million passages extracted from web documents, and 400,000 positively labeled query-passage pairs for training, it is one of the few large datasets available today for benchmarking deep neural IR methods. Secondly, the challenge leaderboard[2]—with 18 entries as of March 3, 2019—is a useful catalog of approaches that show state-of-the-art performance on this task. Conveniently, several of these high-performing models include public implementations for the ease of reproducibility.

Two comparable benchmarks include the TREC CAR [8, 414] and the Google Natural Questions [415] datasets. However, we note that the queries in the former dataset are synthetically generated—from Wikipedia page titles and section headings. The latter dataset was released fairly recently and does not list many IR methods that have been evaluated on that benchmark—limiting our options to select appropriate baselines for the study. Therefore, we adopt the MS MARCO benchmark for this work.

The MS MARCO passage ranking task comprises of one thousand passages per query that the IR model, being evaluated, should re-rank. Corresponding to every query, one or few passages have been annotated by human editors as containing the answer relevant to the query. The rank list produced by the model is evaluated using the MRR metric against the ground truth annotations. We use the MS MARCO training dataset to train all baseline and treatment models, and report their performance on the publicly available development set which we consider—and hereafter refer to—as the test set for our experiments. This test set contains about seven thousand queries which we posit is sufficient for reliable hypothesis testing.

---

[2]`http://www.msmarco.org/leaders.aspx`

Note that the thousand passages per query were originally retrieved using BM25 from a collection that is provided as part of the MS MARCO dataset. This allows us to also use this dataset in a retrieval setting—in addition to the re-ranking setting used for the official challenge. We take advantage of this in our study.

### 5.4.2 Baseline models

We begin by identifying models listed on the MS MARCO leaderboard that can serve as baselines for our work. We only consider the models with public implementations. We find that a number of top performing entries—*e.g.*, [167]—are based on recently released large scale language model called BERT [327]. The BERT based entries are followed in ranking by Duet and CKNRM. Therefore, we limit this study to BERT, Duet, and CKNRM.

**BERT** Nogueira and Cho [167] report state-of-the-art retrieval performance on the MS MARCO passage re-ranking task by fine tuning BERT [327] pretrained models. In this study, we reproduce the results from their paper corresponding to the BERT Base model and use it as our baseline. Under the term independence assumption, we evaluate the BERT model once per query term—wherein we input the query term as sentence A and the passage as sentence B.

**Duet** We employ the Duet variant described in Section 4.4.1 for this study.

**CKNRM** The CKNRM model [392] combines kernel pooling based soft matching [393] with a convolutional architecture for comparing *n*-grams. CKNRM uses kernel pooling to extract ranking signals from interaction matrices of query and passage *n*-grams. Under the query term independence assumption, the model considers one query term at a time—*i.e.*, the interactions between individual query unigrams and passage *n*-grams. We use a public implementation[3] of the model in our study.

## 5.5 Results

Table 5.1 compares the BERT, the Duet, and the CKNRM models trained under the query term independence assumption to their original counterparts on the pas-sage re-ranking task. During this study, we observed some variance in relevance metrics corresponding to different training runs for the CKNRM model using different random seeds. To control for this variance we train eight different clones of the CKNRM model and report mean and median MRR. Similarly, the metrics corresponding to the Duet model is based on five separate training runs, although we observe negligible variance in the context of this architecture. For the BERT based models, due to long training time we only report results based on a single training and evaluation run. As table 5.1 shows, we observe no statistically significant difference in effectiveness from incorporating the query term independence assumptions in either Duet or CKNRM. The query term independent BERT model performs slightly worse than its original counterpart on MRR but the performance is still superior to other non-BERT based approaches listed on the public leaderboard. Note that all three models emphasize on early interactions between query and document representations—unlike other prior work [164, 412] where the interaction is limited to the final stage. Under the QTI assumption, we allow early interaction between individual query terms and document, but delay the full query-document interaction till the end. Our observation that delaying the query-document interaction has no significant impact on effectiveness of these interaction-based models is a key finding of this study.

---
[3]https://github.com/thunlp/Kernel-Based-Neural-Ranking-Models

**Table 5.1:** Comparing ranking effectiveness of BERT, Duet, and CKNRM with the query independence assumption (denoted as "Term ind.") with their original counterparts (denoted as "Full"). The difference between the median MRR for "full" and "term ind." models are *not* statistically significant based on a student's t-test ($p < 0.05$) for Duet and CKNRM. The difference in MRR is statistically significant based on a student's t-test ($p < 0.05$) for BERT (single run). The BM25 baseline (single run) is included for reference.

| Model | MRR@10 | | |
|---|---|---|---|
| | Mean | ($\pm$ Std. dev) | Median |
| **BERT** | | | |
| Full | 0.356 | | 0.356 |
| Term ind. | **0.333** | | **0.333** |
| **Duet** | | | |
| Full | 0.239 | ($\pm$0.002) | 0.240 |
| Term ind. | 0.244 | ($\pm$0.002) | 0.244 |
| **CKNRM** | | | |
| Full | 0.223 | ($\pm$0.004) | 0.224 |
| Term ind. | 0.222 | ($\pm$0.005) | 0.221 |
| BM25 | 0.167 | | 0.167 |

**Table 5.2:** Comparing Duet (with QTI assumption) and BM25 under the full retrieval settings. The differences in recall and MRR between Duet (term ind.) and BM25 are statistically significant according to student's t-test ($p < 0.01$).

| Model | Recall@1000 | MRR@10 |
|---|---|---|
| BM25 | 0.80 | 0.169 |
| BM25 + Duet | 0.80 | 0.212 |
| **Duet (term ind.)** | **0.85** | **0.218** |

We posit that models with query term independence assumption—even when slightly less effective compared to their full counterparts—are likely to retrieve better candidate sets for re-ranking. To substantiate this claim, we conduct a small-scale retrieval experiment based on a random sample of 395 queries from the test set. We use the Duet model with the query term independence assumption to pre-compute the term-passage scores constrained to (i) the term appears at least once in the passage, and (ii) the term does not appear in more than 5% of the passage collection. Table 5.2 compares Duet and BM25 on their effectiveness as a first stage retrieval method in a potential telescoping setting [111]. We observe a 6.25% improvement in recall@1000 from Duet over the BM25 baseline. To perform similar retrieval from the full collection using the full Duet model, unlike its query-term-independent counterpart, is prohibitive because it involves evaluating the model on every passage in the collection against every incoming query.

## 5.6 Conclusion

The emergence of compute intensive ranking models, such as BERT, motivates rethinking how these models should be evaluated in large scale IR systems. The approach proposed in this paper moves the burden of model evaluation from the query evaluation stage to the document indexing stage. This may have further consequences on computational efficiency by allowing batched model evaluation that more effectively leverages GPU (or TPU) parallelization.

This preliminary study is based on three state-of-the-art deep neural models on a public passage ranking benchmark. The original design of all three models—BERT, Duet, and CKNRM—emphasize on early interactions between query and passage representations. However, we observe that limiting the interactions to passage and individual query terms has reasonably small impact on their effectiveness. These results are promising as they support the possibility of dramatically speeding up query evaluation for some deep neural models, and even employing them to retrieve from the full collection. The ability to retrieve—and not just re-rank—using deep models has significant implications for neural IR research. Any loss in retrieval effectiveness due to incorporating strong query term independence assumptions may be further recovered by additional stages of re-ranking in a telescoping approach [111].

This study is focused on the passage ranking task. The trade-off between effectiveness and efficiency may be different for document retrieval and other IR tasks. Traditional IR methods in more complex retrieval settings—*e.g.*, when the document is represented by multiple fields [146]—also observe the query term independence assumption. So, studying the query term independence assumption in the context of corresponding neural models—*e.g.*, [389]—may also be appropriate. We note these as important future directions for our research.

The findings from this study may also be interpreted as pointing to a gap in our current state-of-the-art neural IR models that do not take adequate advantage of term proximity signals for matching. This is another finding that may hold interesting clues for IR researchers who want to extract more retrieval effectiveness from deep neural methods.

# Chapter 6

# Stochastic learning to rank for target exposure

Retrieval systems mediate what information users are exposed to and consume. A typical large collection may contain several documents that are relevant, albeit to varying degrees, to a user's query. Because users rarely inspect all retrieved results exhaustively, the IR system must prioritize what documents are exposed more than others to maximize the chances of user satisfaction. The need for this prioritization is often operationalized by formulating retrieval as a ranking task, as we have also assumed in previous chapters. Consequently, this assumption that the system produces a ranked list of results is often also baked into the design of many information access interfaces. A common form of presentation involves displaying a vertical (or sometimes horizontal) result list. Even sophisticated visual interfaces, such as grid layouts, or non-visual interaction modes, as in the case of voice-based search, may assume that the backend retrieval system returns a ranked list of results which determines how prominently they should be displayed. Across these different modalities, the probability that the user inspects a certain result depends on its display position [73, 113] and size [114] among other factors, which in turn may be determined by the document's rank in the results list.

A static ordering by estimated relevance makes sense if we assume: (i) the IR system is only concerned about satisfying the user performing the search, and (ii) all relevant documents are equivalent from the user's perspective and therefore the user's interests are best served by ordering retrieved documents strictly by their estimated relevance. In many real-life IR scenarios, however, the system must also care about document and producer-side fairness [115–117]. For example, in web search we may want the IR system to give equal exposure to documents of comparable relevance—which may directly impact their monetization and other value that producers can extract from content exposure. When documents correspond to different demographics like gender or race—*e.g.*, candidate profiles on a job application website—parity of exposure across demographics may be important from fairness and legal concerns. In scenarios where the system produces a ranking of service providers, such as booking a hotel or hailing a ride [416], distributing exposure across multiple providers may be necessary to avoid producer starvation or overload. When retrieved documents have comparable relevance but contain different information, then balanced exposure may increase diversity of consumption and help mitigate phenomenon like *filter bubbles* [417].

In these scenarios, a single fixed ranking makes less sense. Instead, it may be more meaningful for the system to present different randomized permutations of comparably relevant documents to distribute exposure more fairly among them. Such stochastic ranking policies provide a framework for optimizing how exposure is distributed in expectation. In Chapters 4 and 5, we adopted the narrow view that it is sufficient to learn a relevance model whose estimates are appropriate for generating single static ordering of results. In contrast, in this chapter we shift our focus to optimizing models that produce relevance estimates appropriate for generating different permutations of results that minimizes the deviation of expected exposure from a specified target distribution. Our main contribution here is to adapt the learning to rank [39] framework for direct optimization towards target exposure.

## 6.1   Related work

In the learning to rank [39] literature, several optimization objectives have been proposed that can be broadly categorized into: (i) pointwise, (ii) pairwise, and (iii) listwise loss functions. Because the exposure of a document is a function of its rank in the result list, our optimization goals are better served by the listwise formulation. Several listwise loss functions [277, 278] operationalize the idea of deriving the probability of a rank ordering given the score distribution over documents using the Plackett-Luce model [274, 275]. It is noteworthy, that enumerating all distinct document permutations can be computationally challenging even for a moderately sized set of candidates. More recently, Bruch et al. [418] demonstrated a mechanism for sampling rankings from the Plackett-Luce distribution using the reparameterization trick [395] that is amenable to gradient-based optimization. Their approach involves adding independently drawn noise samples from the Gumbel distribution [419] and then deriving the approximate rank of the document following the method proposed by Qin et al. [420] and Wu et al. [421]. While not developed in the context of deploying stochastic ranking models, we adopt a similar methodology herein in our framework.

Our work is at the intersection of learning to rank optimization and expected exposure metrics. For the latter, we operationalize the framework proposed by Diaz et al. [14]. In Section 6.2, we provide a brief primer on this topic.

## 6.2 Expected exposure metrics

Given an information need $q$, Diaz et al. [14] define the expected exposure $\varepsilon_d$ of document $d$ as:

$$\varepsilon_d = \mathbb{E}_{\sigma \sim \pi_q}[\mu(d|\sigma)] \tag{6.1}$$

Where, $\sigma$ is a ranking of documents in the collection, sampled from $\pi_q$, a probability distribution over all possible permutations of documents conditioned on $q$. We use $\mu(d|\sigma)$ to denote the conditional probability of exposure of document $d$ given ranking $\sigma$. To compute $\mu(d|\sigma)$, we can adopt any arbitrary user behavior model [422] that defines how the user interacts with the presented rank list. For example, the rank-biased precision (RBP) [423] metric assumes that a user's probability of visiting a position decreases exponentially with rank.

$$\mu_{\text{RBP}}(d|\sigma) = \gamma^{(\rho_{\sigma,d}-1)} \tag{6.2}$$

Where, $\rho_{\sigma,d}$ is the rank of the document $d$ in $\sigma$—and $\gamma$ is the *patience parameter* that controls how deep in the ranking the user is likely inspect. We adopt this RBP user behavior model in this study but note that this analysis can be easily extended to more elaborate browsing models like the cascade model [424].

Plugging in the RBP user-model into Equation 6.1 we get:

$$\varepsilon_d = \mathbb{E}_{\sigma \sim \pi_q} \left[ \gamma^{(\rho_{\sigma,d}-1)} \right] \tag{6.3}$$

Diaz et al. [14] further define a metric that quantifies the deviation between the expected exposure vector $\varepsilon$ corresponding to all documents in the collection under a retrieval system from a specified target distribution $\varepsilon^*$:

$$\text{EE}(\pi, q) = \|\varepsilon - \varepsilon^*\|_2^2 \tag{6.4}$$

$$= \underbrace{\|\varepsilon\|_2^2}_{\text{EE-D}} - \underbrace{2\varepsilon^\intercal \varepsilon^*}_{\text{EE-R}} + \|\varepsilon^*\|_2^2 \tag{6.5}$$

Equation 6.5 factorizes the expected exposure metric into *expected exposure disparity* (EE-D) and *expected exposure relevance* (EE-R). EE-D measures the inequity in exposure distribution over all documents which we want to minimize when optimizing the parameters of the ranking policy. In contrast, EE-R quantifies how much of the exposure is on relevant documents which a good ranking policy should maximize. This leads to a natural trade-off between disparity (EE-D) and relevance (EE-R) which often relates to the degree of randomization applied by a stochastic policy. A deterministic policy may achieve the highest relevance at the cost of high disparity. Similarly, a policy that randomly samples documents from the collection with uniform probability achieves lowest disparity but also lowest relevance. In our experiments, we plot a disparity-relevance curve by controlling the degree of randomization and report the area under this curve (EE-AUC).

The target exposure $\varepsilon^*$ specifies the ideal behaviour we desire from our retrieval system. One way to compute this is by assuming some oracle ranking policy. For example, in this work we adopt the principle of equal expected exposure defined by Diaz et al. [14] as:

> *Given a fixed information need, no item should be exposed (in expectation) more or less than any other item of the same relevance.*

Under this ideal policy, documents always appear in ranking above other documents of lower grades, and documents in the same grade are permuted with uniform probability. Let $m_g$ be the number of documents with relevance grade $g$ and $m_{>g}$ the number of documents with relevant grade strictly larger than $g$. Given an RBP browsing model, the optimal exposure for a document $d$ with grade $g$ is,

$$\varepsilon_d^* = \frac{1}{m_g} \sum_{\rho \in [1, m_g]} \gamma^{(\rho + m_{>g})} \tag{6.6}$$

$$= \frac{\gamma^{m_{>g}} \cdot (1 - \gamma^{m_g})}{m_g(1 - \gamma)} \tag{6.7}$$

We refer the readers to the original paper for more detailed derivation and discussion of this individual exposure parity target.

If we associate the documents in our collection with a set $\mathscr{A}$ of $k$ attributes, then we can also define a group notion of exposure parity. These attributes may reflect, for example, demographic information about the content producer or some topical grouping by content. Let **A** be a $n \times k$ binary matrix mapping each of the $n$ documents in the collection to their group identity. We can then compute the total exposure for all documents with an attribute as $\xi = \mathbf{A}^\mathsf{T} \varepsilon$. We recover equal exposure across groups, by enforcing $\xi$ to be uniform. We can replace $\varepsilon$ with $\xi$ in Equation 6.4 to define as a measure of equal exposure across groups. Other notions of demographic and group fairness can be similarly derived.

## 6.3 Optimizing for target exposure

Following the Plackett-Luce model [274, 275], given some arbitrary score distribution **y** over documents, we can sample different rankings by iteratively sampling documents without replacement based on the following softmax distribution.

$$P_{\text{PL}}(d|q) = \frac{\exp(\mathbf{y}_d)}{\sum_{\bar{d}} \exp(\mathbf{y}_{\bar{d}})} \tag{6.8}$$

Under the assumption of binary relevance and a perfect relevance estimator, Plackett-Luce randomization should perform optimally. However, learning to rank models are not perfect estimators of relevance. Therefore, we believe there should be some advantage to optimizing directly for expected exposure. We leverage recent results in optimization of relevance-based objectives computed over a distribution over rankings [418]. Our results can be seen as an extension of this framework to individual and group exposure objectives.

We focus on a shared model architecture with varying loss functions in order to measure differences due to the objective alone, instead of artifacts resulting from the functional form of the models. We begin by describing how we optimize for expected exposure before proceeding to our experiment design and empirical results.

### 6.3.1 Individual exposure parity

Although optimizing for pointwise or pairwise loss has been well-studied in the information retrieval community, directly optimizing for a metric based on a distribution over rankings has received less attention.

We begin by defining an appropriate loss function for our model. Turning to Equation 6.5, we can drop the constant term and add a hyperparameter to trade-off disparity and relevance,

$$\ell_\lambda(\varepsilon, \varepsilon^*) = \lambda \|\varepsilon\|_2^2 - (1-\lambda)\varepsilon^\mathsf{T} \varepsilon^* \tag{6.9}$$

Where $\varepsilon^*$ is based on graded relevance.

Let $f_\theta : D \rightarrow \mathbb{R}$ be a document scoring function parameterized by $\theta$. Given a query, **y** is a $n \times 1$ vector of document scores for the entire collection such that, $\mathbf{y}_d = f_\theta(d)$. Using a Plackett-Luce model, we can translate the raw scores into sampling probabilities as in Equation 6.8. This allows us to construct a ranking $\sigma$ by sampling documents sequentially. Unfortunately, this sampling process is non-differentiable and, therefore, prohibitive to a large class of models, including those that learn by gradient descent. We address this by adopting the method proposed by Bruch et al. [418]. To construct a sampled ranking $\sigma$, we reparameterize the probability distribution by adding independently drawn noise samples $G$ from the Gumbel distribution [419] to **y** and sorting documents by the "noisy" probability distribution $\tilde{p}$,

$$\tilde{p}(d_i) = \frac{\exp(\mathbf{y}_{d_i} + G_i)}{\sum_{d_j \in D} \exp(\mathbf{y}_{d_j} + G_j)} \qquad (6.10)$$

Where $G_i$ is sample from the Gumbel distribution.

$$G_i = -\log(-\log U_i) \qquad (6.11)$$
$$U \sim \text{Uniform}(0,1) \qquad (6.12)$$

Given the perturbed probability distribution $\tilde{p}$, we compute each document's smooth rank [420, 421] as,

$$\overline{\sigma}_d = \sum_{d' \in D/d} \left(1 + \exp\left(\frac{\tilde{p}(d) - \tilde{p}(d')}{\tau}\right)\right)^{-1} \qquad (6.13)$$

The smooth rank is sensitive to the temperature $\tau$ [425]. At high temperatures the smooth rank is a poor approximation of the true rank and at low temperatures may result in vanishing gradients. To rectify this issue, we employ the straight-through estimator [426] to compute the true ranks in forward pass but differentiate the gradients with respect to the smooth ranks during backpropagation.

Using the estimated ranks and a specified user model we compute the exposure for each document. For example, assuming RBP as the user model the exposure of document $d$ from a single ranking $\sigma$ is given by $\varepsilon_d = \gamma^{(\rho_{\sigma,d}-1)}$. We compute expected exposure by averaging over $n_{\text{train}}$ different rankings—each generated by independently sampling different Gumbel noise in Equation 6.10.

We use this expected exposure vector $\varepsilon$ in Equation 6.9 to compute the loss that we minimize through gradient descent, as shown in Figure 6.1. The relevance grades are not used for training beyond computing target exposure. We set $\tau$ in Equation 6.13 to 0.1.

### 6.3.2　Group exposure parity

We can also adapt this model to optimize fpr group-level exposure parity. To do so, we replace $\|\varepsilon\|_2^2$ with $\|\xi\|_2^2$ in Equation 6.9 to define an optimization objective that trades-off relevance and group parity.

$$\ell_{\text{group},\lambda} = \lambda \|\xi\|_2^2 - (1-\lambda)\varepsilon^\intercal \varepsilon^* \tag{6.14}$$

This loss function assumes that the ideal policy distributes exposure equally across all groups. Optimization objectives corresponding to other group exposure criteria can be derived similarly in future work.

## 6.4　Experiments

### 6.4.1　Models

We restrict our choice of baselines to neural networks so that the exposure-based optimization can be compared to baseline ranking loss functions with respect to the same model. Our base model consists of a fully-connected neural network with two hidden layers of size 256 nodes per layer and rectified linear unit for activation function. We choose a learning rate of 0.001 and a dropout rate of 0.1 and perform early-stopping for all models based on validation sets. Baseline stochastic rankings are derived by employing Plackett-Luce sampling over two deterministic policies (pointwise and pairwise models) with varying softmax temperatures to obtain different trade-off points between disparity and relevance. We set $n_{\text{train}}$ to 20 for our model and $n_{\text{test}}$ to 50 for all models.

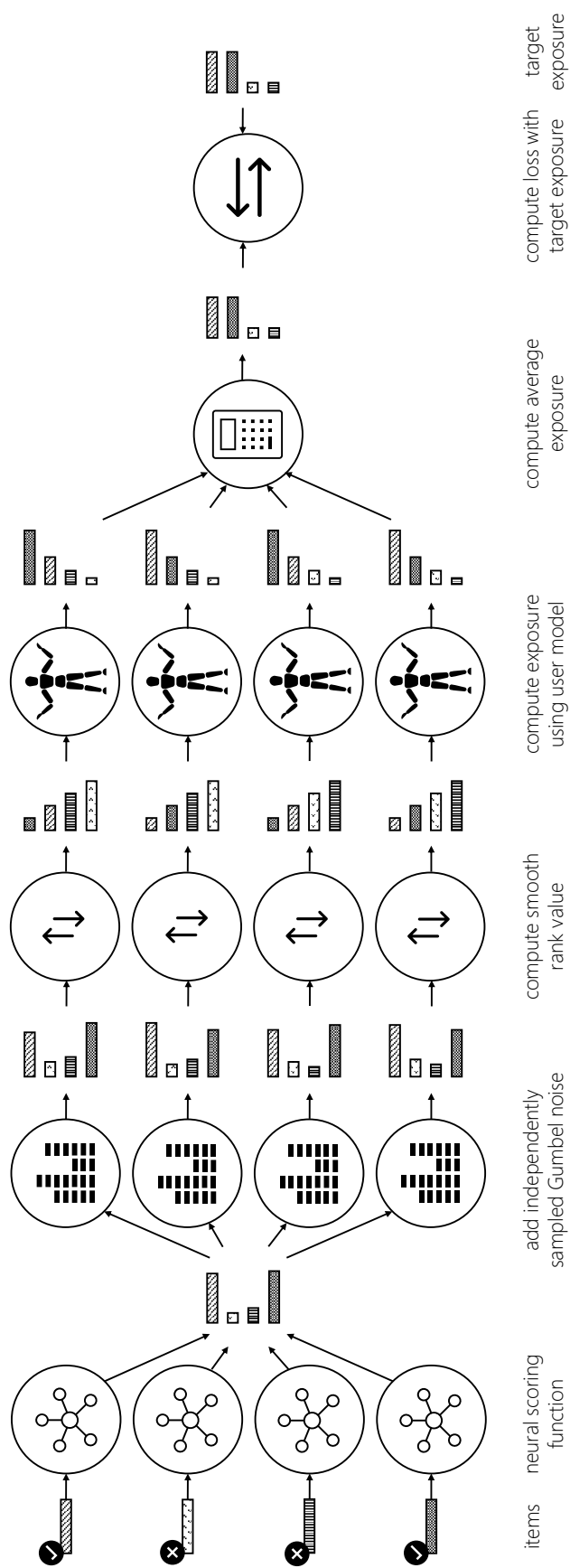

**Figure 6.1:** To sample multiple rankings proportional to the softmax distribution over document scores, we first add independently sampled noise from the Gumbel distribution to the scores and then estimate corresponding smooth rank values. We then compute exposure of each document for a given ranking based on a preselected user model. Next, we estimate the expected exposure of a document by averaging across multiple rankings. Finally, we compute the loss between the predicted and the target expected exposure vectors, which can be then minimized using gradient-based methods as every step in the above process is differentiable.

We consider three training objectives in our experiments. The pointwise model [240] minimizes the squared error between the model prediction and true relevance. The pairwise model [235] minimizes misclassified preferences using a cross-entropy loss. The expected exposure model minimizes the loss in Equation 6.9 and, in our group parity experiments, Equation 6.14.

### 6.4.2 Data

Our experiments use the MSLR-WEB10k dataset [427], a learning-to-rank dataset containing ten thousand queries. We perform five-fold cross validation (60/20/20 split between training, validation, and testing sets). Each query-document pair is represented by a 136-dimensional feature vector and graded according to relevance on a five-point scale.

For the group parity experiments, as there are no obvious appropriate group attributes in the MSLR-WEB10k dataset, we discretize the PageRank feature in the ranges <1000, 1000–10000, and ≥10000 and treat it as a group attribute. The choice of using discretized PageRank as a group attribute is rather arbitrary, but we confirm that this discretization scheme is reasonable as roughly 70% of the queries have at least one document corresponding to each group with a relevance grade greater than one.

### 6.4.3 Evaluation

We use a $\gamma = 0.50$ for all of our experiments, as consistent with standard TREC evaluation protocol. RBP is evaluated at depth 20.

## 6.5 Results

We present the results of our experiments in Table 6.1. In terms of expected exposure, we do not observe a difference in performance between pointwise and pairwise models. However, directly optimizing for expected exposure resulted in a 3.9% improvement in EE-AUC over the pointwise and pairwise models. We confirm that the difference in EE-AUC follows a normal distribution and accordingly perform a paired student's t-test to check their statistical significance. The EE-AUC differences between our proposed method and the baselines are statistically significant ($p < 0.01$).

Table 6.1: Results for optimizing towards individual and group parity using different ranking objectives. We report average EE-AUC for both tasks and highlight the best performance for each in bold. Optimizing directly for individual and group parity using our proposed method achieves best performance in both cases.

| Loss function | AUC | |
| --- | --- | --- |
| | **Individual parity** | **Group parity** |
| Pointwise loss | 0.229 | 0.112 |
| Pairwise loss | 0.229 | 0.108 |
| **Our methods** | | |
| Expected exposure loss (Eqn. 6.9) | **0.238** | 0.141 |
| Group parity loss (Eqn. 6.14) | | **0.178** |

In terms of group parity, we observe a difference in performance between pointwise and pairwise models. Moreover, directly optimizing for expected exposure results in improved performance while directly optimizing for group parity further boosts performance. The gap in EE-AUC between all pairs of models are statistically significant ($p < 0.01$).

These results, while based on a limited study, indicate that direct optimization for expected exposure metrics is both viable in the learning to rank framework as well as useful for optimization under fairness constraints.

## 6.6 Conclusion

An exposure-based view of retrieval explicitly codifies the role that IR systems play as intermediary in two-sided marketplaces consisting of users seeking information and documents (or their producers). Stochastic ranking policies allow for more balanced distribution of exposure over multiple rankings. In this work, we demonstrate that these policies can be directly optimized to reduce deviation from specified target exposure distribution. While our work is grounded in parity of individual and group exposure, the framework described is flexible enough to incorporate any arbitrary target exposure policy beyond fairness constraints—*e.g.*, based on topical diversity considerations or to maximize monetization in the context of paid search.

Our definition of target exposure in this work is based on a universal notion of relevance. If the relevance of a document instead changes based on the searcher (*i.e.*, personalization) or other context (*e.g.*, location), then our framework needs to be appropriately extended. Exposure can also be nuanced by user attributes. For example in commercial searches, exposure to users with an intent to purchase may be weighted differently than to users who may be casually browsing. We believe that there is a rich space for exploring different extensions of our proposed framework.

Deploying stochastic ranking policies may also come with its own unique challenges. For example, randomized rankings may have unintended consequences on the system's caching mechanisms. It may also make it harder for users to re-find information [428] they have previously discovered for a query. More detailed studies are also necessary to understand the differential impact of stochastic policies on queries of varying difficulty, especially on queries for which the model's relevance estimates are highly uncertain.

# Chapter 7

# Learning to Rank for Query Auto-Completion

In this chapter, we discuss the application of deep architectures to the query auto-completion task that presents different challenges than ad hoc retrieval. Query auto-completion helps the user of a search system to formulate their information request by recommending queries based on their partially typed query. The query auto-completion system typically considers the user history, the task context the location and temporal context, and other information to make more relevant recommendation. The ranking task, in case of query auto-completion, therefore, involves ranking either query suffixes or full query candidates in response to a query prefix. In this chapter, we discuss work in which we employ deep neural networks for that ranking task.

## 7.1 Query Auto-Completion for Rare Prefixes

As users enter their query into the search box, most modern search engines provide a ranked list of query suggestions based on the current prefix already typed by the user. In a typical approach used by many query auto-completion (QAC) systems, candidate queries are identified by doing an exact prefix lookup against a fixed set of popular queries, using a data structure such as a prefix tree [143]. The candidates are then ranked by their expected likelihood, which is typically computed as a function of its past popularity (commonly referred to as the *MostPopularCompletion* (MPC) model [429]). Such a system can only suggest queries with enough historic popularity to make it into the prefix tree.

**Table 7.1:** Synthetic QAC candidates generated by the suffix-based approach and ranked using only the CDSSM similarity feature. The CDSSM model projects both the prefix and the suffix to a common 128-dimensional space allowing us to rank according to prefix-suffix cosine similarity. One of the lower quality synthetic candidates "cheapest flights from seattle to airport" is ranked seventh in the second list.

| |
|---|
| what to cook with chicken and broccoli and |
| what to cook with chicken and broccoli *and bacon* |
| what to cook with chicken and broccoli *and noodles* |
| what to cook with chicken and broccoli *and brown sugar* |
| what to cook with chicken and broccoli *and garlic* |
| what to cook with chicken and broccoli *and orange juice* |
| what to cook with chicken and broccoli *and beans* |
| what to cook with chicken and broccoli *and onions* |
| what to cook with chicken and broccoli *and ham soup* |
| cheapest flights from seattle to |
| cheapest flights from seattle *to dc* |
| cheapest flights from seattle *to washington dc* |
| cheapest flights from seattle *to bermuda* |
| cheapest flights from seattle *to bahamas* |
| cheapest flights from seattle *to aruba* |
| cheapest flights from seattle *to punta cana* |
| cheapest flights from seattle *to airport* |
| cheapest flights from seattle *to miami* |

We propose an additional candidate generation strategy for QAC by mining popular query *suffixes*. Candidate suffixes are popular *n*-grams that appear at the ends of queries. By appending such *n*-grams suffixes to a user's query prefix we can generate synthetic suggestion candidates that have never been observed in the historical query logs. Table 7.1 contains examples of such suggestions. We further propose a supervised framework for ranking these synthetic queries alongside the traditional full-query suggestion candidates. We also explore new ranking signals in this framework, based on the query *n*-gram statistics and a deep CDSSM [362].

### 7.1.1   Related work

Most modern browsers, search engines, text editors and command shells implement some form of an auto-completion feature to aid users in faster text entry. In Web search, *pre-computed auto-completion* systems are popular, where the suggestions are typically filtered by exact prefix matching from a pre-selected set of candidates and ranked according to past popularity. Ranking suggestions by past frequency is commonly referred to as the *MostPopularCompletion* (MPC) model and can be regarded as a maximum likelihood approximator [429]. Given a prefix *p* and a set of all unique queries *Q* from the search logs,

$$MPC(p) = \underset{\bar{q} \in \text{pc}(p)}{\arg\max} \frac{\text{lf}(\bar{q})}{\sum_{q_i \in Q} \text{lf}(q_i)} \qquad (7.1)$$

pc returns the set of queries that qualify as completions for the query *q* and lf is the frequency of the query in the search logs.

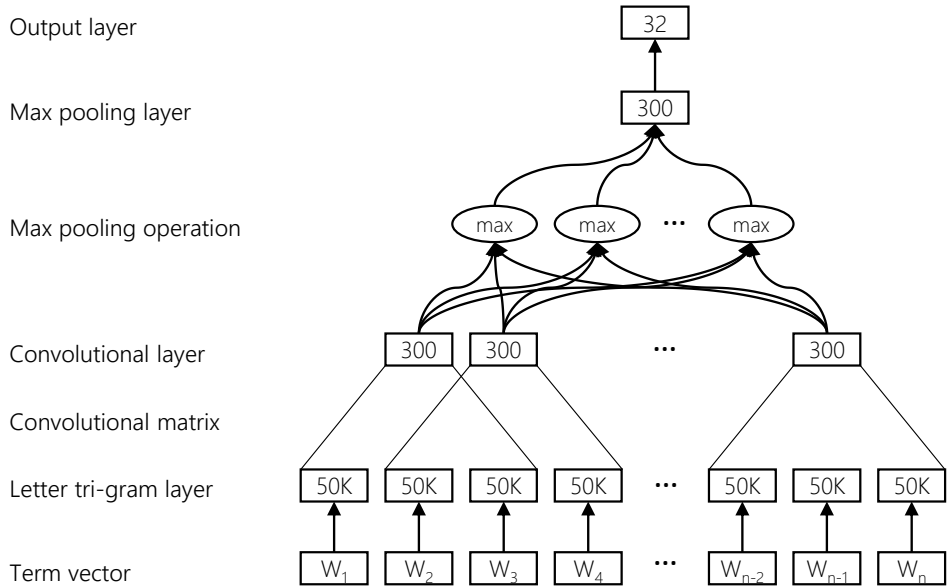

**Figure 7.1:** Architecture of the CDSSM. The model has an input layer that performs the *word hashing*, a convolutional layer, a max pooling layer, and an output layer that produces the final semantic vector representation of the query.

Language modelling based approaches for *sentence completion* have been studied in the context of e-mail and document authoring[430–433]. In Web search, White and Marchionini [434] and Fan et al. [435] proposed models for term recommendations to aid users in their query formulation process. Bhatia et al. [436] extracted frequently occurring phrases from document corpus and used them to gen-erate suggestion candidates in the absence of a query log. Duan and Hsu [437] have studied the problem of online spelling correction for query auto-completion and Hawking and Griffiths [438] have explored mechanisms for generating query suggestions in the enterprise settings. Our proposed approach generates synthetic query suggestion candidates by combining the input prefix with popular query *suffixes* to augment the regular full-query QAC suggestions. Within our proposed supervised framework, we explore CDSSM [231, 362] as a ranking signal.

**Table 7.2:** Comparing the nearest neighbours for "seattle" and "taylor swift" in the CDSSM embedding spaces when the model is trained on query-document pairs vs. query prefix-suffix pairs. The former resembles a Topical notion of similarity between terms, while the latter is more Typical in the definition of inter-term similarities.

| seattle | | taylor swift | |
|---|---|---|---|
| Query-Document | Prefix-Suffix | Query-Document | Prefix-Suffix |
| weather seattle | chicago | taylor swift.com | lady gaga |
| seattle weather | san antonio | taylor swift lyrics | meghan trainor |
| seattle washington | denver | how old is taylor swift | megan trainor |
| ikea seattle | salt lake city | taylor swift twitter | nicki minaj |
| west seattle blog | seattle wa | taylor swift new song | anna kendrick |

### 7.1.2 Model

For document retrieval, Shen et al. [362] demonstrated that discriminatively training a deep neural network model with a convolutional-pooling structure on clickthrough data can be effective for modelling query-document relevance. We adopt the CDSSM by training on a *prefix-suffix* pairs dataset (instead of query-document titles). The training data for the CDSSM is generated by sampling queries from the search logs and splitting each query at every possible word boundary. For example, from the query "breaking bad cast" we generate the two pairs ("breaking", "bad cast") and ("breaking bad", "cast"). The architecture shown in Figure 7.1 is used on both the prefix and the suffix side of the CDSSM model.

　　It is important to emphasize that our earlier discussion in Section 3.2.2 on different notions of similarity between terms that can be learnt by shallow embedding models is also relevant in the context of these deeper architectures. In the case of CDSSM [231], the notion of similarity being modelled depends on the choice of the paired data that the model is trained on. When the CDSSM is trained on query and document title pairs [231] then the notion of similarity is more *Topical* in nature. However, when the same CDSSM architecture is trained on query prefix-suffix pairs—as described in this section—it captures a more *Typical* notion of similarity, as shown in Table 7.2.

Table 7.3: Most popular query suffixes extracted from the publicly available AOL logs.

| Top suffixes | Top 2-word suffixes | Top 3-word suffixes |
|---|---|---|
| com | for sale | federal credit union |
| org | yahoo com | new york city |
| net | myspace com | in new york |
| gov | google com | or no deal |
| pictures | new york | disney channel com |
| lyrics | real estate | my space com |
| edu | of america | in new jersey |
| sale | high school | homes for sale |
| games | new jersey | department of corrections |
| florida | space com | chamber of commerce |
| for sale | aol com | bath and beyond |
| us | s com | in las vegas |

### 7.1.3 Method

We propose two key ideas in this section. Firstly, we generate synthetic query suggestion candidates for QAC using popular query *suffixes*. We then introduce *n*-gram and CDSSM based features in a supervised learning setting to rank these synthetic suggestions alongside the full-query suggestion candidates.

**Candidate Generation** From every query in the search engine logs we generate all possible *n*-grams from the end of the query. For example, from the query "bank of america" we generate the suffixes "america", "of america" and "bank of america". By aggregating across all queries we identify the most popular suffixes. Table 7.3 shows the most frequently observed query suffixes in the publicly available AOL logs [85].

Next, for a given prefix we extract the *end-term* as shown in Figure 7.2. We match all the suffixes that start with the end-term from our precomputed set. These selected suffixes are appended to the prefix to generate synthetic suggestion candidates. For example, the prefix "cheap flights fro" is matched with the suffix "from seattle" to generate the candidate "cheap flights from seattle". Note that many of these synthetic suggestion candidates are likely to not have been observed by the search engine before.

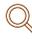

**Figure 7.2:** Examples of fully or partially typed end-terms extracted from the input prefixes. The end-term is used for selecting the set of candidate suffixes for the generation of synthetic query suggestions.

We merge these synthetic suggestions with the set of candidates selected from the list of historically popular queries. This combined set of candidates is used for ranking as we will describe in Sec 7.1.4.

**Ranking Features** For every prefix and suggestion candidate (synthetic or previously observed), we compute a set of common features for the supervised ranking model. We describe these features in this section, focusing on the *n*-gram and CDSSM features that we propose for this setting.

- *N-gram based features*. Given a candidate suggestion $q$, we compute features based on the frequency of *n*-grams $f_{\text{ngram}_i}$ of different lengths (for $i = 1$ to 6).

$$f_{\text{ngram}_i} = \sum_{g \in \text{ng}_i(q)} \text{lf}(g) \qquad (7.2)$$

where, $\text{ng}_i(q)$ is the set of all *n*-grams in query $q$ of length $i$. $\text{lf}(g)$ is the observed frequency of the *n*-gram $g$ in the historical query logs. These *n*-gram features model the likelihood that the candidate suggestion is generated by the same language model as the queries in the search logs.

- *CDSSM based features*. Given a prefix $p$ and a suggestion candidate $c$, we extract a normalized prefix $\bar{p}$ by removing the *end-term* from the prefix. Then a normalized suffix $\bar{s}$ is extracted by removing $\bar{p}$ from the query $c$. Then we use the trained CDSSM model to project the normalized prefix and the normalized suffix to a common 128-dimensional space and compute a $f_{\text{cdssm}}$ feature.

$$f_{\text{cdssm}}(\bar{p}, \bar{s}) = cosine(\vec{v}_{\bar{p}}, \vec{v}_{\bar{s}}) = \frac{\vec{v}_{\bar{p}}^\intercal \vec{v}_{\bar{s}}}{\|\vec{v}_{\bar{p}}\|\|\vec{v}_{\bar{s}}\|} \quad (7.3)$$

where $\vec{v}_{\bar{p}}$ and $\vec{v}_{\bar{s}}$ are the CDSSM vector outputs corresponding to $\bar{p}$ and $\bar{s}$, respectively. Table 7.1 shows examples of synthetic suggestion candidates ranked by the $f_{\text{cdssm}}$ feature alone.

- *Other features*. Other features used in our model includes the frequency of the candidate query in the historical logs, length based features (length of the prefix, the suffix and the full suggestion in both characters and words) and a boolean feature that indicates whether the prefix ends with a space character.

### 7.1.4 Experiments

Our experiment setup is based on the learning to rank framework proposed by Shokouhi [439]. We generate all possible prefixes[1] from each query impression to use for training, validation and testing. For each prefix we identify the set of candidate suggestions as described in Section 7.1.3. We associate a positive relevance judgment with the candidate that matches the original query from which the prefix was extracted. To accurately measure the coverage impact of our approach we retain all prefix impressions where the submitted query is not in the list of candidates available for ranking.

We train LambdaMART [440] models for ranking the suggestions using features described in Section 7.1.3. We limit our ranking task to instances where the prefix contains at least one complete word, since completions with very short prefixes is already well solved by our popularity-based features and we are focusing on rare prefixes. We always train 300 trees (with early stopping using a validation set)

---

[1] Mitra et al. [72] showed that users use QAC more at word boundaries but for simplicity we sample the prefixes with equal probability.

and evaluate the model performances on the test set using the mean reciprocal rank (MRR) metric.

We conduct all our experiments on the publicly available AOL query logs [85] and reproduce the same results on the large-scale query logs of the Bing search engine. We refer to these two datasets hereafter as the AOL testbed and the Bing testbed, respectively.

The query impressions on both the testbeds are divided into four temporally separate partitions (background, training, validation and test). On the AOL testbed we use all the data from 1 March, 2006 to 30 April, 2006 as the background data. We sample queries from the next two weeks for training, and from each of the following two weeks for validation and test, respectively. On the Bing testbed we sample data from the logs from April, 2015 and use the first week of data for background, the second week for training, the third for validation and the fourth for testing. We normalize all the queries in each of these datasets by removing any punctuation characters and converting them to lower case.

For candidate generation, both the list of popular queries and suffixes are mined from the background portion of the two testbeds. We use 724,340 and 1,040,674 distinct queries on the AOL testbed and the Bing testbed, respectively, as the set of full-query candidates. We evaluate our approach using 10K and 100K most frequent suffixes. We limit the number of full-query candidates per prefix to ten and compute the final reciprocal rank by considering only the top eight ranked suggestions per model. Finally, the CDSSM models are trained using 44,558,631 and 212,854,198 prefix-suffix pairs on the AOL and the Bing testbeds, respectively.

### 7.1.5 Results

Table 7.4 summarizes the experiment results and clearly demonstrates the improvements from the synthetic suggestion over the MPC model. All the LambdaMART models with different feature sets when combined with the suffix-based candidates show an improved MRR over the popularity based baseline. The models however perform no better, and in most cases worse, compared to the MPC baseline when only the full-query based candidates are considered. This is expected as the models

**Table 7.4:** Comparison of all models on the AOL and the Bing testbeds. Due to the proprietary nature of the Bing dataset, we only report MRR improvements relative to the MPC model for this testbed. Statistically significant differences by the t-test ($p < 0.01$) are marked with "*". Top three highest MRR values per testbed are bolded.

| Models | AOL MRR | AOL % Improv. | Bing % Improv. |
|---|---|---|---|
| **Full-query based candidates only** | | | |
| MostPopularCompletion | 0.1446 | - | - |
| LambdaMART Model ($f_{\text{ngram}_i}$ = no, $f_{\text{cdssm}}$ = no) | 0.1445 | -0.1 | -1.7* |
| LambdaMART Model ($f_{\text{ngram}_i}$ = yes, $f_{\text{cdssm}}$ = no) | 0.1427 | -1.4* | -1.2* |
| LambdaMART Model ($f_{\text{ngram}_i}$ = no, $f_{\text{cdssm}}$ = yes) | 0.1445 | -0.1 | -1.2* |
| LambdaMART Model ($f_{\text{ngram}_i}$ = yes, $f_{\text{cdssm}}$ = yes) | 0.1432 | -1.0* | -1.5* |
| **Full-query based candidates + Suffix based candidates (Top 10K suffixes)** | | | |
| MostPopularCompletion | 0.1446 | - | - |
| LambdaMART Model ($f_{\text{ngram}_i}$ = no, $f_{\text{cdssm}}$ = no) | 0.2116 | +46.3* | +32.8* |
| LambdaMART Model ($f_{\text{ngram}_i}$ = yes, $f_{\text{cdssm}}$ = no) | 0.2326 | +60.8* | +42.6* |
| LambdaMART Model ($f_{\text{ngram}_i}$ = no, $f_{\text{cdssm}}$ = yes) | 0.2249 | +55.5* | +40.1* |
| LambdaMART Model ($f_{\text{ngram}_i}$ = yes, $f_{\text{cdssm}}$ = yes) | 0.2339 | **+61.7*** | +43.8* |
| **Full-query based candidates + Suffix based candidates (Top 100K suffixes)** | | | |
| MostPopularCompletion | 0.1446 | - | - |
| LambdaMART Model ($f_{\text{ngram}_i}$ = no, $f_{\text{cdssm}}$ = no) | 0.2105 | +45.5* | +39.9* |
| LambdaMART Model ($f_{\text{ngram}_i}$ = yes, $f_{\text{cdssm}}$ = no) | 0.2441 | **+68.7*** | **+54.2*** |
| LambdaMART Model ($f_{\text{ngram}_i}$ = no, $f_{\text{cdssm}}$ = yes) | 0.2248 | +55.4* | **+48.9*** |
| LambdaMART Model ($f_{\text{ngram}_i}$ = yes, $f_{\text{cdssm}}$ = yes) | 0.2453 | **+69.6*** | **+55.3*** |

are trained with the suffix-based candidates in the training data.

The models with the $f_{\text{cdssm}}$ feature perform better than the corresponding models without the feature across all experiments. However, in general the $f_{\text{ngram}_i}$ features seems to be showing higher improvements compared to the CDSSM based feature. We hypothesize that the $f_{\text{cdssm}}$ feature is less precise than the $f_{\text{ngram}_i}$ features. For example, we can see in Table 7.1 that the CDSSM based feature ranks a suffix highly that generates a semantically meaningless query suggestion "cheapest flight from seattle to airport". While "airport" is a location that you can take a flight to, in the context of the given prefix it is clearly an inappropriate suggestion. It is possible that the prefix-suffix pairs based training of the CDSSM can be further improved. We believe that this is an important area for future investigations given that the CDSSM holds certain other advantages over *n*-gram models. For example, the CDSSM has limited storage requirements[2], and because of the *word hashing* technique the CDSSM may be more robust to morphological variations and spelling errors in the input prefix compared to the *n*-gram based models.

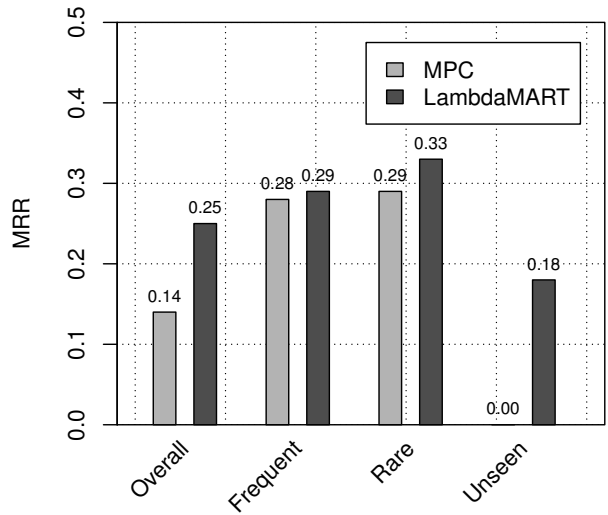

**Figure 7.3:** MRR improvements by historical popularity of the input prefix on the AOL testbed. The LambdaMART model uses *n*-gram and $f_{\text{cdssm}}$ features and includes suffix-based suggestion candidates. Any prefix in the top 100K most popular prefixes from the background data is considered as *Frequent*. There are 7622, 6917 and 14,135 prefix impressions in the *Frequent*, *Rare* and *Unseen* segments, respectively. All reported differences in MRR with the MPC model are statistically significant by the t-test ($p < 0.01$).

Figure 7.3 analyses the improvements by segmenting the prefixes based on their historical popularity. The improvements from the suffix-based candidates are expectedly higher for the rarer prefixes. Interestingly, the absolute MRR values for both models are higher for rare prefixes than for the frequent ones. One factor in this is that rare prefixes tend to be longer and therefore more specific, giving fewer candidates to rank and making it easier to achieve good MRR.

---

[2]The CDSSM model itself needs to be stored in memory but has no data storage requirements, unlike the *n*-gram models.

### 7.1.6 Conclusion

We proposed a novel candidate generation technique for query auto-completion by mining and ranking popular query *suffixes*. Our empirical study shows that this is an effective strategy for significantly improving MRR for rare and unseen prefixes. The supervised ranking framework proposed in this paper is generic and can be employed in any QAC system that combines multiple sources of candidates. We described features based on *n*-gram language models and convolutional neural networks with demonstrable improvements.

While we have shown significant improvements in MRR using synthetic candidate generation, we have not measured how often this approach generates semantically meaningless synthetic suggestions and have not quantified the effect of showing synthetic suggestions to search users. A user study on this aspect is left as future work. There is also further scope for exploring other language models (such as recurrent neural networks) in the context of this task.

## 7.2 Session Context Modelling for Query Auto-Completion

Short-term user history provides useful cues about the user intent that an IR system can consider to improve the relevance of retrieved results [132]. In QAC systems, in particular, when only a few characters have been entered the search engine has little understanding of the actual information need of the user and the generic suggestions provided by a non-contextual QAC system typically perform poorly [429]. The high ambiguity associated with short prefixes makes QAC a particularly interesting candidate for leveraging any additional information available about the user's current task. The same study also showed that 49% of Web searches are preceded by a different search which can be used to gain additional insights into the user's current information need.

The majority of previous work [123, 441] on using short-term user history for search personalization has been focused on modelling the topical relevance of the candidate results (documents or query suggestions) to the previous queries and viewed documents in the same search session. Using such implicit feedback has been shown to be a very attractive strategy for improving retrieval performance when the user intent is ambiguous. For example, knowing that the user's previous query was "guardians of the galaxy" can help to inform a QAC system to promote the query "imdb" in ranking over "instagram" when the user has just typed "i" in the search box. Query reformulation behaviours within search sessions have also been studied but are mostly limited to taxonomy based classifications [442, 443] and models based on syntactic changes [444]. A quick study of a sample of Bing's search engine logs reveal that users frequently search for "san francisco 49ers" and "san francisco weather" immediately after searching for "san francisco". Similarly, the query "detroit" is often followed by the queries "detroit lions" and "detroit weather". Intuitively, "san francisco" → "san francisco 49ers" represents a similar shift in user's intent as "detroit" → "detroit lions". We can see many such frequently occurring patterns of reformulations in large scale search logs. Modelling these reformulations using lexical matching alone is difficult. For example, we understand that "movies" → "new movies" is not the same intent shift as "york" → "new york" even though in both cases the same term was added to both the queries by the user. On the other hand, "london" → "things to do in london" and "new york" → "new york tourist attractions" are semantically similar although the two reformulations involve the addition of completely disjoint sets of new terms to the queries.

In text processing, Mikolov et al. [204] demonstrated that the distributed representation of words learnt by continuous space language models are surprisingly good at capturing syntactic and semantic relationships between the words. Simple algebraic operations on the word vectors have been shown to produce intuitive results. For example, $\vec{v}_{king} - \vec{v}_{man} + \vec{v}_{woman}$ results in a vector that is in close proximity to $\vec{v}_{queen}$. In Section 7.2.2, we show that the embeddings learnt by CDSSM [362] exhibit similar favourable properties and hence provide an intuitive mechanism to represent query reformulations as the offsets between query vectors.

Our empirical study, described in Section 7.2.3, demonstrate that the vector representations of queries and reformulations can be useful for capturing session context for the retrieval of query suggestions. The CDSSM is trained to map queries (and documents) with similar intents to the same neighbourhood in the semantic space. Therefore they are suitable for measuring the topical similarity between candidate suggestions and the user's recent queries. In addition, our experiments show that the vector representation of the reformulation, from the user's previous query to the candidate suggestion, can also be a useful signal for predicting the relevance of the suggestion. We present our results in Section 7.2.5 that demonstrate that session context features based on these vector representations can significantly improve the QAC ranking over the supervised ranking baseline proposed by Shokouhi [439].

The main contributions of the work described in this section are,

- Demonstrating that query reformulations can be represented as low-dimensional vectors which map syntactically and semantically similar query changes close together in the embedding space.

- Using features based on the distributed representations of queries and reformulations to improve upon a supervised ranking baseline for session context-aware QAC ranking. Our experiments on the large-scale query logs of the Bing search engine and the publicly available AOL query logs [85] show that these features can improve MRR by more than 10% on these testbeds.

- Demonstrating that CDSSM trained on session query pairs performs significantly better for the contextual QAC ranking task compared to the CDSSM model trained on clicked query-document pairs.

Next, we review related work that are relevant to this study.

### 7.2.1 Related work

In Web search, Bennett et al. [132] investigated the impact of short-term and long-term user behaviour on relevance prediction, and showed that short-term user history becomes more important as the session progresses. Li et al. [445] evaluated DSSM and CDSSM for modelling session context for Web search. Besides the primary IR task, QAC as opposed to Web ranking, our work differs from this study by going beyond computing the topical similarity using the existing models and explicitly modelling query reformulations as vectors. We also show the benefits of optimizing a CDSSM model directly for capturing session context by training on session query pairs.

Yan et al. [446] proposed an approach that maps queries and clicks to latent search intents represented using Open Directory Project[3] categories for making context-aware query recommendations. Cao et al. [441] and Liao et al. [447] have explored session context using latent concept clusters from click-through bipartite graphs, while Guo et al. [448] represented the user's previous queries using a regularized topic model. Zhang et al. [449] proposed a task-centric click model for characterizing user behaviour within a single search session. Cao et al. [450] learnt a variable length *Hidden Markov Model* from large scale search logs, whereas Boldi et al. [451] studied *random walks* on query-flow graphs for improved recommendations.

Previous studies on the relationships between neighbouring queries from a search session have been mostly focused on categorizing the reformulations based on broad manually defined taxonomies (e.g., generalization, specialization, error correction and parallel move) [452] or understanding the user goals behind common actions (e.g., addition, removal or substitution of terms) [453]. Motivated by the broad manually identified reformulation categories Xiang et al. [131] and Jiang et al. [454] designed simple features for supervised retrieval models. Finally, Guan et al. [444] use reinforcement learning for modifying term weights in response to the observed modifications made to the query by the user.

While clearly using session context for Web search is a well-studied topic, context-sensitive query auto-completion has been discussed less thoroughly in the literature. Weber and Castillo [455] and Shokouhi [439] showed how query distributions change across different user demographics and argued that QAC systems based on personalization features can significantly outperform popularity-based baselines. Ranking suggestions based on temporal context has also been explored

---

[3] http://www.dmoz.org/

[456, 457].

The two QAC related studies most relevant to our work have been done by Shokouhi [439] and Kharitonov et al. [458]. To capture short-term context, Shokouhi [439] relied on letter *n*-gram matches between the previous queries and the candidates, and trained a supervised ranking model for combining them with MPC and other non-contextual and user demographic features. Kharitonov et al. [458] proposed a unified framework for contextualizing and diversifying the ranking of QAC suggestions. Their empirical evaluations show that by considering the user's previous query alone more than 96% of the improvements can be achieved, as compared to additionally considering the document examination history and diversification context. Given the previous query, their proposed model computes the expected probability of a given completion as follows,

$$p(q_1|q_0) = p(c=0|q_0)p(q_1) + p(c=1|q_0)p(q_1|c=1,q_0) \qquad (7.4)$$

Where c is an indicator variable whose value is 1 if the user continues the current task, and 0 otherwise. The two primary components of the above equation are $P(q_1)$ and $P(q_1|c=1,q_0)$, which correspond to the probability of observing the query $q_1$ globally and in the context of the query $q_0$, respectively, in the query logs.

For our evaluation, we implement the supervised ranking framework proposed by Shokouhi and include the *n*-gram similarity, the query frequency and the query pairwise frequency features among others as described in Section 7.2.3.

### 7.2.2  Model

We adopt the CDSSM architecture proposed by Shen et al. [362] for our study. Unless specified otherwise, for all models in this paper the window size for the convolutional layer is set to three and the dimensions of the output vector to 32.

The training data for the CDSSM models consists of source-target text pairs. The original DSSM [164] and CDSSM [362] models were trained on clickthrough data which consists of pairs of queries and document titles, corresponding to clicked

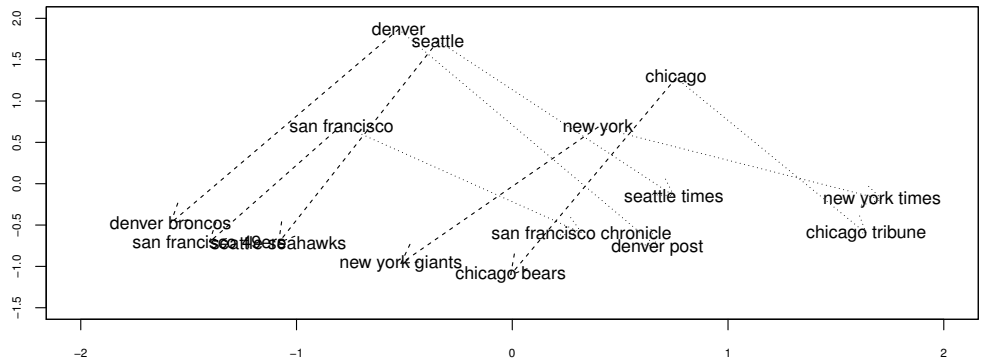

**Figure 7.4:** A two-dimensional PCA projection of the 32 dimensional CDSSM output vectors shows how intuitively similar intent transitions, represented by the directed edges, are automatically modelled in the embedding space. The CDSSM model used for this illustration is trained on the symmetric session pairs dataset.

results.

In addition to clickthrough data, we also train the CDSSM models on sampled pairs of queries from search logs that were observed in succession during user sessions. In the rest of this paper, we refer to this as the *session pairs* dataset. For a pair of observed queries $q_1$ and $q_2$, if the dataset includes both the ordering $q_1 \rightarrow q_2$ and $q_2 \rightarrow q_1$ then we refer to it as the *symmetric* session pairs dataset, otherwise as *asymmetric*. The symmetric session pairs data is further randomly sub-sampled by half to keep the count of the training pairs in both the datasets comparable.

The session pairs datasets are extracted from the exact same user sessions from which the clickthrough data is generated. While this does not imply that the actual count of training pairs in these two types of datasets are equal, it does make the comparison more meaningful as it assumes the same amount of raw log data is examined for training both the types of models. In practice, however, we did observe the data sizes to be comparable across all three datasets during this study.

All the CDSSM models in this study are trained using mini-batch based stochastic gradient descent, as described by Shen et al. [362]. Each mini-batch consists of 1024 training samples (source-target pairs) and for each positive pair 100 negative targets are randomly sampled from the data for that source that were not originally paired.

**Table 7.5:** *k*-means clustering of 65K in-session query pairs observed in search logs. Examples from five of the top ten biggest clusters shown here. The first and the second clusters contain examples where the follow up query is a different formulation of the exact same intent. The third and the fourth clusters contain examples of *narrowing* intent, in particular the fourth cluster contains reformulations where the additional specification is based on location disambiguation. Finally, the last cluster contains examples of intent jumps across tasks.

| | | |
|---:|:---:|:---|
| soundcloud | → | www.soundcloud.com |
| coasthills coop | → | www.coasthills.coop |
| american express | → | www.barclaycardus.com login |
| duke energy bill pay | → | www.duke-energy.com pay my bill |
| cool math games | → | www.coolmath.com |
| majesty shih tzu | → | what is a majesty shih tzu |
| hard drive dock | → | what is a hard drive dock |
| lugia in leaf green | → | where is lugia in leaf green |
| red river log jam | → | what is th red river log jam |
| prowl | → | what does prowl mean |
| rottweiler | → | rottweiler facebook |
| sundry | → | sundry expense |
| elections | → | florida governor race 2014 |
| pleurisy | → | pleurisy shoulder pain |
| elections | → | 2014 rowan county election results |
| cna classes | → | cna classes in lexington tennessee |
| container services inc | → | container services ringgold ga |
| enclosed trailers for sale | → | enclosed trailers for sale north carolina |
| firewood for sale | → | firewood for sale in asheboro nc |
| us senate race in colorado | → | us senate race in georgia |
| siol | → | facebook |
| cowboy bebop | → | facebook |
| mr doob | → | google |
| great west 100 west 29th | → | facebook |
| avatar dragons | → | youtube |

The CDSSM models project the queries to an embedding space with fixed number of dimensions. The semantic similarity between two queries $q_1$ and $q_2$ in this semantic space is defined by,

$$Sim(q_1, q_2) = cosine(\vec{v}_{q_1}, \vec{v}_{q_2}) = \frac{\vec{v}_{q_1}^\intercal \vec{v}_{q_2}}{\|\vec{v}_{q_1}\| \|\vec{v}_{q_2}\|} \quad (7.5)$$

Where $\vec{v}_{q_1}$ and $\vec{v}_{q_2}$ are the CDSSM vector outputs corresponding to the two queries,

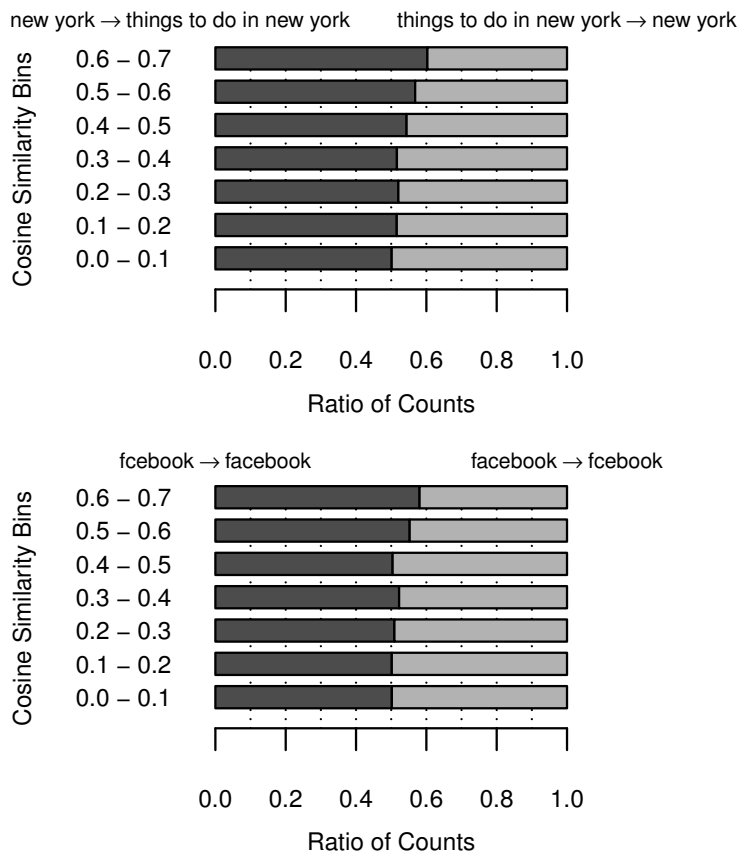

**Figure 7.5:** Visualization of the cosine similarity scores of a given reformulation with respect to a set of 100,000 other reformulations randomly sampled from Bing's logs. The similarity scores are binned and the ratio of the counts are shown above. The counts corresponding to bins with cosine similarity greater than 0.7 were too small, hence excluded.

respectively. A close examination of the CDSSM output vectors reveal that the learnt distributed representations hold useful information about inter-query relationships. Figure 7.4 illustrates how the offset vectors between pairs of queries, represented by the directed edges, are directionally similar in the embedding space for similar intent transitions. This matches the observations made by Mikolov et al. [191] on continuous space language models for text processing, and gives us an intuitively understandable representation of query reformulations as their offset vectors in the embedding space. More specifically, we define the *reformulation* from query $q_1$ to $q_2$ as,

$$\text{ref}(q_1, q_2) = \vec{v}_{q_2} - \vec{v}_{q_1} = \frac{\vec{v}_{q_2}}{\|\vec{v}_{q_2}\|} - \frac{\vec{v}_{q_1}}{\|\vec{v}_{q_1}\|} \quad (7.6)$$

Where $\vec{v}_{q_1}$ and $\vec{v}_{q_2}$ are the CDSSM vector embeddings of the two queries, respectively. This explicit vector representation provides a framework for studying frequently occurring query reformulation patterns. To illustrate this, we randomly sample approximately 65K pairs of queries that were observed in succession in Bing's logs. For each pair, we compute the offset vector using a CDSSM model. We then run a simple *k*-means clustering ($k = 100$) and examine the top clusters. Example reformulations from five of the biggest clusters are shown in Table 7.5.

A further study of these reformulation vectors can reveal important insights about user behaviour, such as the popularity of certain reformulation patterns. For example, we randomly sampled 100,000 adjacent pairs of queries from Bing's logs that were observed in search sessions. Our analysis show that there are more pairs similar to the *narrowing* reformulation "new york" → "things to do in new york" in the sampled set, than its inverse. Similarly, the misspelling "fcebook" followed by "facebook" is a more commonly observed pattern than the other way around, as illustrated in Figure 7.5.

Next, we list qualitative examples in Table 7.6 to demonstrate the predictive aspect of these reformulation vectors. Similar to the analogy based test proposed by Mikolov et al. [191], these examples show that we can obtain intuitively understandable results by performing simple algebraic operations in the embedding space. For example, we compute the vector sum of the projections (normalized to their unit norm) of the queries "new york" and "newspaper".

$$\vec{v}_{target} = \vec{v}_{newyork} + \vec{v}_{newspaper} = \frac{\vec{v}_{newyork}}{\|\vec{v}_{newyork}\|} + \frac{\vec{v}_{newspaper}}{\|\vec{v}_{newspaper}\|} \quad (7.7)$$

Then from a fixed set of candidates we find the query whose embedding has the highest cosine similarity with $\vec{v}_{target}$. For our analysis we picked the top one mil-lion most popular queries from one day of Bing's logs as the candidate set. In this query set, the closest query vector to $\vec{v}_{target}$ corresponds to the query "new york times". Similarly, the nearest neighbour search for $\vec{v}_{\text{how old is gwen stefani}} - \vec{v}_{\text{gwen stefani}} + \vec{v}_{\text{meghan trainor}}$ yields a vector close to $\vec{v}_{\text{how old is meghan trainor}}$. These examples show that the vector representation captures simple syntactic as well as semantic relationships. We intentionally also include some examples where the nearest neighbour search yields unexpected results (e.g., $\vec{v}_{\text{beyonce}} + \vec{v}_{\text{net worth}}$) to highlight that these predictions are often noisy.

**Table 7.6:** Examples of simple syntactic and semantic relationships in the query embedding space. The nearest neighbour search is performed on a candidate set of one million most popular queries from one day of Bing's logs.

| Query vector | Nearest neighbour |
|---|---|
| $\vec{v}_{\text{chicago}} + \vec{v}_{\text{newspaper}}$ | $\vec{v}_{\text{chicago suntimes}}$ |
| $\vec{v}_{\text{new york}} + \vec{v}_{\text{newspaper}}$ | $\vec{v}_{\text{new york times}}$ |
| $\vec{v}_{\text{san francisco}} + \vec{v}_{\text{newspaper}}$ | $\vec{v}_{\text{la times}}$ |
| $\vec{v}_{\text{beyonce}} + \vec{v}_{\text{pictures}}$ | $\vec{v}_{\text{beyonce images}}$ |
| $\vec{v}_{\text{beyonce}} + \vec{v}_{\text{videos}}$ | $\vec{v}_{\text{beyonce videos}}$ |
| $\vec{v}_{\text{beyonce}} + \vec{v}_{\text{net worth}}$ | $\vec{v}_{\text{jaden smith net worth}}$ |
| $\vec{v}_{\text{www.facebook.com}} - \vec{v}_{\text{facebook}} + \vec{v}_{\text{twitter}}$ | $\vec{v}_{\text{www.twitter.com}}$ |
| $\vec{v}_{\text{www.facebook.com}} - \vec{v}_{\text{facebook}} + \vec{v}_{\text{gmail}}$ | $\vec{v}_{\text{www.googlemail.com}}$ |
| $\vec{v}_{\text{www.facebook.com}} - \vec{v}_{\text{facebook}} + \vec{v}_{\text{hotmail}}$ | $\vec{v}_{\text{www.hotmail.xom}}$ |
| $\vec{v}_{\text{how tall is tom cruise}} - \vec{v}_{\text{tom cruise}} + \vec{v}_{\text{tom selleck}}$ | $\vec{v}_{\text{how tall is tom selleck}}$ |
| $\vec{v}_{\text{how old is gwen stefani}} - \vec{v}_{\text{gwen stefani}} + \vec{v}_{\text{meghan trainor}}$ | $\vec{v}_{\text{how old is meghan trainor}}$ |
| $\vec{v}_{\text{how old is gwen stefani}} - \vec{v}_{\text{gwen stefani}} + \vec{v}_{\text{ariana grande}}$ | $\vec{v}_{\text{how old is ariana grande 2014}}$ |
| $\vec{v}_{\text{university of washington}} - \vec{v}_{\text{seattle}} + \vec{v}_{\text{chicago}}$ | $\vec{v}_{\text{chicago state university}}$ |
| $\vec{v}_{\text{university of washington}} - \vec{v}_{\text{seattle}} + \vec{v}_{\text{denver}}$ | $\vec{v}_{\text{university of colorado}}$ |
| $\vec{v}_{\text{university of washington}} - \vec{v}_{\text{seattle}} + \vec{v}_{\text{detroit}}$ | $\vec{v}_{\text{northern illinois university}}$ |

### 7.2.3 Experiments

Our empirical evaluations are based on the learning to rank framework proposed by Shokouhi [439] for personalized query auto-completions. In this setup, we learn a supervised ranking model based on training data generated from implicit user feedback. The output of the CDSSM models, described in the previous section, are used to generate additional features for this supervised ranking model. The baseline ranking model (henceforth referred to simply as the *baseline model*) contains both the non-contextual and the (non-CDSSM based) contextual features. We compare all models using the MRR metric, and the study is repeated on two different testbeds to further confirm the validity of the results.

**Testbeds** We conduct our experiments on a large scale search query dataset sampled from the logs of the Bing search engine. We also reproduce our results using the publicly available AOL query logs [85]. In the rest of this paper we refer to these two datasets as the *Bing testbed* and the *AOL testbed*, respectively.

- *Bing testbed* Bing's logs contain a record of all the queries submitted by its users associated with the corresponding anonymized user IDs, timestamps and any clicked Web results[4] (the URL and the displayed title). We sampled queries from these logs for the duration of the last week of October, 2014 and use this as the background data, for computing the feature values and training the CDSSM models. From the first week of November, we sampled 175,392 queries from two consecutive days for training the supervised ranking models, and from the following two individual days we sampled 79,000 queries for validation and 74,663 queries for testing, respectively.

- *AOL testbed* This dataset contains queries sampled between 1 March, 2006 and 31 May, 2006. For each query, the data includes an anonymized user ID and a timestamp. If a result was clicked then the rank of the clicked item and the domain portion of its URL are also included. In aggregate, the data contains 16,946,938 query submissions and 36,389,567 document clicks by 657,426 users.

    We consider all queries before 1 May, 2006 as the background data. All queries from the next two weeks of data are used for training the supervised ranking models, and the remaining two sets, consisting of one week of data each, is used for validation and testing, respectively.

---

[4]For impressions with multiple clicked results we consider only the last clicked document.

To have a separation of users in training and test datasets, on both the testbeds we use only the users with even user IDs for training and validation, and those with odd numbered user IDs for testing. Also, in all the datasets the queries are lower-cased and the punctuations are removed.

**Learning to rank** To generate the training, the validation and the test sets we sample query impressions from the corresponding portions of the logs. For each query impression, a prefix is generated by splitting the query at a randomly selected position[5]. For each prefix a positive relevance judgment is assigned to the suggestion candidate that matches the final submitted query and all the others are labelled as irrelevant.

The training data collected in the above process consists of labelled prefix-query pairs. With respect to the choice of learning-to-rank algorithms, we chose LambdaMART [440], a boosted tree version of LambdaRank [459], that won the Yahoo! Learning to Rank Challenge (2010) [460] and is considered as one of the state-of-the-art learning algorithms. We train 500 trees across all our experiments with the same set of fixed parameters tuned using standard training and validation on separate sets.

We consider the top 10 million most popular queries in the background data as the pre-computed list of suggestion candidates and filter out all the impressions where the final submitted query is not present in this list. For each impression in the training, the validation and the test sets we retain a maximum of 20 suggestion candidates - the submitted query as the positive candidate and 19 other most frequently observed queries from the background data that starts with the same prefix, as the negative examples. Furthermore, for each impression up to 10 previous queries from the same session are made available for computing the session context features. Similar to other previous work [87, 461] we define the end of a session by a 30 minute window of user inactivity.

For our final evaluation we report the Mean Reciprocal Rank of the submitted query averaged over all sampled impressions on each of the two testbeds.

---

[5]The prefixes in our study are strictly shorter than the original query and limited to no more than 30 characters in length.

**Table 7.7:** Comparison of QAC ranking models trained with CDSSM based features against the MPC model and the supervised baseline ranker model. All the reported MRR improvements are statistically significant by the t-test ($p < 0.01$) over the MPC baseline and the baseline model. Additionally, corresponding to each of the different CDSSM models, the ranking model containing both the *similarity* and the *reformulation* features shows statistically significant ($p < 0.01$) improvements in MRR over the model containing only the *similarity* features on both the testbeds. The three highest MRR improvements per testbed are shown in bold below.

|  | Bing | AOL | |
|---|---|---|---|
| Models | % Improv. | MRR | % Improv. |
| **Baselines** | | | |
| MostPopularCompletion | - | 0.5110 | - |
| Baseline Model | +48.6 | 0.7983 | +56.2 |
| **CDSSM (query-document pairs)** | | | |
| All features | +55.9 | - | - |
| Reformulation features | +54.3 | - | - |
| Similarity features | +55.3 | - | - |
| **CDSSM (Asymmetric session query pairs)** | | | |
| All features | **+58.0** | 0.8775 | **+71.7** |
| Reformulation features | **+57.4** | 0.8747 | **+71.2** |
| Similarity features | +54.2 | 0.8580 | +67.9 |
| **CDSSM (Symmetric session query pairs)** | | | |
| All features | **+59.0** | 0.8801 | **+72.2** |
| Reformulation features | +57.2 | 0.8744 | +71.1 |
| Similarity features | +55.8 | 0.8636 | +69.0 |

### 7.2.4 Features

The baseline contextual and non-contextual features, as well as the features based on the CDSSM outputs are described in this section.

- *Non-contextual features* The *MostPopularCompletion* (MPC) model is one of the baselines for our study. We also use the output of this model as a feature for the supervised ranking model. Other non-contextual features include the prefix length (in characters), the suggestion length (in both characters and words), the vowels to alphabets ratio in the suggestion, and a boolean feature indicating whether the suggestion contains numeric characters.

- *N-gram similarity features* We compute the character *n*-gram similarity (*n*=3) between the suggestion candidate and the previous queries from the same user session. This is an implementation of the *short history* features described by Shokouhi [439]. A maximum of 10 previous queries are considered.

- *Pairwise frequency feature* From the background data, we generate the top 10 million most popular adjacent pairs of queries observed in search sessions. For a given impression, the previous query and the suggestion candidate pair is matched against this dataset and the corresponding frequency count is used as the feature value. If no matches are found, then the feature value is set to zero.

- *CDSSM topical similarity features* The CDSSM models are trained as described in Section 7.2.2 using the background portion of the data on each testbed. The cosine similarity between the CDSSM vectors corresponding to the suggestion candidate and a maximum of previous 10 queries from the same session are computed and used as 10 distinct features in the QAC ranking model.

  Training on the session query pairs data produces a pair of *pre-post* CDSSM models. When trained on the asymmetric data, the *pre-* model is used for projecting the user's previous queries and the *post-* model is used for projecting the suggestion candidates for the cosine similarity computation. For the symmetric data however, both the *pre-* and the *post-* models are equivalent, and hence we use only the *pre-* model in our experiments.

  The AOL logs contains only the domain portion of the clicked results. Hence we are unable to get the corresponding document titles. Therefore we only train the session pairs based CDSSM models on this testbed and report those results in this paper.

- *CDSSM reformulation features* We compute the $n$-dimensional ($n$=32) vector representation of the reformulation from the previous query to the suggestion candidate. The raw values from this vector are used as $n$ distinct features into the supervised ranking model. For both the session pair based models, the *pre-* model is used for projecting the suggestion candidates, as well as the previous query.

## 7.2.5 Results

Table 7.7 compares the results of training the supervised QAC ranking model with the different CDSSM based session context features. Due to the proprietary nature of Bing's data, we report only relative improvements of each of the models over the MPC baseline for this testbed. On the AOL testbed, however, we report both the absolute MRR values and the relative improvements for all the models.

On both the testbeds, the baseline model which also contains session context features (the *n*-gram similarity and the pairwise frequency) shows a large improvement over the MPC baseline, which is expected. All the models trained with the CDSSM based contextual features show further statistically significant improvements over the baseline model. Both the CDSSM models trained on session pairs perform better than the models trained on clickthrough data, with the model trained on the symmetric session pairs performing slightly better overall. Table 7.9 lists examples of cases from one of the test sets where the ranking model with the CDSSM based contextual features perform better compared to both the baselines.

The supervised ranking models trained with both the CDSSM based similarity features and the CDSSM based reformulation features perform better than the corresponding models trained with the similarity features alone. The improvements are statistically significant and demonstrate the additional information provided by the reformulation features to the ranking model over the CDSSM based similarity features. The reformulation features perform particularly superior when the CDSSM model has been trained on the session pairs dataset.

Table 7.8 shows the impact of considering different number of previous queries in the session for computing the CDSSM based similarity features. The results indicate that considering the previous query alone achieves most of the improvements observed from these similarity features.

We also compare the improvements from the different models based on the length of the input prefixes. Bar-Yossef and Kraus [429] have previously reported that non-contextual QAC systems generally perform poorly when the user has typed only a few characters due to the obvious ambiguity in user intent. Figure 7.6 illus-trates this behaviour on the AOL testbed. Both the supervised ranking models, the baseline and the model with the CDSSM features, show significantly large improvements over the MPC baseline on short prefixes. After the user has typed a few more characters in the search box, the set of suggestion candidates reduce significantly and the performance of the MPC model improves. Therefore the improvements on the longer prefixes are smaller for both the supervised ranking models. The supervised ranking model with the CDSSM features, however, show statistically significant better MRR compared to both the MPC baseline and the supervised baseline ranking model on all the prefix length based segments. Finally, Figure 7.7 shows that better MRR can be achieved by training the CDSSM model with a higher number of output dimensions.

**Table 7.8:** Comparison of QAC ranking models with CDSSM similarity features computed considering different maximum number of previous queries in the same session. The results show that most of the improvements from short-term history similarity features can be achieved by considering just the immediately previous query.

| Models | Bing % Improv. | AOL MRR | AOL % Improv. |
|---|---|---|---|
| **Baselines** | | | |
| MostPopularCompletion | - | 0.5110 | - |
| Baseline Model | +48.6 | 0.7983 | +56.2 |
| **CDSSM (Symmetric session query pairs)** | | | |
| Previous 1 query | +55.2 | 0.8631 | +68.9 |
| Previous 3 queries | +56.1 | 0.8639 | +69.1 |
| Previous 5 queries | +56.1 | 0.8642 | +69.1 |
| Previous 10 queries | +55.8 | 0.8636 | +69.0 |

### 7.2.6 Discussion

We demonstrated significant improvements in the query auto-completion ranking task using the CDSSM based session context features. We now discuss potential implications of these vector representations on session modelling and list some of the assumptions and limitations of the evaluation framework used in this study.

**Implications for session modelling** The distributed representation of queries and query reformulations provides an interesting framework for thinking about sessions and task context. The sequence of queries (and documents) in a search session can be considered as a directed path in the embedding space. What are the common attributes shared by these session paths? What properties of these paths vary depending on the type of the user task or information need? These are examples of research questions that may be interesting to study under the distributed representation framework. Hassan et al. [462], for example, studied long search sessions and compared user behaviours when the user is struggling in their information task to when they are exploring. Features based on the CDSSM projections of queries and documents, such as the types of user reformulations in the session and the similarity between submitted queries and viewed documents, can be explored to improve the prediction accuracy for such session classification tasks.

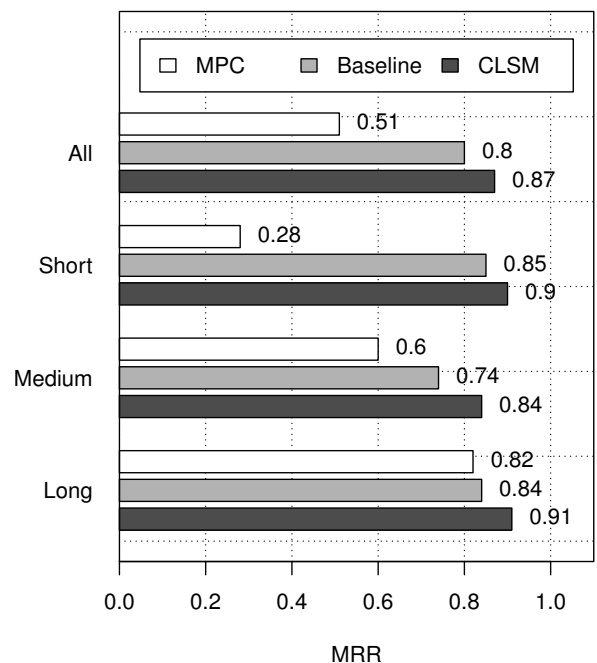

**Figure 7.6:** Comparison of the MPC model, the baseline ranker model and the experimental ranker model with the CDSSM based features (the CDSSM model considered here is trained on symmetric session pairs with all features) across different prefix lengths on the AOL testbed. Prefixes less than 4 characters are considered as short, 4 to 10 characters as medium, and greater than 10 characters as long. Both the supervised ranking models contain contextual features (CDSSM based or otherwise) and hence show large improvements on the short prefixes where the ambiguity is maximum. Across all prefix lengths the model with CDSSM based features out-perform the baseline ranking model. All reported differences in MRR are statistically significant by the t-test ($p < 0.01$).

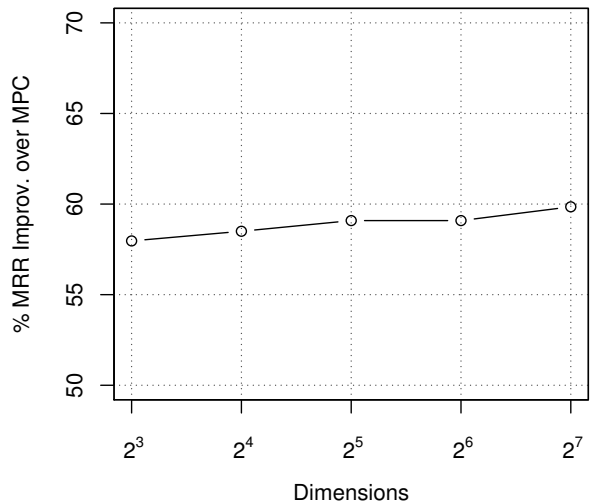

**Figure 7.7:** Evaluation of the impact of training the CDSSM models with different number of dimensions. Except for the pair of CDSSM models trained with 32 and 64 dimensions, all other reported differences in MRR are statistically significant by the t-test ($p < 0.01$).

In this paper we have examined individual query reformulations. Studying reformulation chains may teach us further about how user intents evolve during a session and support the design of future models for session search. For example, White and Huang [463] have explored the value of search trails, over the origins and the destinations. While we have only examined the representation of queries and reformulations in this paper, CDSSM also allows for documents to be represented in the same embedding space. A unified study of queries, reformulations and viewed (searched or browsed) documents using the vector representation framework is an area for future work.

In the query change retrieval model (QCM) proposed by Guan et al. [444], we can explore using the reformulation vectors for representing the user agent's actions. Similarly, we may be able to gain further insights by conducting a similar study as Hollink et al. [453] by examining query changes under the vector representation framework.

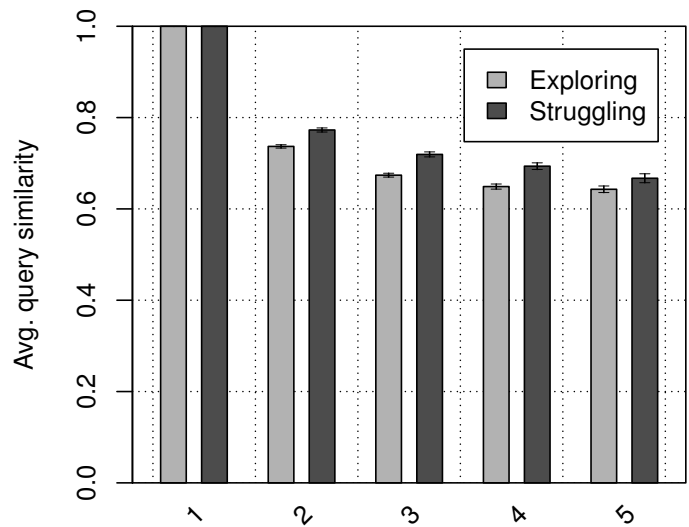

**Figure 7.8:** Average similarity between the first five queries to the first query in search sessions annotated by crowdsourcing judges as exploring or struggling. The similarity was computed using the distributed representation learnt by the CDSSM model trained on the symmetric session query pairs data. All differences are statistically significant at the $p < 0.05$ level according to a two-tailed t-test.

Generating a distributed representation of users based on their search and other online activities is also an interesting problem. Other potential directions for future studies using the vector framework includes examining how query reformulations differ based on the search expertise of the user and the kind of device the search is performed on.

**Assumptions and limitations** We have based our empirical study on the supervised ranking framework proposed by Shokouhi [439]. In doing so, we inherit some of the assumptions in the designs of that framework. Firstly, we assume that the user has a pre-determined query in mind for input and would be satisfied if it appears in the QAC suggestions list. However Hofmann et al. [73] have shown that due to the high examination bias towards top-ranked results, sub-optimal QAC ranking can negatively affect the quality of the query submitted by the user. As many Web search engines implement some form of an auto-completion feature, it is likely that those QAC systems influenced the actual query observed in the logs. We ignore this effect in the generation of our training and test sets.

**Table 7.9:** Examples from the win-loss analysis on one of the test sets. For a given prefix and the previous query from the same user session, the top ranked suggestion by the different models are shown below. The actual submitted query is denoted by the checkmark (✓). The CDSSM features include both the similarity and the reformulation features and the CDSSM model is trained on the symmetric session pairs dataset.

| | | |
|---|---|---|
| Previous query | the fighter | airline tickets |
| Prefix | amer | amer |
| MPC | american express | american express |
| Supervised baseline | american express | american express |
| Supervised \w CDSSM Features | american psycho movie ✓ | |
| | | |
| Previous query | usairways | 2007 toyota yaris |
| Prefix | us | us |
| MPC | us elections 2014 predictions | us elections 2014 predictions |
| Supervised baseline | usps.com | usaa |
| Supervised \w CDSSM Features | usairways.com ✓ | used cars ✓ |

The generation of the prefixes also assumes that each query was typed completely by the user in a strictly left-to-right progression and the user is equally likely to examine and engage with the QAC system after each character is typed. In practice, however, users are often aided in the query formulation process (partially or completely) by various features of the search engine, such as QAC or related query recommendations. Users also often correct already entered text during the query formulation process. In these cases the generation of all possible prefixes from the submitted query does not accurately reflect the actual prefixes typed by the user.

Li et al. [464] and Mitra et al. [72] have also shown that user engagement with QAC varies with different factors such as whether the user is at a word boundary or the distance of the next character to be typed on the keyboard. This suggests that prefixes should be sampled with different importance depending on the likelihood that the user would examine the QAC suggestions for that prefix. Li et al. [464] proposed a two-dimensional click model for QAC, demonstrating that in the presence of keystroke level logging of QAC sessions the click model can be used to filter out prefix impressions with low expected probability of examination. However, as the testbeds we consider for this study do not all have the keystroke level granularity of records, we do not pursue this line of experimentation.

Lastly, Shokouhi [439] generates all the possible prefixes of each query in the log data. This results in an obvious over-representation of long prefixes in the generated datasets. To avoid this issue we extract a single prefix per query by splitting at a random position within the query.

Despite the different underlying assumptions, the framework proposed by Shokouhi [439] provides a reasonable setup to learn a baseline context-aware ranking model for QAC, and hence we adopt it for this study.

### 7.2.7 Conclusion

We have demonstrated that the distributed representation of queries by the CDSSM holds useful information about inter-query relationships. The reformulation vectors exhibit regularities that makes them interesting for modelling session context for query suggestion tasks. Our experiments show that using features based on the reformulation vectors improves MRR for QAC ranking over using features based on the query vectors alone. The best improvements, however, are achieved by the combination of features based on both these vector representations. We have also demonstrated that training the latent semantic models on session query pairs produces further improvements over the model trained on query-document pairs. While the biggest improvements are observed on short prefixes, the ranking model containing the CDSSM based features perform better than the supervised ranking baseline on all the prefix length based segments. We have also studied the effects of considering different number of previous queries within the session for context and the number of dimensions used to represent the query and reformulation vectors on the model performance. While we evaluate these models on the query auto-completion ranking task, the features we described in this paper may also be useful for generating context sensitive related query recommendations and query rewriting. Furthermore, by projecting documents to this same embedding space, future studies may be able to extend these contextual features to document ranking in Web search.

Lastly, the reformulation vectors provide an interesting framework for studying sessions and intent progressions. We anticipate that these distributed representations of queries, documents and reformulations will become more frequently used as tools for future studies on search personalization and session search.

# Chapter 8

# Benchmarking for neural IR

Neural IR is an emerging field. In recognition of the significant impact of deep learning on other application areas, we organized a workshop titled Neu-IR [373, 465] (pronounced "*new IR*") at SIGIR 2016. The purpose was to provide a forum for new and early work relating to deep learning and other neural approaches to IR, and discuss the main challenges facing this line of research. Since then, research publication in the area has been increasing (see Figure 8.1 and [466]), along with relevant workshops [467–469], tutorials [3–6, 470], and plenary talks [471, 472].

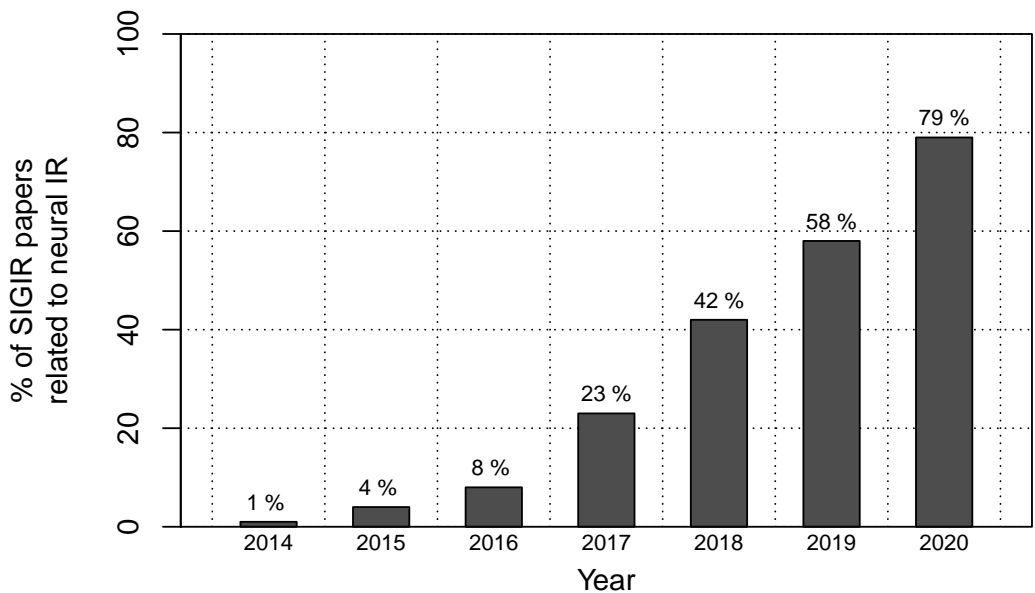

**Figure 8.1:** The percentage of neural IR papers at the ACM SIGIR conference—as determined by a manual inspection of the papers—shows a clear trend in the growing popularity of the field.

While there has been significant interest in deep learning for ad-hoc ranking [1], the work till recently has largely been done with small data, proprietary data or synthetic data. With small data, there has been some discussion about whether deep learning methods really outperform strong traditional IR baselines [473]. Using a proprietary set of document ranking data with 200,000 training queries we beat a traditional IR baseline in 2017, as reported in Chapter 4, but it was impossible for others to follow up on the work without a data release. Dietz et al. [414] have a TREC task with enough training data to investigate such findings, but on synthetic rather than human-labeled data.

Since significant questions remain about baselines and the required volume of human-labeled data, we initiated an effort to benchmark IR models in the presence of large scale training data at TREC 2019. TREC provides a good forum to study such issues. The IR community can submit strong baselines at TREC and there is a blind one-shot evaluation to avoid overfitting. We present our findings from the TREC 2019 Deep Learning track [15] in this Chapter.

## 8.1 TREC Deep Learning track

The TREC 2019 Deep Learning Track has two tasks: Document retrieval and passage retrieval. Each task has a dataset that is new to TREC, although the passage task is similar to the MS MARCO passage ranking leaderboard [52], but with a new test set in the TREC version with more comprehensive labeling. Both tasks are ad-hoc retrieval, meaning that there is a fixed document set, and the goal of the information retrieval system is to respond to each new query with results that would satisfy the querying user's information need. Ad-hoc retrieval is a very common scenario in real-world search applications and in TREC.

The main goals of the track are: (i) To provide large reusable datasets for training and evaluation of deep learning and traditional ranking methods in a large training data regime, (ii) To perform a rigorous blind single-shot evaluation, where test labels don't even exist until after all runs are submitted, to compare different ranking methods, and (iii) To study this in both a traditional TREC setup with end–to-end retrieval and in a re-ranking setup that matches how some models may be deployed in practice.

The track has two tasks: Document retrieval and passage retrieval. Participants were allowed to submit up to three runs per task, although this was not strictly enforced. Participants were provided with an initial set of 200 test queries, then NIST later selected 43 queries during the pooling and judging process, based on budget constraints and with the goal of producing a reusable test collection. The same 200 queries were used for submissions in both tasks, while the selected 43 queries for each task were overlapping but not identical.

When submitting each run, participants also indicated what external data, pre-trained models and other resources were used, as well as information on what style of model was used. Below we provide more detailed information about the document retrieval and passage retrieval tasks, as well as the datasets provided as part of these tasks.

**Document retrieval task** The first task focuses on document retrieval—with two subtasks: (i) Full retrieval and (ii) top-100 reranking.

In the full retrieval subtask, the runs are expected to rank documents based on their relevance to the query, where documents can be retrieved from the full document collection provided. This subtask models the end-to-end retrieval scenario. Note, although most full retrieval runs had 1000 results per query, the reranking runs had 100, so to make the MAP and MRR results more comparable across subtasks we truncated full retrieval runs by taking the top-100 results per query by score.

In the reranking subtask, participants were provided with an initial ranking of 100 documents, giving all participants the same starting point. The 100 were retrieved using Indri [156] on the full corpus with Krovetz stemming and stopwords eliminated. Participants were expected to rerank the candidates *w.r.t.* their estimated relevance to the query. This is a common scenario in many real-world retrieval systems that employ a telescoping architecture [111, 112]. The reranking subtask allows participants to focus on learning an effective relevance estimator, without the need for implementing an end-to-end retrieval system. It also makes the reranking runs more comparable, because they all rerank the same set of 100 candidates.

For judging, NIST's pooling was across both subtasks, and they also identified additional documents for judging via classifier. Further, for queries with many relevant documents, additional documents were judged. These steps were carried out to identify a sufficiently comprehensive set of relevant results, to allow reliable future dataset reuse. Judgments were on a four-point scale:

- [3] **Perfectly relevant:** Document is dedicated to the query, it is worthy of being a top result in a search engine.

- [2] **Highly relevant:** The content of this document provides substantial information on the query.

- [1] **Relevant:** Document provides some information relevant to the query, which may be minimal.

- [0] **Irrelevant:** Document does not provide any useful information about the query.

**Passage retrieval task** Similar to the document retrieval task, the passage retrieval task includes (i) a full retrieval and (ii) a top-1000 reranking tasks.

In the full retrieval subtask, given a query, the participants were expected to retrieve a ranked list of passages from the full collection based on their estimated likelihood of containing an answer to the question. Participants could submit up to 1000 passages per query for this end-to-end retrieval task.

In the top-1000 reranking subtask, 1000 passages per query query were provided to participants, giving all participants the same starting point. The sets of 1000 were generated based on BM25 retrieval with no stemming as applied to the full collection. Participants were expected to rerank the 1000 passages based on their estimated likelihood of containing an answer to the query. In this subtask, we can compare different reranking methods based on the same initial set of 1000 candidates, with the same rationale as described for the document reranking subtask.

For judging, NIST's pooling was across both subtasks, and they also identified additional passages for judging via classifier. Further, for queries with many rel-evant passages, additional passages were judged. These steps were carried out to identify a sufficiently comprehensive set of relevant results, to allow reliable future dataset reuse. Judgments were on a four-point scale:

[3] **Perfectly relevant:** The passage is dedicated to the query and contains the exact answer.

[2] **Highly relevant:** The passage has some answer for the query, but the answer may be a bit unclear, or hidden amongst extraneous information.

[1] **Related:** The passage seems related to the query but does not answer it.

[0] **Irrelevant:** The passage has nothing to do with the query.

## 8.2 Datasets

Both tasks have large training sets based on human relevance assessments, derived from MS MARCO. These are sparse, with no negative labels and often only one positive label per query, analogous to some real-world training data such as click logs.

In the case of passage retrieval, the positive label indicates that the passage contains an answer to a query. In the case of document retrieval, we transferred the passage-level label to the corresponding source document that contained the passage. We do this under the assumption that a document with a relevant passage is a relevant document, although we note that our document snapshot was generated at a different time from the passage dataset, so there can be some mismatch. Despite this, in the document retrieval task machine learning models seem to benefit from using the labels, when evaluated using NIST's non-sparse, non-transferred labels. This suggests the transferred document labels are meaningful for our TREC task.

The passage corpus is the same as in MS MARCO passage retrieval leaderboard. The document corpus is newly released for use in TREC. Each document has three fields: (i) URL, (ii) title, and (iii) body text.

Table 8.1 provides descriptive statistics for the datasets. More details about the datasets—including directions for download—is available on the TREC 2019 Deep

**Table 8.1:** Summary of statistics on TREC 2019 Deep Learning Track datasets.

|  | **Document retrieval** | **Passage retrieval** |
| --- | --- | --- |
| **File description** | # of records | # of records |
| Collection | 3,213,835 | 8,841,823 |
| Train queries | 367,013 | 502,939 |
| Train qrels | 384,597 | 532,761 |
| Validation queries | 5,193 | 6,980 |
| Validation qrels | 5,478 | 7,437 |
| Test queries | 200 → 43 | 200 → 43 |

**Table 8.2:** Summary of statistics of runs for the two retrieval tasks at the TREC 2019 Deep Learning Track.

|  | **Document retrieval** | **Passage retrieval** |
| --- | --- | --- |
| Number of groups | 10 | 11 |
| Number of total runs | 38 | 37 |
| Number of runs w/ category: nnlm | 15 | 18 |
| Number of runs w/ category: nn | 12 | 8 |
| Number of runs w/ category: trad | 11 | 11 |
| Number of runs w/ category: rerank | 10 | 11 |
| Number of runs w/ category: fullrank | 28 | 26 |

Learning Track website[1]. Interested readers are also encouraged to refer to [52] for details on the original MS MARCO dataset.

## 8.3 Results and analysis

**Submitted runs** A total of 15 groups participated in the TREC 2019 Deep Learning Track, with an aggregate of 75 runs submitted across both tasks.

Based run submission surveys, we classify each run into one of three categories:

- **nnlm:** if the run employs large scale pre-trained neural language models, such as BERT [327] or XLNet [474]

- **nn:** if the run employs some form of neural network based approach—*e.g.*, Duet or using word embeddings [394]—but does not fall into the "nnlm" category

---

[1]https://microsoft.github.io/TREC-2019-Deep-Learning/

- **trad:** if the run exclusively uses traditional IR methods like BM25 [80] and RM3 [160].

We placed 33 (44%) runs in the "nnlm" category (32 using BERT and one using XLNet), 20 (27%) in the "nn" category, and the remaining 22 (29%) in the "trad" category.

We further categorize runs based on subtask:

- **rerank:** if the run reranks the provided top-*k* candidates, or

- **fullrank:** if the run employs their own phase 1 retrieval system.

We find that only 21 (28%) submissions fall under the "rerank" category—while the remaining 54 (72%) are "fullrank". Table 8.2 breaks down the submissions by category and task.

We also encouraged some participants to run strong traditional IR baselines, and submit them as additional runs under the "BASELINE" group.

**Overall results** Our main metric in both tasks is Normalized Discounted Cumulative Gain (NDCG)—specifically, NDCG@10, since it makes use of our 4-level judgments and focuses on the first results that users will see. To analyse if any of the fullrank runs recall more relevant candidates in phase 1 compared to those provided for the reranking subtask, we also report Normalized Cumulative Gain (NCG) at rank 100 and 1000 for the document and passage ranking tasks, respectively. We choose to report NCG because it discriminates between recalling documents with different positive relevance grades and is a natural complement to NDCG, our main metric. Although NCG is not officially supported by trec_eval, we confirm that it correlates strongly with the recall metric for these analysed runs.

**Deep learning *vs.* traditional ranking methods** An important goal of this track is to compare the performance of different types of model, using large human-labeled training sets, for the core IR task of ad-hoc search. Indeed this is the first time a TREC-style blind evaluation has been carried out to compare state-of-the-art neural and traditional IR methods.

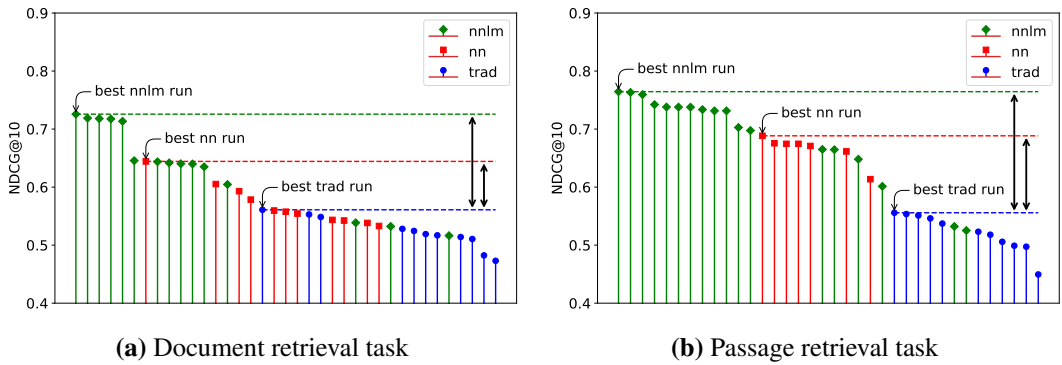

**Figure 8.2:** NDCG@10 results, broken down by run type. Runs of type "nnlm", meaning they use language models such as BERT, performed best on both tasks. Other neural network models "nn" and non-neural models "trad" had relatively lower performance. More iterations of evaluation and analysis would be needed to determine if this is a general result, but it is a strong start for the argument that deep learning methods may take over from traditional methods in IR applications.

Figure 8.2a plots the NDCG@10 performance of the different runs for the document retrieval task, broken down by model type. In general, runs in the category "nnlm" outperform the "nn" runs, which outperform the "trad" runs. The best performing run of each category is indicated, with the best "nnlm" and "nn" models outperforming the best "trad" model by 29.4% and 14.8% respectively.

The passage retrieval task reveals similar pattern. In Figure 8.2b, the gap between the best "nnlm" and "nn" runs and the best "trad" run is larger, at 37.4% and 23.7% respectively. One explanation for this could be that vocabulary mismatch between queries and relevant results is more likely in short text, so neural methods that can overcome such mismatch have a relatively greater advantage in passage retrieval. Another explanation could be that there is already a public leaderboard, albeit without test labels from NIST, for the passage task. Some TREC participants may have submitted neural models multiple times to the public leaderboard, and are well practiced for the passage ranking task.

In query-level win-loss analysis for the document retrieval task (Figure 8.3) the best "nnlm" model outperforms the best "trad" run on 36 out of 43 test queries (*i.e.*, 83.7%). Passage retrieval shows a similar pattern in Figure 8.4. Neither task has a large class of queries where the "nnlm" model performs worse. However, more iterations of rigorous blind evaluation with strong "trad" baselines, plus more scrutiny of the benchmarking methods, would be required to convince us that this is true in general.

Next, we analyze the runs by representing each run as a vector of 43 NDCG@10 scores. In this vector space, two runs are similar if their NDCG vectors are similar, meaning they performed well and badly on the same queries. Using t-SNE [475] we then plot the runs in two dimensions, which gives us a visualization where similar runs will be closer together and dissimilar results further apart. This method of visualizing inter-model similarity was first proposed by Mitra et al. [7] and we employ it to generate the plots in Figure 8.5.

On both document and passage retrieval tasks, the runs appear to be first clustered by group—see Figures 8.5b and 8.5d. This is expected, as different runs from the same group are likely to employ variations of the same approach. In Figures 8.5a and 8.5c, runs also cluster together based on their categorization as "nnlm", "nn", and "trad".

**End-to-end retrieval *vs.* reranking.** Our datasets include top-$k$ candidate result lists, with 100 candidates per query for document retrieval and 1000 candidates per query for passage retrieval. Runs that simply rerank the provided candidates are "rerank" runs, whereas runs that perform end-to-end retrieval against the corpus, with millions of potential results, are "fullrank" runs. We would expect that a "fullrank" run should be able to find a greater number of relevant candidates than we provided, achieving higher NCG@$k$. A multi-stage "fullrank" run should also be able to optimize the stages jointly, such that early stages produce candidates that later stages are good at handling.

According to Figure 8.6, "fullrank" did not achieve much better NDCG@10 performance than "rerank" runs. While it was possible for "fullrank" to achieve better NCG@$k$, it was also possible to make NCG@$k$ worse, and achieving significantly higher NCG@$k$ does not seem necessary to achieve good NDCG@10.

Specifically, for the document retrieval task, the best "fullrank" run achieves only 0.9% higher NDCG@10 over the best "rerank' run. For the passage retrieval

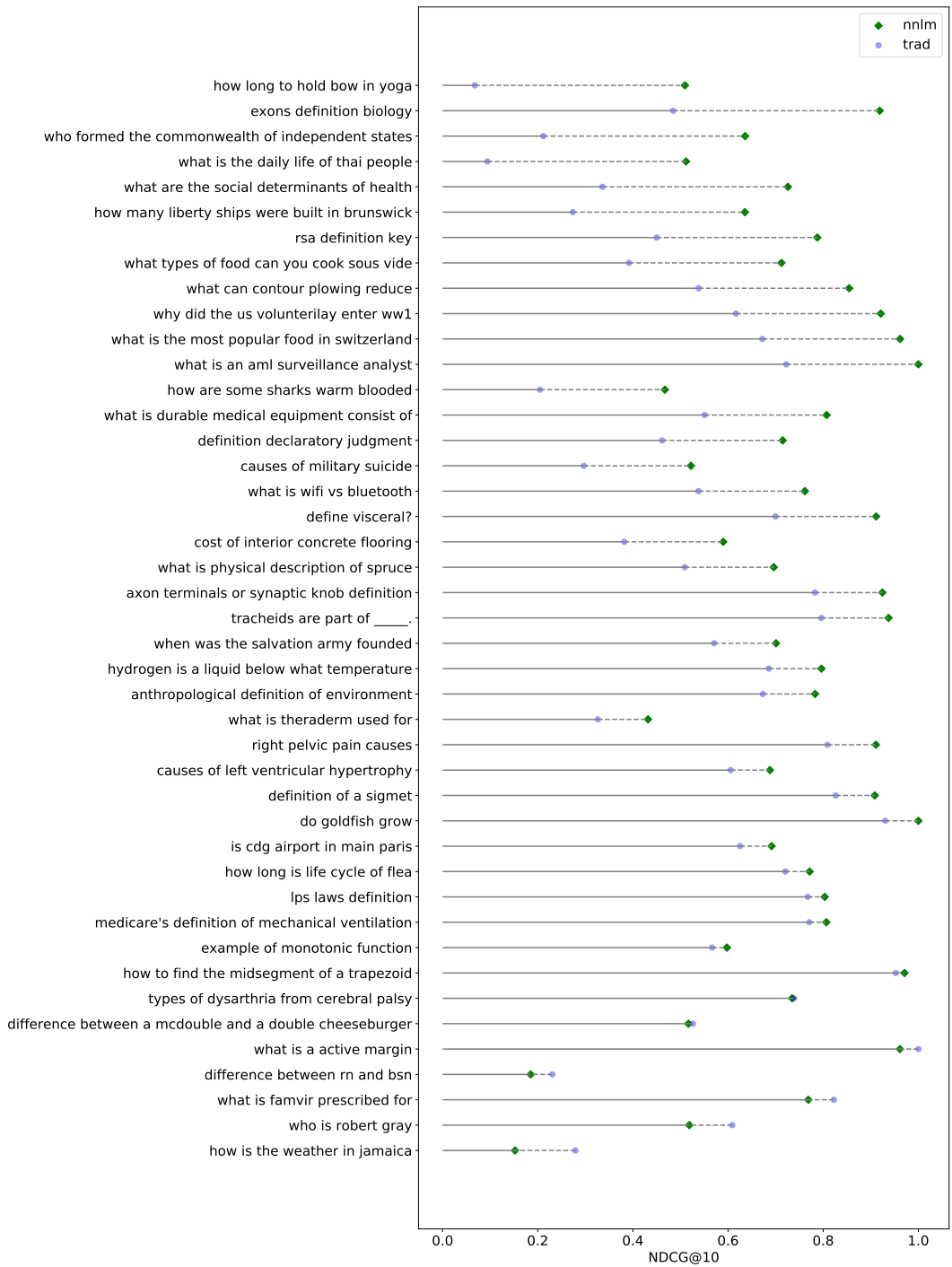

**Figure 8.3:** Comparison of the best "nnlm" and "trad" runs on individual test queries for the document retrieval task. Queries are sorted by difference in mean performance between "nnlm" and "trad" runs. Queries on which "nnlm" wins with large margin are at the top.

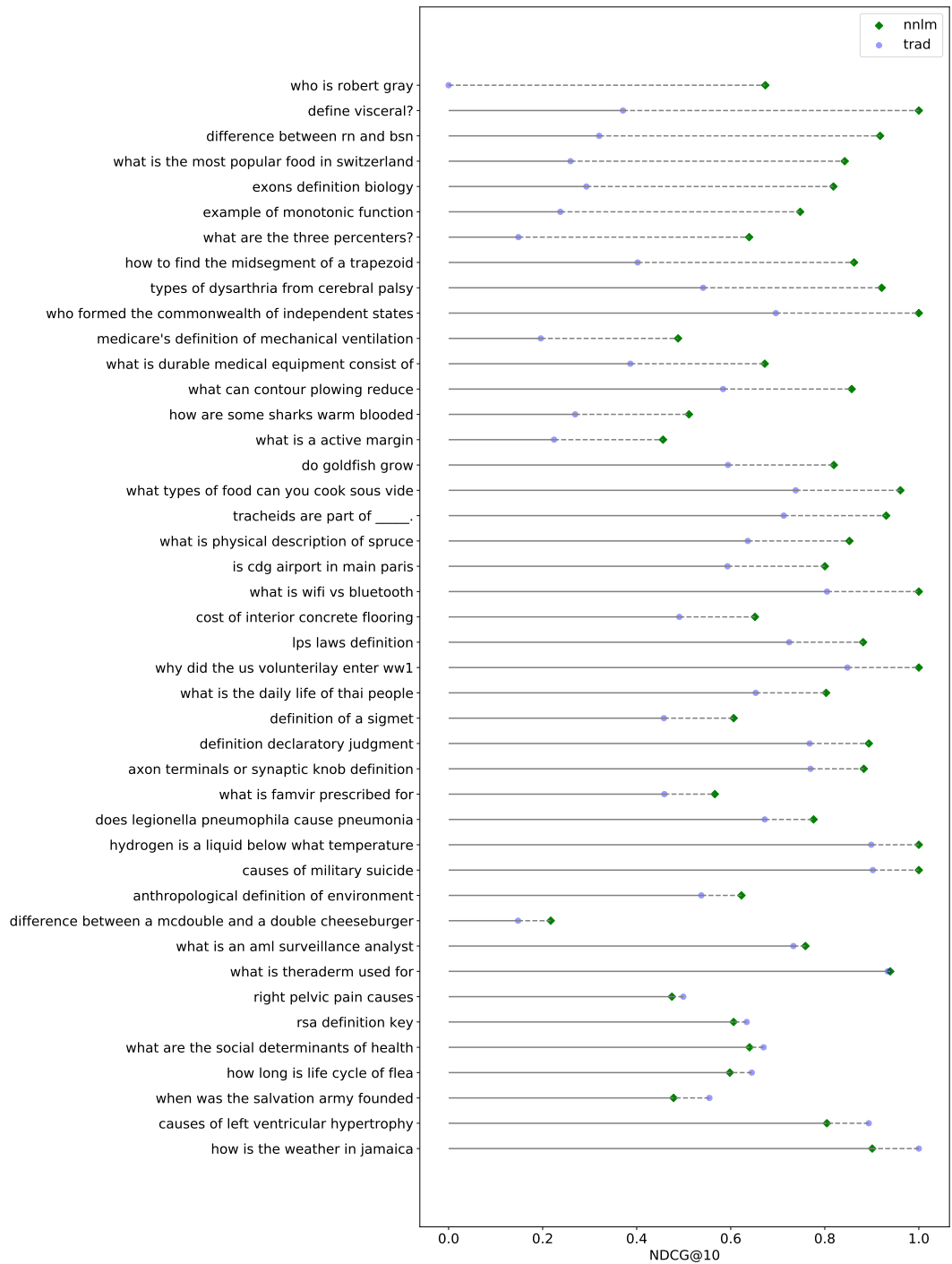

**Figure 8.4:** Comparison of the best "nnlm" and "trad" runs on individual test queries for the passage retrieval task. Queries are sorted by difference in mean performance between "nnlm" and "trad" runs. Queries on which "nnlm" wins with large margin are at the top.

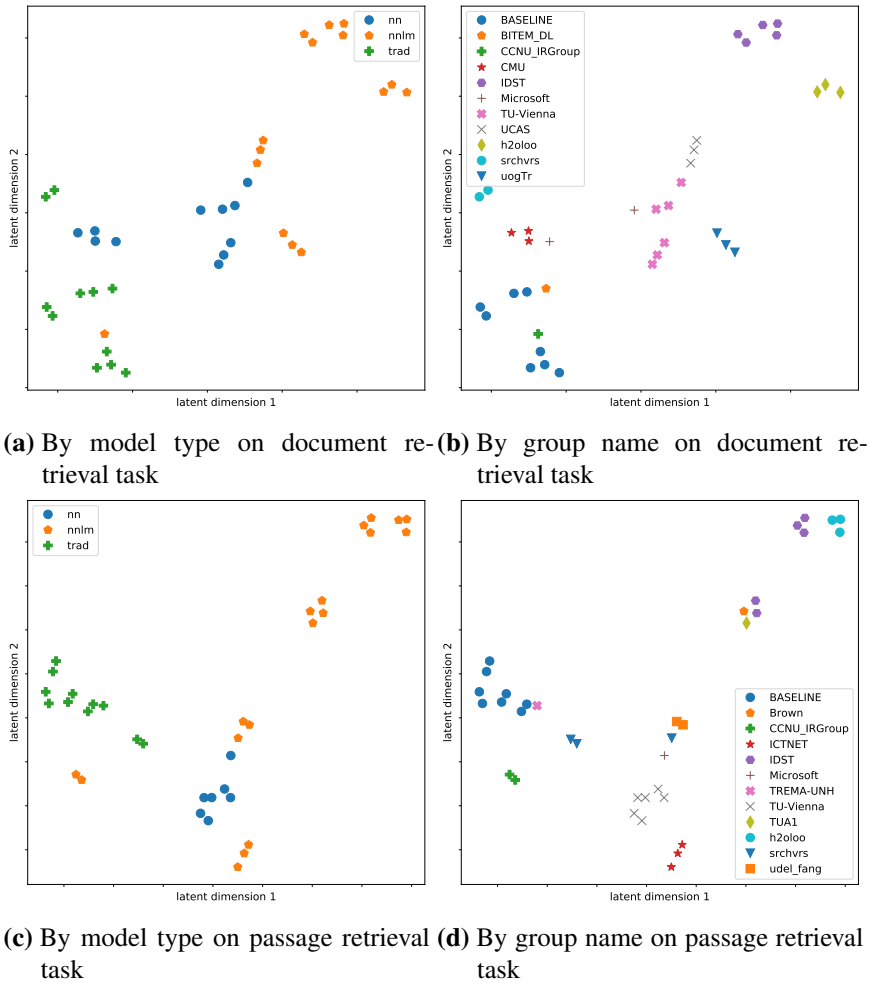

(a) By model type on document retrieval task

(b) By group name on document retrieval task

(c) By model type on passage retrieval task

(d) By group name on passage retrieval task

**Figure 8.5:** Visualizing inter-run similarity using t-SNE. Each run is represented by a 43-dimensional vector of NDCG@10 performance on corresponding 43 test queries. The 43-dimensional vector is then reduced to two-dimensions and plotted using t-SNE. Runs that are submitted by the same group generally cluster together. Similarly, "nnlm", "nn", and "trad" runs also demonstrate similarities.

task, the difference is 3.6%.

The best NCG@100 for the document retrieval task is achieved by a well-tuned combination of BM25 [80] and RM3 [160] on top of document expansion using doc2query [476]—which improves by 22.9% on the metric relative to the set of 100 candidates provided for the reranking task. For the passage retrieval task, the best NCG@1000 is 20.7% higher than that of the provided reranking candidate set.

Given this was the first ever Deep Learning Track at TREC, we are not yet seeing a strong advantage of "fullrank" over "rerank". However, we hope that as the body of literature on neural methods for phase 1 retrieval (*e.g.*, [13, 389, 405, 476])

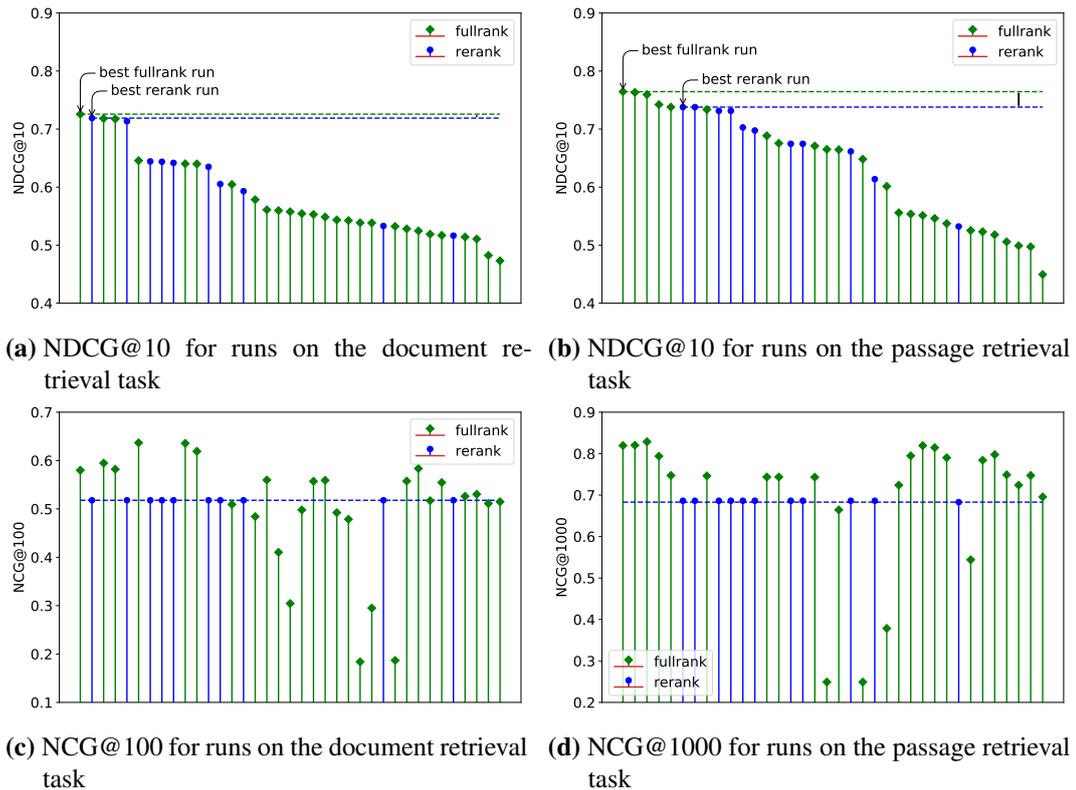

**(a)** NDCG@10 for runs on the document retrieval task

**(b)** NDCG@10 for runs on the passage retrieval task

**(c)** NCG@100 for runs on the document retrieval task

**(d)** NCG@1000 for runs on the passage retrieval task

**Figure 8.6:** Analyzing the impact of "fullrank" *vs.* "rerank" settings on retrieval performance. Figure (a) and (b) show the performance of different runs on the document and passage retrieval tasks, respectively. Figure (c) and (d) plot the NCG@100 and NCG@1000 metrics for the same runs for the two tasks, respectively. The runs are ordered by their NDCG@10 performance along the $x$-axis in all four plots. We observe, that the best run under the "fullrank" setting outperforms the same under the "rerank" setting for both document and passage retrieval tasks—although the gaps are relatively smaller compared to those in Figure 8.2. If we compare Figure (a) with (c) and Figure (b) with (d), we do not observe any evidence that the NCG metric is a good predictor of NDCG@10 performance.

grows, we would see a larger number of runs with deep learning as an ingredient for phase 1 in future editions of this TREC track.

**NIST labels *vs.* Sparse MS MARCO labels.** Our baseline human labels from MS MARCO often have one known positive result per query. We use these labels for training, but they are also available for test queries. Although our official evaluation uses NDCG@10 with NIST labels, we now compare this with reciprocal rank (RR) using MS MARCO labels, and MRR using NIST labels. Our goal is to understand how changing the labeling scheme and metric affects the overall results of the track,

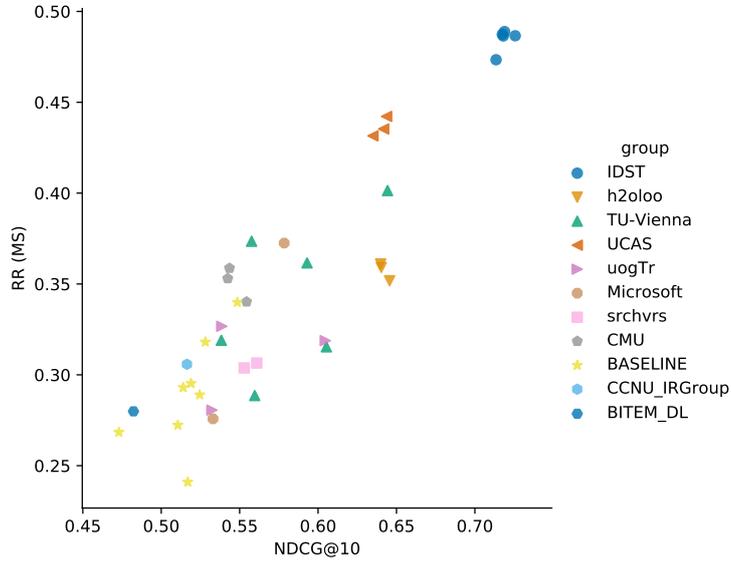

**(a)** Document retrieval task.

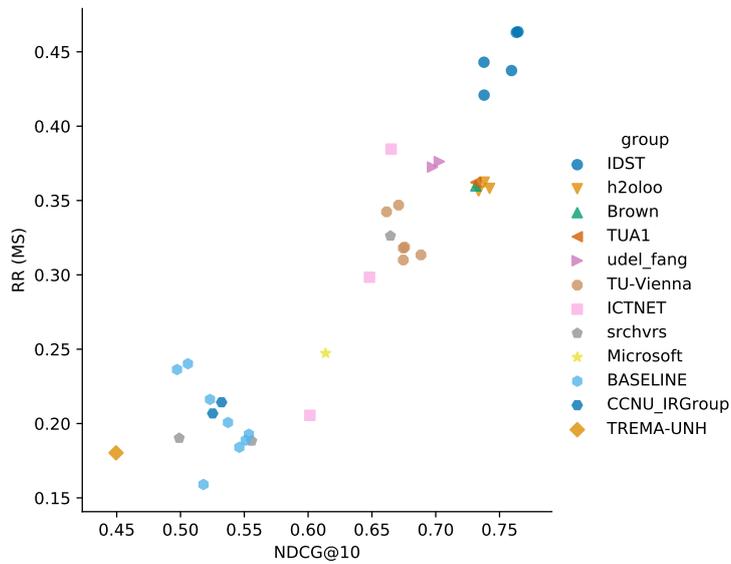

**(b)** Passage retrieval task.

**Figure 8.7:** Metrics agreement scatter plot, broken down by group. MRR (MS) is reciprocal rank calculated with the sparse MS MARCO labels, while NDCG@10 is calculated using NIST labels.

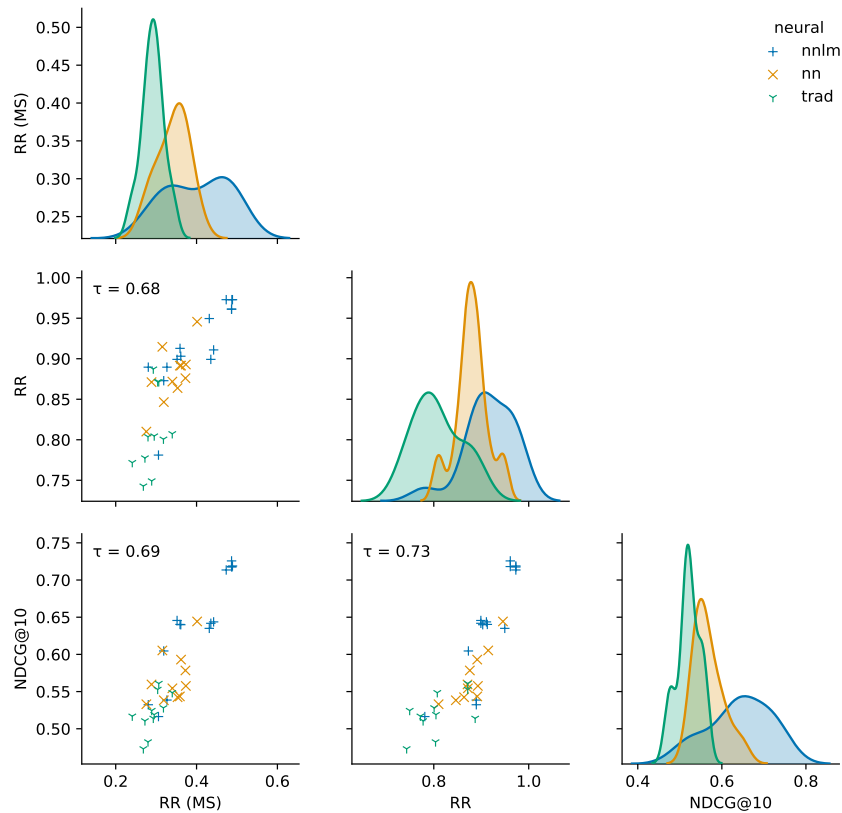

**Figure 8.8:** Metrics agreement analysis, broken down by model type, for the document retrieval task. Kendall correlation ($\tau$) indicates agreement between metrics on system ordering. MRR (MS) is calculated using MS MARCO sparse labels, while MRR and NDCG@10 are calculated using NIST labels.

but if there is any disagreement we believe the NDCG results are more valid, since they evaluate the ranking more comprehensively and a ranker that can only perform well on labels with exactly the same distribution as the training set is not robust enough for use in real-world applications, where real users will have opinions that are not necessarily identical to the preferences encoded in sparse training labels.

In Figure 8.8 and 8.9, We observe general agreement between results using MS MARCO and NIST labels–*i.e.*, runs that perform well on MS MARCO-style evaluation also tends to achieve good performance when evaluated under traditional TREC settings, and vice versa. This is good news, validating the MS MARCO leaderboard results are at least somewhat indicative of results that are found with pooled judging.

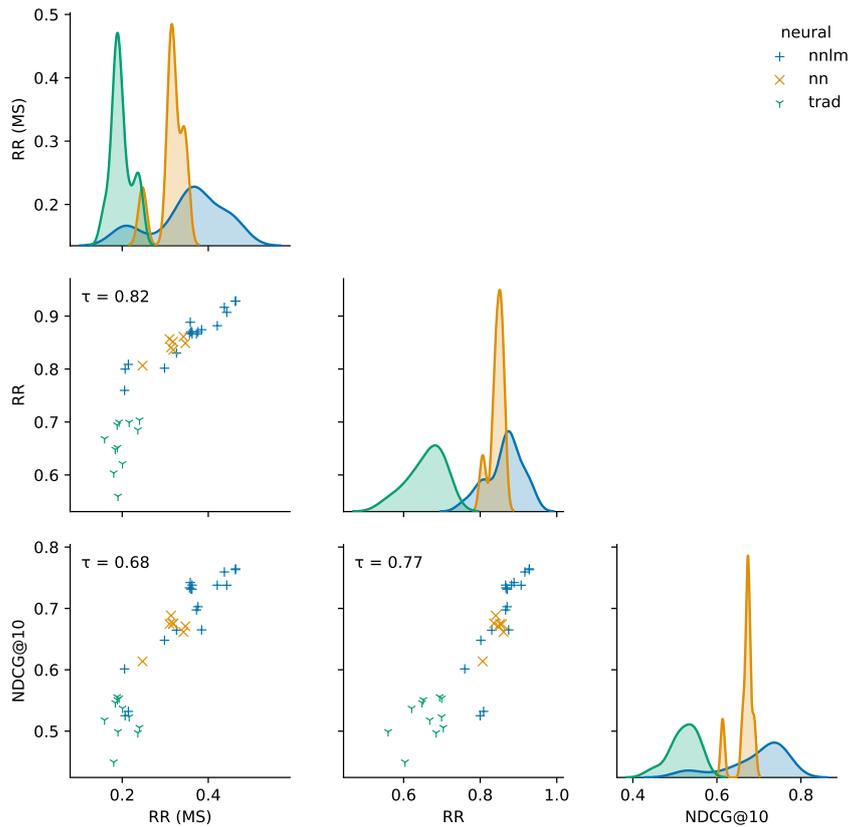

**Figure 8.9:** Metrics agreement analysis, broken down by model type, for the passage retrieval task. Kendall correlation ($\tau$) indicates agreement between metrics on system ordering. MRR (MS) is calculated using MS MARCO sparse labels, while MRR and NDCG@10 are calculated using NIST labels.

## 8.4 Conclusion

The TREC 2019 Deep Learning Track introduced two large training datasets, for a document retrieval task and a passage retrieval task, generating two ad hoc test collections with good reusability. For both tasks, in the presence of large training data, non-neural network runs were outperformed by neural network runs. Among the neural approaches, the best-performing runs tended to use transfer learning, employing a pretrained language model such as BERT. In future it will be interesting to confirm and extend these results, understanding what mix of data and multi-stage training lead to the best overall performance.

We compared reranking approaches to end-to-end retrieval approaches, and there was not a huge difference, with some runs performing well in both regimes. This is another result that would be interesting to track in future, since we would expect that end-to-end retrieval should perform better if it can recall documents that are unavailable in a reranking subtask.

In the first year of the track there were not many non-neural runs, so it would be important in subsequent year's track to see more runs of all types, to further understand the relative performance of different approaches. Although the test collections are of high quality, meaning that they are likely to give meaningful results when reused, overfitting can still be a problem if the test set is used multiple times during the development of a new retrieval approach. The most convincing way to show that a new approach is good is to submit TREC runs. There is no chance of overfitting, or any kind of repeated testing, because the test labels are not generated until after the submission deadline. Through a combination of test collection reuse (from past years) and blind evaluation (submitting runs) the Deep Learning Track is offering a framework for studying ad hoc search in the large data regime.

# Chapter 9

# General Conclusions

> হাঁস ছিল, সজারু, (ব্যাকরণ মানি না),
>
> হয়ে গেল 'হাঁসজারু' কেমনে তা জানি না।
>
> Was a duck, porcupine (to grammar I bow not)
>
> Became Duckupine, but how I know not.
>
> ― *Sukumar Ray, Khichuri*
>
> *(Translation by Prasenjit Gupta)*

Unlike traditional IR methods, where relevance is estimated largely by counting occurrences of query terms in document text, the neural methods described in this thesis focus on learning useful text representations guided by optimization objectives that correspond to tasks such as ranking and language modeling. Based on the empirical evidence presented in this thesis—and the substantial body of neural IR literature that has been emerging over the recent years—it is safe to conclude that these representation learning methods are able to demonstrate sizeable improvements over traditional IR methods in the presence of large training corpora. Ongoing new research efforts in this area may be concerned with further improving result quality (*effectiveness*) while lowering compute and memory costs (*efficiency*), and even coming up with more elaborate measures of successful retrieval outcomes (*e.g.*, *exposure*-based metrics) that these models can be optimized towards.

However, this emerging family of neural methods may be causing more fundamental shifts in the field of IR. For example, we argue that after at least two decades of largely unsuccessful attempts at leveraging models and artifacts from NLP to improve IR tasks [477–481], we are now witnessing surprisingly huge benefits from applications of deep NLP models in retrieval. These new NLP artifacts, however, are not in the form thesauri or parts of speech tags, but rather in the form of pre-trained language models and latent text representations. While, these black box language models may pick up certain linguistic regularities from training on large corpora, it is also possible, if not likely, that these learned latent representations encode relationships and attributes that are very different to our own notion of linguistic properties. By simply modeling observed regularities in unlabeled corpora, a language model may in fact learn that "duck" and "porcupine" are similar given they appear in similar contexts—such as, "how much does a duck weigh?" and "how much does a porcupine weigh?". If our goal is to maximize some averaged relevance metrics for a query auto-completion task, it may indeed be reasonable that "duck" and "porcupine" have similar latent representations. Similarly, the latent space may be able to encode seemingly nonsensical concepts such as a "duckupine" even if it has no meaningful counterpart in the real world, except may be in literary fiction (see Figure 9.1).

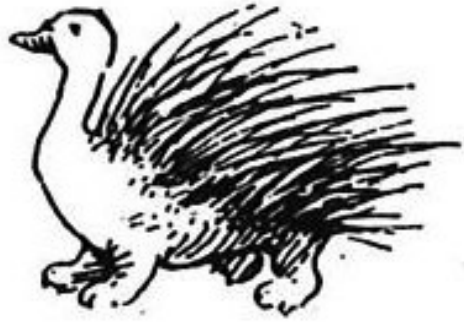

**Figure 9.1:** Sukumar Ray's illustration of a ``হাঁসজারু'' (pronounced: "haashjaru") or a duck-upine, a fictional animal from his poem "Khichuri".

This poses an interesting challenge for the research community. While, we are reasonably good at measuring how effective these black box models are at improving retrieval, it is significantly harder to articulate exactly what knowledge and world view these models encode (and do not encode), and even more difficult to quantify the progress the IR community is making with regards to better understanding of retrieval tasks from the application of these models. This is not to imply that the learned latent representations must be perfectly interpretable to qualify as scientific progress, but rather we are making a case for viewing the contributions of neural IR through a much broader lens that encourages its usage to aid the development of new IR theory and improved understanding of retrieval tasks.

On that note, we conclude this thesis by summarize the contribution of our own work, as described in the earlier chapters, in Section 9.1, and identifying key future challenges and opportunities for the field in Section 9.2.

## 9.1 A summary of our contributions

This thesis summarizes a substantial body of work on neural methods for text retrieval. We ground our contributions by presenting a thorough survey of the field. We highlight the challenges that are unique to IR and use them to motivate novel learning approaches and model architectures.

We begin with Duet—a neural model that gathers evidence of a document's relevance to a query by inspecting patterns of query term matches in the document as well as learning latent query and document representation for matching. The proposed model achieves state-of-the-art performance on several public and proprietary benchmarks—on IR tasks that involve ranking long text documents or short passages. The performance of the model is particularly promising when large quantities of examples are available for training.

The scope of impact of neural IR models is limited, if restricted only to late stage re-ranking. Therefore, we incorporate a query term independence assumption to re-design the Duet model. The re-architected model is amenable to full precomputation while retaining all the effectiveness of the original Duet architecture. This opens the opportunity to employ deep neural models, like Duet and BERT-based ranking, for efficient retrieval from the full collection.

While, learning to rank methods traditionally focus on producing a static rank-ing, we also explore an optimization strategy for stochastic ranking. We argue that in real world retrieval systems, it makes sense to measure and optimize towards expected exposure of retrieved items, in the pursuit of fairness and diversity related outcomes.

We demonstrate the usefulness of deep neural network based approaches to IR tasks beyond document and passage retrieval, such as query auto-completion and session modeling. Finally, we initiate a large-scale benchmarking effort for neural IR methods at TREC and report our key findings.

The body of work described in this thesis was not conducted in isolation. We conducted several other studies, in collaboration, focused on neural IR that we do not describe here. These efforts focused on exploring schemes for explicit regularization [141, 144] during model training, studying reinforcement learning based approaches to retrieval [79], designing neural ranking models for structured documents [389], prototyping proactive retrieval systems [62], and even contributing to general purpose neural toolkits [339].

## 9.2 The Future of neural IR

An ideal IR model would be able to infer the meaning of a query from context. Given a query about the Prime Minister of UK, for example, it may be obvious from context whether it refers to John Major or Teresa May—perhaps due to the time period of the corpus, or it may need to be disambiguated based on other context such as the other query terms or the user's short or long-term history. If the model learns a representation that encodes this context, perhaps making Prime Minister close to Teresa May in a latent space, it is like a *library*. To scale to a large corpus, this memorization would need to cover a massive number of connections between entities and contexts, which could potentially be limited by model capacity. Memorization could also cause update problems—*e.g.*, if there is a new Prime Minister but the model still refers to the old one—or encode problematic societal biases [482]. To avoid these problems, another design could avoid memorizing connections in the corpus, and instead perform some per-query process that reads the corpus and perhaps even reasons about the content, like a *librarian*.

Many of the breakthroughs in deep learning have been motivated by the needs of specific application areas. Convolutional neural networks, for example, are commonly employed by the vision community, whereas recurrent architectures find more applications in speech recognition and NLP. It is likely that the specific nature of IR tasks and data will inform our choice of neural architectures and drive us towards new designs. Future IR explorations may also be motivated by developments in related areas, such as NLP. Neural architectures that have been evaluated on non-IR tasks [483–487] can be investigated in the retrieval context. New methods for training neural IR models—*e.g.*, using reinforcement [79, 488, 489] or adversarial learning [141, 289]—may also emerge as important directions for future explorations. In particular, large scale unsupervised training of language models—*e.g.*, BERT [327]—have already demonstrated significant jump in retrieval performance on public benchmarks [15].

However, given the pace at which the area of deep learning is growing, in terms of the number of new architectures and training regimes, we should be wary of the combinatorial explosion of trying every model on every IR task. We should not disproportionately focus on maximizing quantitative improvements and in the process, neglect theoretical understanding and qualitative insights. It would be a bad outcome for the field if these explorations do not grow our understanding of the fundamental principles of machine learning and information retrieval. Neural models should not be the hammer that we try on every IR task, or we may risk reducing every IR task to a nail.[1] Rather, these new models should also be the lens through which researchers gain new insights into the underlying principles of IR tasks. This may imply that sometimes we prefer neural models that, if not interpretable, then at least are amenable to analysis and interrogation. We may elicit more insights from simpler models while more sophisticated models may achieve state-of-the-art performances. As a community, we may need to focus on both to achieve results that are both impactful as well as insightful.

---

[1] `https://en.wikipedia.org/wiki/Law_of_the_instrument`

IR also has a role in the context of the ambitions of the machine learning community. Retrieval is key to many one-shot learning approaches [490, 491]. Ghazvininejad et al. [492] proposed to "search" external information sources in the process of solving complex tasks using neural networks. The idea of learning local representations proposed by Diaz et al. [139] may be applicable to non-IR tasks. While we look at applying neural methods to IR, we should also look for opportunities to leverage IR techniques as part of—or in combination with—neural and other machine learning models.

We must also renew our focus on the fundamentals, including benchmarking and reproducibility. An important prerequisite to enable the "neural IR train" to steam forward is to build shared public resources—*e.g.*, large scale datasets for training and evaluation, and repository of shared model implementations—and to ensure that appropriate bindings exist (*e.g.*, [348, 349]) between common IR frameworks and toolkits from the neural network community. At the time of writing this thesis, we are preparing for the third edition of the TREC Deep Learning track [15, 16] using the MS MARCO [52] dataset. On-going analysis of runs submitted to the TREC Deep Learning track and the MS MARCO leaderboards not only reveal how different approaches perform competitively, but also provide more nuanced insights into different failure modes [493]. We hope that these datasets and others would assume the same critical role in fueling neural IR progress as the ImageNet database [494] in the computer vision community.

The emergence of new IR tasks also demands rethinking many of our existing metrics. The metrics that may be appropriate for evaluating document ranking systems may be inadequate when the system generates textual answers in response to information seeking questions. In the latter scenario, the metric should distinguish between whether the response differs from the ground truth in the information content or in phrasing of the answer [53, 495, 496]. As multi-turn interactions with retrieval systems become more common, the definition of task success will also need to evolve accordingly. A good Neural IR research agenda should not only focus on novel techniques, but also encompass all these other aspects.

# Appendix A

# Published work

The first three chapters of this thesis has been separately published as the following peer-reviewed book.

1. **Bhaskar Mitra** and Nick Craswell. An introduction to neural information retrieval. Foundations and Trends® in Information Retrieval, Now Publishers, 2018.

In addition, Bhaskar Mitra published the following papers, listed in reverse chronological order, that forms the broader foundation of this PhD:

2. Jaime Arguello, Adam Ferguson, Emery Fine, **Bhaskar Mitra**, Hamed Zamani, and Fernando Diaz. Tip of the Tongue Known-Item Retrieval: A Case Study in Movie Identification. In Proc. CHIIR, 2021 (to appear).

3. Fernando Diaz, **Bhaskar Mitra**, Michael Ekstrand, Asia J. Biega, and Ben Carterette. Evaluating Stochastic Rankings with Expected Exposure. In Proc. CIKM, 2020. **Best long paper nominee.**

4. Hamed Zamani, **Bhaskar Mitra**, Everest Chen, Gord Lueck, Fernando Diaz, Paul Bennett, Nick Craswell, and Susan Dumais. Analyzing and Learning from User Interactions for Search Clarification. In Proc. SIGIR, ACM, 2020.

5. Sebastian Hofstätter, Hamed Zamani, **Bhaskar Mitra**, Nick Craswell, and Allan Hanbury. Local Self-Attention over Long Text for Efficient Document Retrieval. In Proc. SIGIR, ACM, 2020.

6. Emine Yilmaz, Nick Craswell, **Bhaskar Mitra**, and Daniel Campos. On the Reliability of Test Collections to Evaluating Systems of Different Types. In Proc. SIGIR, ACM, 2020.